%% file: paper.tex
\documentclass[twoside]{IEEEtran}

\usepackage{amsmath}
\usepackage{amssymb}

\usepackage{tikz}
\usepackage{float}
\usepackage{multirow}
\usepackage{tabularx}	
\usepackage{pgfplots}
\pgfplotsset{compat=newest}
\pgfplotsset{plot coordinates/math parser=false}
\newlength\figureheight
\newlength\figurewidth

\usepackage{optidef}
\usepackage{algorithm}
\usepackage{algpseudocode}

\usepackage{amsthm}
\newtheorem{remark_new}{Remark}
\newtheorem{assumption_new}{Assumption}

\usepackage{comment}

\usepackage{acro}

\include{acronyms}

\usepackage[sort&compress,square,comma,numbers]{natbib}

\usepackage[colorlinks=true,urlcolor=blue,citecolor=blue,linkcolor=blue,bookmarks=true]{hyperref}
 
\begin{document}
\title{Energy Aware and Safe Path Planning for Unmanned Aircraft Systems}

\author{Sebastian~Gasche, Christian~Kallies, Andreas~Himmel, and Rolf~Findeisen\thanks{S. Gasche and C. Kallies are with the Institute of Flight Guidance, German Aerospace Center, Brunswick, Germany.}\thanks{S. Gasche and R. Findeisen are with the Technical University of Darmstadt, Darmstadt, Germany.}}

\markboth{PREPRINT: This work has been submitted to the IEEE for possible publication. Copyright may be transferred without notice.}%
{S. Gasche \MakeLowercase{\textit{et al.}}: Energy Aware and Safe Path Planning for Unmanned Aircraft Systems}

\maketitle

\begin{abstract}
This paper proposes a path planning algorithm for multi-agent unmanned aircraft systems (UASs) to autonomously cover a search area, while considering obstacle avoidance, as well as the capabilities and energy consumption of the employed unmanned aerial vehicles. The path planning is optimized in terms of energy efficiency to prefer low energy-consuming maneuvers. In scenarios where a UAS is low on energy, it autonomously returns to its initial position for a safe landing, thus preventing potential battery damage. 
To accomplish this, an energy-aware multicopter model is integrated into a path planning algorithm based on model predictive control and mixed integer linear programming. Besides factoring in energy consumption, the planning is improved by dynamically defining feasible regions for each UAS to prevent obstacle corner-cutting or over-jumping.
\end{abstract}


\IEEEpeerreviewmaketitle
\section{Introduction}
\label{cha:Introduction}

In recent years, \acp{uas} have attracted rising interest due to the wide range of scientific and commercial applications. These include among other things the fields of surveillance, search-and-rescue \citep{Hildmann(2019),Mirzaei(2011),Sampedro(2019)}, inspection \citep{Steich(2016)}, meteorological monitoring \citep{Korolkov(2018)}, mapping \citep{Luo(2008)}, as well as the transport of packages, data, or passengers \citep{Aurambout(2019)}. \acp{uas} are mainly used for applications that are too tedious, dangerous, dirty, or expensive to operate with manned aircraft. Especially rotary-wing \acp{uav}, e.g. multicopters or helicopters, are highly suitable for most of the mentioned applications due to their high maneuverability and hovering capabilities. In the upcoming years as the world advances (from fossil fuels) toward clean energy, \acp{uas} will be especially important due to their zero-emission potential. 

However, \acp{uas} face critical obstacles, with the most prominent one being the energy performance. The limited onboard energy of a \ac{uav}, whose source is commonly a \acl{lib}, strongly restricts the class of missions a \ac{uas} can successfully carry out since it limits the \ac{uav}'s endurance, flight time, range, and payload. Especially rotary-wing \acp{uav} consume large amounts of energy to remain in flight. Therefore, research on enhancing the energy efficiency of \acp{uas} is essential to occupy their full ecological and economic potential.
\citet{Karydis(2017)} review several approaches for enhancing the energy efficiency of small-scale \acp{uav}. They point out that energy performance optimization is achieved either by hardware-based or algorithm-based optimization. \textit{Hardware-based optimization} deals with optimizing the structure and design of a \ac{uav} to reduce weight, e.g., by utilizing light-weight manufacturing materials, careful component selection, or structural redesign. Examples for structural redesignes are proposed in \citep{Xiong(2019),Ryll(2015),Morbidi(2018),Driessens(2015)}. \textit{Algorithm-based optimization} deals with energy-aware motion planning and control to reduce energy consumption and extend flight times, e.g., by planning energy-efficient flight trajectories and preferring low energy-consuming maneuvers. Algorithm-based approaches are divided into model-free and model-based approaches.  While model-free approaches are superior in considering hard-to-model and less-known effects, e.g., environmental disturbances, performance changes, or aerodynamical changes due to a payload \citep{Tagliabue(2019),Kreciglowa(2017),DiFranco(2015)}, model-based approaches allow considering vehicle capabilities. For example, the optimal control of quadcopter \acp{uav} is introduced in \citep{Fouad(2017),Li(2022),Lu(2018),Morbidi(2016),Yacef(2017),Yacef(2020)}. Commonly, hardware-based optimization needs extensive development, while algorithm-based optimization is more flexible and can be implemented in existing systems. Driven by ecological and economic considerations, we propose a model- and algorithm-based energy performance optimization by planning energy-efficient paths.

Over the years, several path planning approaches have been proposed, ranging from classical to more advanced optimization-based approaches, which differ in their capabilities, computational efficiency, robustness and how they conceptualize the path planning problem.
\textit{Classical approaches} to path planning, such as potential field methods and graph search algorithms, treat path planning as a purely mathematical problem. While classical methods are well-suited for known static environments, they are computationally expensive, require precise information on the environment and are rigid, failing to adapt well to dynamic or uncertain settings, constraining their usage in real-time applications. 
Potential field methods, for example, model the goals, obstacles, and other boundary conditions by potentials, which are accumulated to define the potential field. The vehicle navigates through the environment by minimizing the potential field's gradient to reach the target, where the lowest potential is located. This approaches are computationally lightweight, simple to implement and configurable. However, they tend to converge into local minima, unless augmented by correction methods like the waterfall or wall-following methods. Potential field methods also require accurate and detailed environmental information, making them more suitable for static, known environments \citep{Patle2019,Mac2016,Strobel(2016)}.
In contrast, graph search methods reduce path planning to a graph traversal problem, where nodes represent positions and edges represent feasible paths between these positions. The path between initial position (initial node) and the target position (target node) is described by a sequence of connected nodes. For graph generation commonly Voronoi graphs, visibility graphs or cell decomposition approaches are used. Popular graph search algorithms such as Dijkstra \citep{Dijkstra1959}, \(A^*\) \citep{Hart1968}, \(D^*\) \citep{Lee2010}, Kruskal \citep{Kruskal1956}, or Prim \citep{Prim1957} are efficient in finding optimal paths in structured environments, such as grid-based or roadmap-based setups. Due to the abstraction of the environment into a graph, these methods only guarantee an optimal solution through the graph. Further, their computational cost can rise significantly with increased environment complexity \citep{Patle2019,Mac2016,Strobel(2016)}.

\textit{Sampling-based} approaches like probabilistic roadmaps \citep{Kavraki1996} and rapidly-exploring random trees \citep{Kuffner2000}, offer a probabilistic perspective on path planning. These methods explore feasible paths by randomly sampling the environment and are particularly effective in high-dimensional spaces, while handling complex environments with motion constraints. While being computationally efficient and scalable to complex environments, they provide a feasible but often non-optimal solution due to their probabilistic nature. Consequently, post-processing for path smoothing is needed, for example, by repeatedly planning the paths, while only keeping the most optimal one. However, this possibility is limited in real-world application due to limited computational resources. Due to their online resampling capabilities, sampling-based methods are effective in dynamic environments, but commonly can not guarantee feasibility in presence of uncertainties \citep{kingston2018}.

\textit{Heuristic approaches} include methods inspired by nature, such as neural networks, fuzzy logic, genetic algorithms, and swarm intelligence optimization, which commonly treat the path planning problem as an optimization problem of some kind. 
Neural networks, for instance, are inspired by the human brain and its learning capability, treating the path planning problem as a learning task, where models are trained with previous experience (training data), using learning methods such as supervised or reinforced learning. The resulting model is capable of decision making and path planning depending on state and environmental information gathered by sensors. Some approaches also allow for online learning, which makes these approaches especially well-suited for dynamic and partially unknown environments. However, neural networks require extensive training data and computational power during the training. Further, their robustness in unseen environments remains a challenge \citep{Patle2019,Mac2016,Popovic2024}. 
Fuzzy logic systems, on the other hand, mimic human reasoning by using linguistic if-then rules for decision making. These systems are especially suitable for uncertain conditions, offering adaptability and robustness in the face of sensor inaccuracies or uncertain environmental conditions. However, designing appropriate rules and membership functions can be complex, especially for high-dimensional spaces where the number of rules increases significantly, decreasing the efficiency of this method \citep{Patle2019,Mac2016}.
Combining human reasoning and learning ability, neuro-fuzzy systems were developed to create a robust and flexible path planning approach. However, the combination of neural networks and fuzzy logic also combines their disadvantages and is computationally intensive \citep{Mac2016}.
Genetic algorithms and swarm intelligence methods such as \ac{pso} and \ac{aco} treat path planning as a population-based search problem. Generic algorithms utilizes evolution theory by evolving a population of candidate paths over multiple generations to find the best solution based on pre-defined selection criteria. Meanwhile \ac{pso} and \ac{aco} are inspired by the swarm behaviour of bird flocks and ants, respectively. Here agents update their positions based on individual and swarm experiences. Together the agents explore the environment and find possible paths, converging to the best solution. While these methods are well-suited for global optimization in static environments, they can struggle with premature convergence and are inefficient in high-dimensional dynamic settings \citep{Patle2019,Mac2016}.

Numerous path planning approaches, treat path planning in similar ways as the mentioned approaches or combine some of them to achieve more robust path planning in dynamic and uncertain environments. However, this often requires more computational resources and introduces more complexity, or other disadvantages, which have to be balanced \citep{Patle2019,Mac2016}.

Representing an advanced \textit{optimization-based approach}, \ac{mpc} treats the path planning problem as a constrained optimization task, generating feasible optimal paths within a moving horizon by minimizing an objective function.  \ac{mpc}'s ability to incorporate constraints, originating from the vehicle dynamics, the environment, limitations and uncertainties as well as the ability to consider multiple objectives, e.g  mission success and safety and energy efficiency, makes it effective for advanced path planning tasks. \ac{mpc} is especially well-suited for dynamic and uncertain environments, offering robustness by continuously updating the solution as new data is received. Moreover, \ac{mpc}'s flexibility enables integration with \ac{mip}, further expanding its applicability to decision-making tasks in complex scenarios, e.g. if the goal is to cover an area instead of reaching a specific position. While early implementations of \ac{mpc} faced computational challenges, recent advances in optimization algorithms and computational power have significantly enhanced its real-time capabilities, which can be further improved by linear and convex reformulation of the optimization problem.  Nonetheless, the safety guarantees and robustness provided by \ac{mpc} make it one of the most promising approaches for real-time path planning \citep{Wei2022,Malyuta2022,Quirynen2024}. Due to the flexibility and safety guarantees of \ac{mpc}, this study introduces a moving-horizon multi-agent \ac{ppa} for area coverage based on \ac{mpc} and \ac{milp} by further developing the \ac{ppa} proposed by \citep{Ibrahim(2020)}. It considers the environment, vehicle dynamics and limitations, while focusing on energy-efficient and safe path planning. 

Our contribution is divided into two parts: First, the modeling of multicopter \acp{uav}, considering their energy consumption, which we derived in \citep{Gasche2024_ECM} and shortly present in Sections \ref{cha:UAV}. Second, the development of the \ac{ppa}, which is presented in Section \ref{cha:PPA}. As application scenario, Section \ref{cha:Simulation} presents simulation results for a search-and-rescue scenario, deploying a \ac{uas} swarm to cover an area after a major flooding event and search for injured people with on-board cameras to assist the rescue team. Lastly, we discuss this work's developments in Section \ref{cha:Discussion} and conclude the contributions in Section \ref{cha:Conclusion}.
 
\section{Energy Aware Multicopter Model}
\label{cha:UAV}
\begin{figure}[H]
\centering
\input{Tikz_Energy_Consumption_0_short}
\vspace{-0.25cm}
\caption{Simplified power train of an electric-propelled \ac{uav} \cite{Gasche2024_ECM}}
\label{fig:EC_0}
\end{figure}
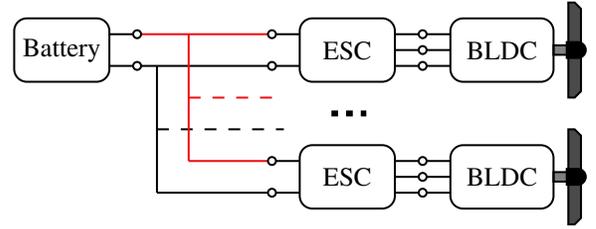

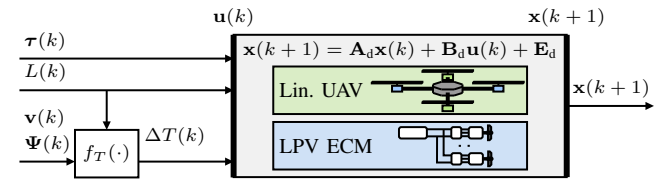
\begin{figure}[H]
\centering
\input{Tikz_Control_Setup_short}
\vspace{-0.25cm}
\caption{Structure of the linear energy aware multicopter model \cite{Gasche2024_ECM}}
\label{fig:C_Setup}
\end{figure}

The terms \ac{uas} and \ac{uav} are often used as acronyms to describe the same system. In this study, we use the term \ac{uas} to identify a system consisting of a \ac{uav}, a ground station, and a communication system to transfer data from the \ac{uav} to the ground station and vice versa. The term \ac{uav} refers to the vehicle itself, which may be piloted by a controller or fly autonomously. Moreover, a \ac{uas} swarm consists of several \acp{uav} that share the ground station and communication system. \acp{uav} come in a variety of designs and aerodynamic configurations. Each type of \ac{uav} is suitable for different applications and has advantages and disadvantages. In the following, we will look at multicopter \acp{uav} because of their high maneuverability. Their ability to take off and land vertically and to hover makes them ideal for surveillance or monitoring missions in cluttered and dynamic environments.
\vspace{-0.25cm}
\begin{figure}[H]
\centering
\input{Tikz_Quadcopter_Frames}
\vspace{-0.25cm}
\caption{Frames of reference (black: inertial frame, red: body-fixed frame); Forces/torques acting on the body's center of mass (blue) \cite{Gasche2024_ECM}}
\label{fig:QC_Frames}
\end{figure}
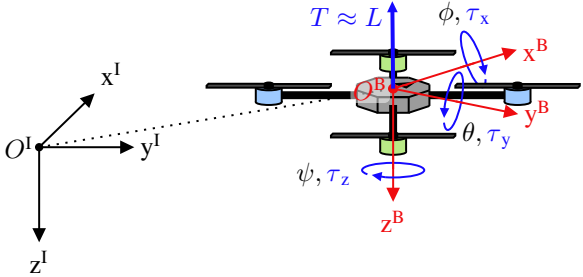
\vspace{-0.25cm}
The discrete-time linear multicopter model considering energy consumption is derived in \cite{Gasche2024_ECM}. Two reference frames are defined, as shown in Fig.~\ref{fig:QC_Frames}. The inertial frame, with origin \(O^\text{I}\), is fixed to earth's surface and its axes are aligned north (\(\text{x}^\text{I}\)), east (\(\text{y}^\text{I}\)), and down (\(\text{z}^\text{I}\)). The body-fixed frame, with origin \(O^\text{B}\) at the multicopter's center of mass, has its axes pointing forward (\(\text{x}^\text{B}\)), right (\(\text{y}^\text{B}\)), and down (\(\text{z}^\text{B}\)). The position \(\mathbf{p} = (x,y,z)^\top\) and velocity \(\mathbf{v} = (v_\text{x},v_\text{y},v_\text{z})^\top\) are defined in the inertial frame, while the orientation, given by Euler angles \(\boldsymbol{\Psi} = (\phi,\theta,\psi)^\top\), represents the rotation between the frames. The angular velocity \(\boldsymbol{\omega} = (\omega_\text{x},\omega_\text{y},\omega_\text{z})^\top\) defines the rotation rates in the body-fixed frame. 
The motion is controlled by thrust \(T\) and torques \(\boldsymbol{\tau} = (\tau_\text{x},\tau_\text{y},\tau_\text{z})^\top\). The model is linearized around the hover state, where thrust \(T\) balances the weight force and the multicopter maintains its position. Due to the decoupling of horizontal and vertical dynamics, the thrust \(T\) is replaced with the lift \(L\), which acts in the \(\text{z}^\text{I}\)-direction and equals \(T\) at the set point. It is then discretized using a Taylor-Lee series with a discretization order of \(N_{\text{dis}} \geq 2\) to account for higher model dynamics. Due to several limitations on the validity of the linearized system dynamics, we state:
\begin{assumption_new}
\label{ass:UAV_1}
The multicopter is axis-symmetric with a nearly spherical body. Its \(N_\text{M}\) identical motors and rotors are arranged around the center of mass equally spaced by \(2\pi/N_\text{M}\).
\end{assumption_new}
For the multicopter's \ac{ecm}, we derive models for the power train components and combine them as shown in Fig.~\ref{fig:EC_0}. These components include the the \acl{lib} (\ac{lib}), the \ac{bldc} motors with attached rotors and the \acp{esc}. The battery dynamics are characterized by the depth of discharge \(\mathrm{DoD}\) and the polarization voltage \(u_\text{th}\), which derives from the Thevenin model, used as \ac{lib} cell model. The current state of the battery is described by the state of charge \(\mathrm{SoC}\), battery voltage \(u_\text{b}\), and current \(i_\text{b}\). To enhance model accuracy, the nonlinear battery discharge curve is approximated with piece-wise linear functions, resulting in a \ac{lpv} model. Although the \ac{ecm} is derived based on the motor speeds \(\Omega_i, \ \ i \in \{1,\dots,N_\text{M}\}\) of the \ac{bldc} motors, the linearization around the hovering state with a fully charged battery allows for reducing the input of the \ac{ecm} to the combined thrust of the rotors \(T\).
\vspace{-0.25cm}
\vspace{-0.25cm}
Fig.~\ref{fig:C_Setup} shows the resulting time-discrete \ac{lpv} model
\begin{equation}
\label{Vehicle_Model}
\mathbf{x}(k+1) = 
\mathbf{A}_\text{d}\mathbf{x}(k) + 
\mathbf{B}_\text{d}\mathbf{u}(k) + 
\mathbf{E}_\text{d}.
\end{equation}
Here, the state 
\begin{equation*}
\mathbf{x} = (x,y,z,v_\text{x},v_\text{y},v_\text{z},\phi,\theta,\psi,\omega_\text{x},\omega_\text{y},\omega_\text{z},\mathrm{DoD}, u_\text{th})^\top
\end{equation*}
includes the multicopter's position \(\mathbf{p}\), velocity \(\mathbf{v}\), orientation \(\boldsymbol{\Psi}\), angular velocity \(\boldsymbol{\omega}\), as well as the battery's depth of discharge \(\mathrm{DoD}\), and polarization voltage \(u_\text{th}\). 
The input 
\begin{equation*}
\mathbf{u} = (L,\tau_\text{x},\tau_\text{y},\tau_\text{z},\Delta T)^\top,
\end{equation*}
contains the multicopter's controllable torques \(\boldsymbol{\tau}\) as well as the lift \(L \approx T - m\,\text{g}\), which represents the deviation of the upwards pointing thrust component from the set point. For the \ac{ecm} the corrected deviation of the thrust \(\Delta T = T - m\,\text{g}\) from the set point is included.
Moreover, the model includes the state-space matrices \(\mathbf{A}_\text{d}\), \(\mathbf{B}_\text{d}\), \(\mathbf{C}_\text{d}\), \(\mathbf{D}_\text{d}\), and the offset matrix \(\mathbf{E}_\text{d}\), which accounts for the energy consumption during hovering. These matrices change depending on the depth of discharge \(\mathrm{DoD}\) to approximate the nonlinear discharge curve of the battery.
\vspace{-0.25cm}
\vspace{-0.25cm}
To represent the vehicle and power train capabilities while protecting the battery from damage, we apply the constraints outlined in \cite{Gasche2024_ECM}. We also introduce a corrected formulation of the thrust \(T\) for the \ac{ecm}, which depends on the lift \(L\), the orientation \(\boldsymbol{\Psi}\), and velocity \(\mathbf{v}\). This correction improves the approximation of efficiency gains in forward flight and captures increased power consumption during aggressive maneuvers or high-velocity flight, which are not represented when using the lift \(L\) as input for the \ac{ecm}.

\section{Path Planning Algorithm for Area Coverage}
\label{cha:PPA}  
This section presents the moving-horizon \acl{ppa} (\acs{ppa}) based on \acl{mpc} (\acs{mpc}) and \acl{milp} (\acs{milp}). This concept was initially presented in \citep{schouwenaars2006safe} and revisited in \citep{Ibrahim(2020)}. \citet{kogel2023safe} proceeded to investigate the concept with the objective of enhancing its robustness and performance in order to facilitate real-time applications. With a similar goal, \citet{elsayed2023generic} reformulated the optimization problem to optimize over a set of generic motion primitives.  The \ac{ppa} proposed in this section is based on \citep{Ibrahim(2020)} and further developed to plan paths for a \ac{uas} swarm in \ac{3D} space, considering the \acp{uas}’ physical capabilities and obstacle/collision avoidance, while preferring low energy-consuming maneuvers. The mission goal is to cover a predefined search area. In reality, this could be a search and rescue mission after a mayor flooding event, where the \ac{uas} swarm utilizes cameras to search for injured people and provide additional information to the rescue teams. Additionally, the path planning is improved in already covered areas and returns \acp{uas} low on energy to their initial positions. Where, they deactivate themselves to avoid damage to the batteries. An earlier version of the \ac{ppa} is used in \citep{kallies2024multi} to plan \ac{2D} paths for indoor applications. Meanwhile, in \citep{hagag2024energy}, the \ac{ppa} is adjusted to plan paths for passenger transportation missions in future airtaxi services in cluttered urban environments. In the following, we provide a brief overview of the fundamentals of \ac{mpc}, followed by introducing special formulations in \ac{milp} that are used to linearize common nonlinear functions. Finally, we present the \ac{ocp} of the \ac{mpc}.

\subsection{Fundamentals of Model Predictive Control}
In the following, we briefly review the fundamental concept of \ac{mpc}, which is an online optimization-based feedback control strategy and part of the optimal control strategies. It employs a mathematical model of the system and an objective, consisting of an objective function and constraints, to formulate a finite horizon \ac{ocp}. The \ac{ocp} has to be solved to predict the system's future behavior over a given prediction horizon \(N \geq 2\) and optimize it by an optimal control sequence \(\mathbf{u}^*(\cdot)\). The \ac{ocp} of a discrete-time linear \ac{mpc} 
\begin{mini}[1]
{\mathbf{u}(\cdot),\mathbf{x}_\text{p}(\cdot)}{J(\mathbf{x}_\text{p}(\cdot),\mathbf{u}(\cdot))}{\label{eqn:MPC_OCP}}{}
\addConstraint{\mathbf{x}_\text{p}(n+1)}{=\mathbf{A}_\text{d}\,\mathbf{x}_\text{p}(n)+\mathbf{B}_\text{d}\,\mathbf{u}(n)}{\ \text{(system dynamics)}}
\addConstraint{\mathbf{x}_\text{p}(0)}{= \mathbf{x}(k)}{\ \text{(initial state)}}
\addConstraint{\mathbf{x}_\text{p}(n)}{\in \mathbb{X}, \quad \mathbf{u}(n) \in \mathbb{U}}{\ \text{(stage constraints)}}
\addConstraint{\mathbf{x}_\text{p}(N)}{\in \mathbb{X}_T}{\ \text{(terminal constraints)}}
\addConstraint{n}{\in\{0,\dots,N-1\}}{\ \text{(prediction horizon).}}
\end{mini}  
with its objective function
\begin{equation*}
J\big(\mathbf{x}_\text{p}(\cdot),\mathbf{u}(\cdot)\big) = \sum_{n=0}^{N-1}l\big(\mathbf{x}_\text{p}(n),\mathbf{u}(n)\big) + E\big(\mathbf{x}_\text{p}(N)\big)
\end{equation*}
has to be solved to optimize the system's future behavior, predicted by a discrete-time linear model of the system dynamics. The objective function consists of two sorts of cost functions. The stage cost function \(l\big(\mathbf{x}_\text{p}(\cdot),\mathbf{u}(\cdot)\big)\) is formulated to achieve the desired performance during the prediction horizon and the terminal cost function \(E\big(\mathbf{x}_\text{p}(\cdot)\big)\) penalizes the state at the end of the prediction horizon. The constraints, depending on the state and input, can be physical limitations of the state and input, e.g. maximum velocity or actuator restrictions, or consider physical values, e.g. fuel/energy restrictions or safety distances. Like the objective function, they are divided up into two types. Stage constraints represent constraints over the prediction horizon and terminal constraints must be satisfied at the end of the prediction horizon. The general procedure of an \ac{mpc} is shown in Algorithm \ref{algo_1}. 
\begin{algorithm}[H]
\caption{Basic MPC Algorithm}\label{algo_1}
\begin{itemize}
    \item[1.] Measure/estimate the state \(\mathbf{x}_\text{p}(0) = \mathbf{x}(k)\) at the current time \(t_k\);
    \item[2.] Solve the OCP \eqref{eqn:MPC_OCP};
    \item[3.] Apply the first element of the resulting optimal control sequences to the system: \(\mathbf{u}(k) = \mathbf{u}^*(0)\);
    \item[4.] k = k + 1;
    \item[5.] Go to step 1;
\end{itemize}
\end{algorithm}
Given is a system whose state \(\mathbf{x}(k) = \mathbf{x}(t_k)\) is measured in discrete time intervals \(t_{k+1} = t_k + \Delta t\). The system dynamics model makes it possible to find a prediction trajectory \(\left(\mathbf{x}_\text{p}(0),...,\mathbf{x}_\text{p}(N)\right)\) for a given control sequence \(\mathbf{u}(\cdot) = \mathbf{u}(0),\dots,\mathbf{u}(N-1)\). Starting with the measured state \(\mathbf{x}_\text{p}(0) = \mathbf{x}(k)\), the optimizer solves the \ac{ocp} to find an optimal control sequence \(\mathbf{u}^*(\cdot)\), which minimizes the objective function, while satisfying the constraints. After that, the first element of the optimal control sequence \(\mathbf{u}(k) = \mathbf{u}^*(0)\) is applied to the system for one time step \(\Delta t\). Then the procedure repeats. Due to this repeating prediction and optimization, the \ac{mpc} is a moving horizon strategy, which can compensate for model inaccuracies and disturbances acting on the system. \citep{Gruene(2017),Ibrahim(2020)}
\subsection{Mixed Integer Linear Programming}
\label{sec:MILP}
The \ac{ocp} of the \ac{ppa} is formulated using \ac{milp} to reduce optimization time. However, this limits the system dynamics, objective function, and constraints to be linear functions, where the following special formulations are used frequently.
\subsubsection{The "big M" Method}
\label{sec:BigM}
The "big M" method adds or subtracts the product of a high-value constant \(M_\text{big}\) and a binary variable \(b \in \{0,1\}\) to activate/tighten or deactivate/relax constraints, depending on the value of the binary expression. For further information about the "big M" method, see \citep{Ibrahim(2020),kong2010cutting}.
\subsubsection{Slack Variables}
\label{sec:SlackApprox}
The absolute value function of a scalar variable \(|a|\) is linearized by employing a slack variable \(a_\text{s}\), which encloses the actual variable \(a\) as absolute value of its lower- and upper-bound:
\begin{align*}
-a_\text{s} \leq a \leq a_\text{s}, \quad  a_\text{s} \geq 0.
\end{align*}
The slack variable can then be penalized in the objective function and is used in linear constraints as 
\begin{align*}
|a| \leq a_\text{max} \quad \rightarrow \quad a_\text{s} \leq a_\text{max}.
\end{align*}

\subsubsection{Polygon Approximation}
\label{sec:MILP_Round}
In many applications, it is necessary to determine whether a vector, originating in the center of a round shape, is within the boundaries of that shape or not. For instance, the Euclidean norm \(\|\mathbf{a}\|\) of a vector \(\mathbf{a} = (a_\text{x},a_\text{y},a_\text{z})^\top\) represents the radius of a sphere enclosing the vector.
Fig.~\ref{fig:Approx_Sphere} illustrates a possible approximation of various round shapes by polyhedrons or polygons defined via a set \(\mathcal{A}:= \big\{f_{\mathrm{poly},h}: h \in \{1,\dots,N_\text{f}\}\big\}\)
of \(N_\text{f}\) linear affine functions \(f_{\mathrm{poly},h}:\mathbb{R}^3 \to \mathbb{R}\). 
Considering a horizontal plane of the polyhedron (see blue area in Fig.~\ref{fig:Approx_Sphere}), the accuracy of the approximation is determined by the even number \(H\) representing the number of polygon sides. By this, we describe three different volumes enclosed by the polygons.

\paragraph*{Infinite high cylinder}
An infinite high cylinder is approximated by an \(H\)-sided polygon. The set \(\mathcal{A}\) consist of \(N_\text{f} = H\) linear affine functions 
\begin{align*}
f_{\mathrm{poly},h}(a_\text{x},a_\text{y},a_\text{z}) =
a_\text{x}\cos(\alpha_h) +
a_\text{y}\sin(\alpha_h), \ \
\alpha_h = \frac{2\,\pi\,h}{H}.
\end{align*}

\paragraph*{Closed cylinder}
For a closed cylinder, we extend \(\mathcal{A}\) by two functions 
\begin{equation*}
\label{eqn:ublb}
\begin{aligned}
f_{\mathrm{poly},H+1}(a_\text{x},a_\text{y},a_\text{z}) &= a_\text{z}, \\
f_{\mathrm{poly},H+2}(a_\text{x},a_\text{y},a_\text{z}) &= -a_\text{z},
\end{aligned}    
\end{equation*}
representing the lower and upper bound of the cylinder height.  

\paragraph*{Sphere}
Lastly, a sphere considers the vertical plane like the horizontal plane by \(H\)-sided polygons. This results in  \(N_\text{f} = H\left(\frac{H}{2}-1\right)+2\) linear affine functions.
The first \(H\left(\frac{H}{2}-1\right)\) functions are given by
\begin{align*}
f_{\mathrm{poly},h}(a_\text{x},a_\text{y},a_\text{z}) = \
&a_\text{x}\cos(\alpha_i)\sin(\beta_j) \ ...\\ &+
a_\text{y}\sin(\alpha_i)\sin(\beta_j) +
a_\text{z}\cos(\beta_j).  
\end{align*}
with the coefficients
\(\alpha_i = \frac{2\,\pi\,i}{H}, \
\forall i \in \{1,\dots,H\}\)
and \linebreak
\(\beta_j  = \frac{2\,\pi\,j}{H}, \
\forall j \in \left\{1,\dots,0.5\,H-1\right\}\). 
Here, the indices are translated by
\begin{align*}
i = \left\lfloor \frac{h - 1}{0.5\,H - 1} \right\rfloor + 1, \quad 
j = \big((h - 1)\,\text{mod}(0.5\,H - 1)\big) + 1,
\end{align*}
where $\left\lfloor \cdot \right\rfloor$ and $\text{mod}$ denote the floor function and the modulo operation, respectively.
The remaining two functions are taken from \eqref{eqn:ublb} representing the lower and upper bound.
\vspace{-0.25cm}
\begin{figure}[H]
\centering
\input{Tikz_Approx_Sphere}
\vspace{-0.25cm}
\caption{Polygon approximation of a sphere with \(H=8\) (left: vertical, right: horizontal)}
\label{fig:Approx_Sphere}
\end{figure}
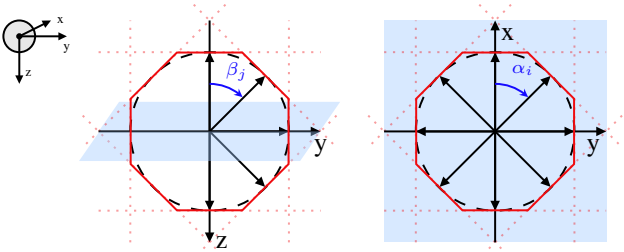
\begin{remark_new}
If the vector consists of slack variables and \(H\) is dividable by \(4\), the number of affine functions \(N_\text{f}\) can be reduced since only \(\alpha_i,\beta_j \in [0, \pi/2]\) has to be considered.
\end{remark_new}
The set \(\mathcal{A}\) of linear affine functions \(f_{\mathrm{poly},h}(a_\text{x},a_\text{y},a_\text{z})\) with \(\forall h \in \{1,\dots,N_\text{f}\} \) allows to formulate the vector's upper bound \linebreak \(\|\mathbf{a}\| \leq a_\text{max}\) as a set of constraints
\begin{equation}
\begin{aligned}
\label{eqn:const_Vec_ub}
f_{\mathrm{poly},h}(a_\text{x},a_\text{y},a_\text{z}) &\leq a_\text{max}\,c_\text{in}, 
\end{aligned}
\end{equation}
For an outer approximation,  \(c_\text{in}\) is set equal to \(1\), while for an inner approximation, \(c_\text{in}=\cos(\pi/H)\) and \(c_\text{in}=\cos(\pi/H)^2\) for a \ac{2D} or \ac{3D} shape, respectively.
The vector's lower bound \(\|\mathbf{a}\| \geq a_\text{min}\) is likewise approximated by the set of constraints
\begin{equation}
\label{eqn:const_Vec_lb}
\begin{aligned}
&f_{\mathrm{poly},h}(a_\text{x},a_\text{y},a_\text{z}) \geq a_\text{min} - M_\text{big}\,b_{h}, \\
&\sum_{h=1}^{N_\text{f}} b_{h} \leq N_\text{f} - 1, \quad b_{h} \in \{0,1\},
\end{aligned}
\end{equation}
where the "big M" method ensures that only one side of the approximated shape is considered. 

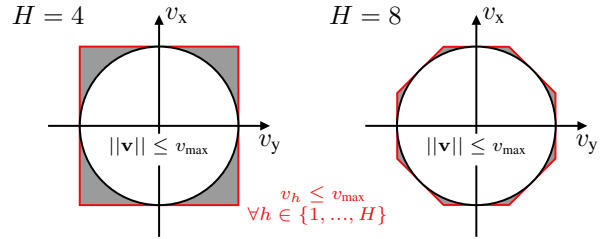
\begin{figure}[H]
\centering
\input{Tikz_PolygonApprox}
\vspace{-0.25cm}
\caption{Velocity limitation employing the Euclidean norm (black) vs. polygon approximation (red)}
\label{fig:PolygonApprox}
\end{figure}

The number of faces, adjusted by \(H\), affects the accuracy of these approximations. For example, Fig. \ref{fig:PolygonApprox} compares a quadratic \ac{2D} velocity constraint to the approximated \ac{2D} velocity constraints with 4 or 8 sides, where the possible approximation errors are colored gray. 

\subsubsection{Convex Hull Approximation}
\label{sec:MILP_Convex}
Likewise, it is possible to determine whether a point is inside the boundary of a convex hull or not. We determine whether a point \(\mathbf{p} = (p_\text{x},p_\text{y},p_\text{z})^\top\) is inside a convex hull by the signed distance \(D(\mathbf{p})\) between the point and the nearest face of the convex hull.
The signed distances \(d_h(\mathbf{p})\) between the point and the \(N_\text{f}\) faces of the convex hull, is derived, using the plane equations of the faces
\begin{align*}
d_h(\mathbf{p}) = c_{\text{x},h}\,p_\text{x} + c_{\text{y},h}\,p_\text{y} + c_{\text{z},h}\,p_\text{z} + c_{\text{0},h}, \ \forall h\in\{1,\dots,N_\text{f}\}.
\end{align*}
Here the plane coefficients \(c_{\text{x},h}, \dots, c_{\text{0},h}\) are normalized by the normal vector of the plane \(h\) pointing away from the convex hull. Based on this definition we state: If all signed distances are negative, the point lies inside the convex hull. 

Thereby, we formulate constraints to ensure that the point lies inside the convex hull \(D(\mathbf{p}) \leq \delta\) by
\begin{align}
\label{eqn:const_CS_in}
d_h(\mathbf{p}) \leq \delta, \quad \forall h\in\{1,\dots,N_\text{f}\}. 
\end{align}
where \(\delta\) can be used to inflate (\(\delta>0\)) or deflate (\(\delta<0\)) the shape, defining a minimum distance to the convex hull.
We ensure that a point lies outside the convex hull \(D(\mathbf{p}) \geq \delta\) by
\begin{equation}
\label{eqn:const_CS_out}
\begin{aligned}
&d_h(\mathbf{p}) \, \geq \delta - M_\text{big}\,b_h, \  \forall h \in\{1,\dots,N_\text{f}\}, \ \ \ \  \\
&\sum_{h=1}^{N_\text{f}} b_h \leq N_\text{f} - 1, \ \ \ \ \ \ \ b_h \in \{0,1\}.
\end{aligned}    
\end{equation}
Here again, the "big M" makes certain that only one side of the convex hull is considered.

\subsection{Path Planning Algorithm}
\label{sec:PPA}
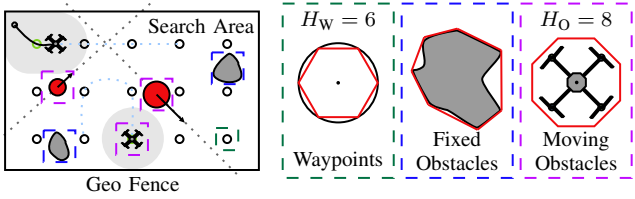
\begin{figure}[H]
\centering
\input{Tikz_PPA}
\caption{Simplified \ac{2D} illustration of a dynamic search area with multiple \acp{uas}, fixed/moving obstacles, and waypoints}
\label{fig:PPA}
\end{figure}

Based on the concept of \ac{mpc} and using the formulations presented in Section \ref{sec:MILP}, we derive a moving-horizon \ac{ppa} for \ac{3D} dynamic environments, as shown in Fig.~\ref{fig:PPA}. Within the search area, waypoints (green) are approximated by spheres. Moving obstacles or other \ac{uas} (violet) utilize the cylindrical approximations. The fixed obstacles (blue) are represented by general convex shapes, which are generated using sampling points of the obstacle and a convex hull algorithm, such as the gift-wrapping algorithm \cite{Preparata1985}.

The energy-efficient \ac{ppa} employs the \ac{ocp}
\begin{mini!}|s|[3]
{\mathbf{u}_\text{s}^i,\mathbf{x}_\text{s}^i,\mathbf{Q}}{J\big(\mathbf{x}_\text{s}^i(\cdot),\mathbf{u}_\text{s}^i(\cdot),\Phi(\cdot),D_\text{target}^i(\cdot)\big) \label{eq:OF}}{}{}
\addConstraint{i}{\in \mathbb{N}_\text{UAS}, \ \ w \in \mathbb{N}_\text{WP} \nonumber}
\addConstraint{\mathbf{x}^i(n+1)}{= \mathbf{A}_\text{d}^i\mathbf{x}^i(n) + \mathbf{B}_\text{d}^i\mathbf{u}^i(n) + \mathbf{E}_\text{d}^i, \ \mathbf{x}^i(0)=\mathbf{x}^i_0 \label{eq:OCP_1}}
\addConstraint{\mathbf{x}^i(n)}{\in \mathbb{X}^i, \quad \mathbf{u}^i(n) \in \mathbb{U}^i \label{eq:OCP_3}}
\addConstraint{\mathbf{x}_\text{s}^i(n)}{\in \mathbb{X}_\text{s}^i, \quad \mathbf{u}_\text{s}^i(n) \in \mathbb{U}_\text{s}^i \label{eq:OCP_4}}
\addConstraint{\mathbf{p}^i(n)}{\in \mathbb{G} \label{eq:OCP_5}}
\addConstraint{\mathbf{p}^i(n)}{\notin \mathbb{O} \label{eq:OCP_6}}
\addConstraint{\|\mathbf{p}^i(n)-\mathbf{p}_\text{target}^i\|}{\leq D_\text{target}^i(n) + c_\text{target}^i(n)  \label{eq:OCP_7}}
\addConstraint{\|\mathbf{p}^w-\mathbf{p}^i(n)\|}{> \delta_\text{WP} \Rightarrow b_\text{W}^{w,i}(n) = 0  \label{eq:OCP_8}}
\addConstraint{\textstyle\sum_{i=1}^{N_\text{UAS}}b_\text{W}^{w,i}(n)}{\leq \Phi^w(n), \quad b_\text{W}^{w,i},\Phi^w(n)\in\{0,1\}\nonumber}
\addConstraint{\Phi^w(n+1)}{=\Phi^w(n)-\sum_{i=1}^{N_\text{UAS}}b_\text{W}^{w,i}(n), \ \ \Phi^w(0)=\Phi^w_0 \nonumber}
\end{mini!}
with the objective function 
\begin{align*}
J(\cdot) &= 
\underbrace{\sum_{n=1}^{N} \sum_{i=1}^{N_\text{UAS}}\mathbf{W}_\text{u}^i\mathbf{u}_\text{s}^i(n) + \mathbf{W}_\text{x}^i\mathbf{x}_\text{s}^i(n)}_{\text{I}}  \\ 
&+ \underbrace{\sum_{n=1}^{N} \sum_{i=1}^{N_\text{UAS}} W_\text{D}^i\,D_\text{target}^i(n)}_{\text{II}} + \underbrace{\sum_{n=1}^{N}\sum_{w=1}^{N_\text{WP}}W_\Phi^w \Phi^w(n) }_{\text{III}}.
\end{align*}
Here, \(\mathbb{N}_\text{UAS} = \{1,\dots,N_\text{UAS}\}\) counts up to the number of \acp{uas} \(N_\text{UAS}\) in the \ac{uas} swarm. Likewise, \(\mathbb{N}_\text{WP} = \{1,\dots,N_\text{WP}\}\) counts up to the number of waypoints \(N_\text{WP}\) in the search area. In the following paragraphs, each part of the objective function \eqref{eq:OF} and each constraint \eqref{eq:OCP_1} - \eqref{eq:OCP_8} are explained in more detail.

\subsubsection{\ac{uav} Dynamics and Capabilities}
\label{sec:UAVDyn}
The \ac{ppa} considers the vehicle dynamics to ensure a physically feasible path generation. For this purpose, the \textit{system dynamics constraints} \eqref{eq:OCP_1} implement the discrete-time linear vehicle dynamics based on \eqref{Vehicle_Model}. Furthermore, we implement a  function that handles parameter-varying models and executes before every \ac{mpc} iteration. If it is necessary this function updates the model parameter depending on the depths of discharge \(\mathrm{DoD}\) to implement the \ac{lpv} system dynamics, described in \cite{Gasche2024_ECM}.

We include in \eqref{eq:OCP_3} the \textit{vehicle physical constraints} to specify the \acp{uav}' capabilities. The needed Euclidean norm functions are approximated by using the MILP formulations in Section \ref{sec:MILP}. For a detailed description of the vehicle model and its capabilities, see \cite{Gasche2024_ECM}.

Since we want to penalize the absolute values of the states and inputs in the objective function, we  define the \textit{slack variable constraints} in \eqref{eq:OCP_4}, based on Section \ref{sec:SlackApprox}, where \(\mathbf{x}_\text{s}^i\) and \(\mathbf{u}_\text{s}^i\) are slack variables of the state \(\mathbf{x}^i\) and input \(\mathbf{u}^i\). Consequently, we penalize the absolute values of the \acp{uav}' inputs and states in the first part of the objective function \eqref{eq:OF}, where the weight is adjusted by the input and state cost coefficient vectors \(\mathbf{W}_\textbf{u}^i\) and \(\mathbf{W}_\textbf{x}^i\). These are chosen to penalize the absolute values of the deviation of the lift \(\Delta L\), the torques \(\boldsymbol{\tau}\), the angular velocities \(\boldsymbol{\omega}\), and the depths of discharge \(\mathrm{DoD}\) to achieve a steady flight and minimize the \acp{uav}' energy consumption. Furthermore, the yaw angle \(\psi\) is penalized to reduce the model inaccuracies. Additionally, the cost coefficient vectors are used to normalize the cost terms depending on the number of \ac{uas}, and the maximum values of the states and inputs.

\subsubsection{Collision Avoidance}
\label{sec:CollisionAvoidance}
To guarantee a minimal distance between the \acp{uas} to avoid collisions, \eqref{eq:OCP_3} includes the \textit{collision avoidance constraints}
\begin{align*}
\|\mathbf{p}^{i}(n)-\mathbf{p}^{j}(n)\| \geq \delta_\text{C,min}^{i,j}, \quad i < j, \quad \forall i,j \in \mathbb{N}_\text{UAS},
\end{align*}
which are implemented employing the formulations introduced in \eqref{eqn:const_Vec_lb}. Here, \(\mathcal{A}\) represents a closed cylinder and the minimal distance between the \acp{uas}, \(\delta_\text{C,min}^{i,j} = \delta_{\text{UAV}}^i + \delta_{\text{UAV}}^j + \delta_\text{safe}\) is the sum of the \(i^\text{th}\) and \(j^\text{th}\) \ac{uav} radii and a safety buffer distance. 

\subsubsection{Geo Fence}
\label{sec:GeoFence}
While operating \acp{uas}, it is important to limit the area within the \acp{uas} are allowed to fly autonomously. We define this area by a convex shape, called the geo fence \(\mathbb{G}\). We restrict the \acp{uas} to only be inside it, employing the \textit{geo fence constraints} in \eqref{eq:OCP_5} 
\begin{align*}
D_\text{G}\big(\mathbf{p}^i(n)\big) \leq \delta_\text{G},
\quad \forall i \in \mathbb{N}_\text{UAS},
\end{align*}
which derive from \eqref{eqn:const_CS_in}. Here, the buffer distance \(\delta_\text{G}\) can be used to relax or narrow the minimal distance to the geo fence.

\subsubsection{Obstacles Avoidance}
\label{sec:CO}
The \acp{uas} further must avoid collisions with obstacles, which are represented by convex shapes. Therefore, we implement in \eqref{eq:OCP_6} the \textit{obstacle avoidance constraints}
\begin{align*}
D_\text{O}^o\big(\mathbf{p}^i(n)\big) \geq \delta_\text{O,min}^{i,o},
\quad \forall i \in \mathbb{N}_\text{UAS}, \quad \forall o \in \mathbb{N}_\text{O},
\end{align*}
which derive from \eqref{eqn:const_CS_out}. Here, \(\mathbb{N}_\text{O} = \{1,\dots,N_\text{O}\}\) counts up to the number \(N_\text{O}\) of obstacles  within the obstacle set \(\mathbb{O}\).
The minimal distance \(\delta_\text{O,min}^{i,o} = \delta_\text{UAV}^{i}  + \delta_\text{safe}^{o}\) between the \(i^\text{th}\) \ac{uas} and the \(o^\text{th}\) obstacle consists of the radius of the \ac{uav} \(\delta_\text{UAV}^i\) and the minimum safe distances \(\delta_\text{safe}^{o}\) of the obstacle. 

The dynamics of moving obstacles are approximated by simple integrator discrete-time models and implemented by 
\begin{align*}
\mathbf{p}^o(n+1) &= \mathbf{p}^o(n) + \mathbf{v}^o(n)\,\Delta\,t, \quad \forall o \in \mathbb{N}_\text{MO}, 
\end{align*}
where \(\mathbb{N}_\text{MO}\) is the index set of the moving obstacles in \(\mathbb{O}\).

\begin{figure}[H]
\centering
\input{Tikz_CC}
\caption{Simplified illustration of the corner cutting constraints. (grey: obstacle, green: next valid area, red: next prohibit area, blue: current area)}
\label{fig:CC}
\end{figure}
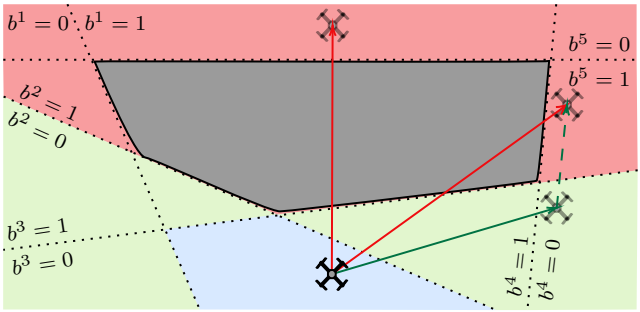

Due to the discrete sampling of the \ac{uas} positions, it has to be avoided that obstacles are jumped over or corners are cutted. The \textit{obstacle avoidance constraints} encode the position of \ac{uas} \(i\) in respect to the face \(h\) of obstacle \(o\) with the binary variable \(b_\text{O}^{i,o,h}\). An inactive constraint is indicated by \(b_\text{O}^{i,o,h} = 1\) since the \ac{uas} and the obstacle are on the same side of the plane, representing the obstacle face. Meanwhile, \(b_\text{O}^{i,o,h} = 0\) indicates an active constraint, because the \ac{uas} and the obstacle are separated by the plane. To define a valid area for the \ac{uas} position in the next time step \(n+1\), we use this encoding in the \textit{corner cutting constraints}
\begin{align*}
&b_\text{O}^{i,o,h}(n) + b_\text{O}^{i,o,h}(n+1) \leq 2\,c_\text{O}^{i,o,h}(n), \\
&\textstyle\sum_{h=1}^{N_\text{f}} c_\text{O}^{i,o,h}(n) \leq N_\text{f}^{o} - 1,  \\
&b_\text{O}^{i,o,h}, c_\text{O}^{i,o,h} \in \{0,1\}, \ \ \forall i \in \mathbb{N}_\text{UAS}, 
\ \ \forall o \in \mathbb{N}_\text{FO}, 
\ \ \forall h \in \mathbb{H}_\text{O}^{o}. 
\end{align*}
 Here, \(\mathbb{N}_\text{FO}\) is the index set of fixed obstacles and \(\mathbb{H}_\text{O}^{o} = \{1,\dots,N_\text{f}^{o}\}\) counts up to the number of faces \(N_\text{f}^{o}\) of the \(o^\text{th}\) obstacle. Further, \(c_\text{O}^{i,o,h}\) indicates whether an active constraint remained active (\(c_\text{O}^{i,o,h} = 0\)) or otherwise (\(c_\text{O}^{i,o,h} = 1\)). This constraint ensures that at least one of the \(N_\text{f}^{o}\) constraints for the \(i^\text{th}\) \ac{uas} and \(o^\text{th}\) obstacle, which is currently active, remains active. As an example, Fig.~\ref{fig:CC} shows the resulting valid area (green) and prohibit area (red) for the \ac{uas} in the next step.

\subsubsection{Waypoint Coverage}
\label{sec:WC}

As it is shown in Fig.~\ref{fig:PPA}, the waypoints are distributed inside the search area, so that by passing (fly-by) all waypoints, all areas of interest should be covered by the sensor range of the \acp{uas} at least once. For the coverage of the search area, we need to decide at time step \(n\) whether waypoint \(w\) is currently covered by \ac{uas} \(i\) or not, which is done by a binary decision variable \(b_\text{W}^{w,i}(n) \in \{0,1\}\). It is constrained to be \(b_\text{W}^{w,i}(n) = 0\) if the \ac{uas} is outside the waypoint with the radius \(\delta_\text{WP}\) or if the waypoint is already covered, which is indicated by the coverage state \(\Phi^w(n) = 0\). If both don't apply the binary variable can be set equal to \(b_\text{W}^{w,i}(n) = 1\), marking the waypoint as covered in the next step \(\Psi^w(n+1) = 0\). All binary decision variables \(b_\text{W}^{w,i}\), are included in the decision matrix \(\mathbf{Q}\in \mathbb{R}^{N \times N_\text{WP} \times N_\text{UAS}}\).
Further, we constrain the binary variables, so that if multiple \ac{uas} are inside the same waypoint, only one \ac{uas} marks it as covered. These dynamics are implemented in \eqref{eq:OCP_8} by the \textit{waypoint currently coverage constraints}
\begin{align*}
&\|\mathbf{p}^{i}(n)-\mathbf{p}^{w}\| \leq \delta_\text{W} + M_\text{big}\big(1-b_\text{W}^{w,i}(n)\big), \\
&\textstyle\sum_{i=1}^{N_\text{UAS}}b_\text{W}^{w,i}(n) \leq \Phi^w(n), \\ &b_\text{W}^{w,i}(n),\Phi^w(n) \in \{0,1\}, \quad i \in \mathbb{N}_\text{UAS}, \quad w \in \mathbb{N}_\text{WP}.
\end{align*}
which are implemented employing the formulations introduced in \eqref{eqn:const_Vec_ub}. Here, \(\mathcal{A}\) represents a sphere. The "big M" method (see Section \ref{sec:BigM}) relaxes or tighten the constraints depending on the binary decision variables \(b_\text{W}^{w,i}\).   

The coverage state \(\Phi^{w}\) declares a waypoint \(w\) as covered (\(\Phi^{w}=0\)) or not (\(\Phi^{w}=1\)) to record whether waypoint \(w\) is already covered and allow the \acp{uas} to visit a waypoint more than once without any cost reduction. Its dynamics are defined in \eqref{eq:OCP_8} by the \textit{waypoint dynamics constraints} 
\begin{align*}
&\Phi^{w}(n+1) = \Phi^{w}(n)-\textstyle\sum_{i=1}^{N_\text{UAS}}b_\text{W}^{w,i}(n), \ \ \Phi^{w}(0) = \Phi^{w}_0, \\
&i \in \mathbb{N}_\text{UAS}, \ \ w \in \mathbb{N}_\text{WP}, 
\end{align*}
where \(\Phi^{w}_0\) is the coverage state at the beginning of the current iteration. Suppose, a currently uncovered waypoint \(w\) (\(\Phi^{w}(n)=1\)) is covered for the first time at time step \(n\), then one of the corresponding binary variables is set to \(b_\text{W}^{w,i}(n)=1\) and the coverage state \(\Phi^{w}(n+1)\) of waypoint \(w\) at the next time step \(n+1\) is set to \(\Phi^{w}(n+1)=0\).

To create an incentive for the \acp{uas} to cover the whole search area, we penalize the number of uncovered waypoints in the third part of the objective function \eqref{eq:OF}, where the coverage cost coefficients \(W_\Phi^w\) can be adjusted to prioritize specific waypoints. 

\subsubsection{Target Distance Dynamics}
\label{sec:RNW}
To improve the path planning in already covered areas and to implement a dynamic, which returns discharged \acp{uas} to their initial positions, we include the second part of the objective function \eqref{eq:OF} and the \textit{target distance constraints} 
\eqref{eq:OCP_7}
\begin{equation*}
\begin{split}
& \|\mathbf{p}^{i}(n)-\mathbf{p}_\text{target}^{i}(n)\| \leq D_\text{target}^{i}(n) + c_\text{target}^i, \quad \forall i \in \mathbb{N}_\text{UAS}.
\end{split}
\end{equation*}
which are implemented employing the formulations introduced in \eqref{eqn:const_Vec_ub} and \(\mathcal{A}\) represents a closed cylinder. Depending on the current operation mode \(\mathrm{op}^{i}\) of  \ac{uas} \(i\), we penalize the distance \(D_\text{target}^i\) between this \ac{uas} and a target at the position \(\mathbf{p}_\text{target}^{i}\) in the obejctive function. This cylindrical distance approximation results in a smooth transit behaviour, when penalized in the objective function.
Furthermore, the relaxation term \(c_\text{target}^i\) is used to relax the constraints, so the target distance \(D_\text{target}^{i}\) will be set equal to zero if certain conditions are fulfilled. 
The \ac{uas}'s four operation modes are introduced in the following paragraphs.

\paragraph{Covering mode (\(\mathrm{op}^{i}=0\))} 
The \ac{uas} is able to cover new waypoints inside the prediction horizon. The target distance dynamics are inactive and nether the corresponding costs or the constraints are included in the \ac{ocp}.\\ 

\paragraph{Transit mode (\(\mathrm{op}^{i}=1\))} 
The \ac{uas} transits to an area with uncovered waypoints. The target distance dynamics are active until the \ac{uas} reaches a selected uncovered waypoint. The variables for the target distance dynamics are given by
\begin{equation*}
\mathbf{p}_\text{target}^{i} = \mathbf{p}^{w}, \quad W_\text{target}^i = W_\text{t}^{i,w}, \quad c_\text{target}^i =  M_\text{big}\big(1-\Phi^{w}(n)\big),
\end{equation*}
where \(\mathbf{p}^{w}\) is the position of the best fitting target waypoint with index \(w\). The cost coefficient \(W_\text{t}^{i,w}\) is normalized by the distance between \ac{uas} \(i\) and the target waypoint \(w\) at the beginning of the iteration. The "big M" method in Section \ref{sec:BigM} and the binary coverage state \(\Phi^w(n)\) are used to relax the constraints, so that the target distance \(D_\text{target}^{i}\) is set equal to zero when the waypoint gets covered. Due to this relaxable constraints, the objective function is minimized by reducing the distance \( D_\text{target}^i\) or by covering the waypoint \(w\).
The position of this uncovered waypoint \(\mathbf{p}^w\) is determined before an \ac{mpc} iteration. For this, we compare all feasible waypoints to determine the best fitting waypoint in the current situation. Waypoints, which are already covered, which will be covered inside the prediction horizon or which are already a target of an other \ac{uas} are declared as infeasible. The remaining waypoints get values for their horizontal distances to the \ac{uas}, for their vertical distances to the \ac{uas} and for the necessary changes of the \ac{uas}'s heading to align it with the waypoints. These values are normalised and weighted before summarizing them to a cost factor. The waypoint with the lowest cost factor is declared as transit target.\\

\paragraph{Return mode (\(\mathrm{op}^{i}=2\))} 
The \ac{uas} is returning to its initial position \(\mathbf{p}_\text{init}^i\) and the target distance dynamics are active until the \ac{uas} reaches this position. The variables for the target distance dynamics are given by
\begin{equation*}
\mathbf{p}_\text{target}^{i} = \mathbf{p}_\text{init}^i, \quad W_\text{target}^i = W_\text{r}^{i}, \quad c_\text{target}^i = 0,
\end{equation*}
where the cost coefficient \(W_\text{r}^i\) increases the target distance costs step by step, depending on the remaining charge. So, the \ac{uas} still explores the search area while returning to its initial position. To prevent the return costs from outweighing the other costs of the objective function, it is normalized by the initial return distance of the current \ac{mpc} iteration.
The return flight is initiated, if the whole area is already covered (\(\Phi^{w} = 0, \ \forall w \in \mathbb{N}_\text{WP}, \)) or the \ac{uas}'s depth of discharge \(\mathrm{DoD}^i\) exceeds the threshold
\begin{align*}
\mathrm{DoD}_\text{r}^i &= \mathrm{DoD}_\text{max}^i-\frac{p_\text{DC,nom}^i\,D_\text{r,max}}{v_\text{cruise}^i\,Q_b^i\,u_\text{b,nom}^i}. \label{eqn:DoD_r}
\end{align*}
It is determined by an overestimated remaining flight distance \(D_\text{r,max}\), a nominal power consumption to hover \(p_\text{DC,nom}^i\), and a constant cruise speed \(v_\text{cruise}^i\), while assuming that the battery voltage is equal to its nominal voltage \(u_\text{b,nom}^i\). The maximum depth of discharge \(\mathrm{DoD}_\text{max}^i\) is chosen to be smaller than the cutoff depth of discharge \(\mathrm{DoD}_\text{cutoff}^i\) to protect the battery from damage. \citep{Gasche2024_ECM} \\

\paragraph{Landed mode (\(\mathrm{op}^{i}=3\))} 
The \ac{uas} is returned to its initial position (\(D_\text{target}^{i} \leq \delta_\text{l}\)), meaning it is inside a radius equal to the landing distance \(\delta_\text{l}\). All costs in the objective function and constraints related to this \ac{uas} are removed from the \ac{ocp}. 

Algorithms \ref{algo_2} and \ref{algo_3} give a brief overview, how the next operation mode \(\mathrm{op}_\text{next}^{i}\)  for the  \(i^\text{th}\) \ac{uas} is determined and how the target distance dynamics are added, updated, or removed since these are not always included in the \ac{ocp} to reduce the optimization time. 

\begin{algorithm}[H]
\caption{Determine Next Operation Mode Function}\label{algo_2}
\begin{algorithmic}[1]
\Function{determine\_next\_op}{}
\State Set \(\mathrm{op}_\text{next}^i = 0\);
\small{\Comment{Default: covering mode}}
\If{\(\mathrm{op}^i = 3\)}
\small{\Comment{Landed mode active}}
    \State Set \(\mathrm{op}_\text{next}^i = 3\);
\ElsIf{\(\mathrm{op} = 2\)}
\small{\Comment{Return mode active}}
    \If{\(D_\text{target}^{i} \leq \delta_\text{l} \ \land \ \mathbf{v}^{i} \approx 0 \)}
    \small{\Comment{At landing site}}
        \State Set \(\mathrm{op}_\text{next}^i = 3\);  
    \Else
    \small{\Comment{Far from landing site}}
        \State Set \(\mathrm{op}_\text{next}^i = 2\); 
    \EndIf
\ElsIf{\(\Phi^{w} = 0, \ \forall w \in \mathbb{N}_\text{WP} \ \lor \ \mathrm{DoD}^i \geq \mathrm{DoD}_\text{r}^i\)}\\
    \small{\Comment{All WPs covered or UAS low on energy}}
    \State Set \(\mathrm{op}_\text{next}^i = 2\);
\ElsIf{\(\mathrm{op}^i = 1\)}
\small{\Comment{Transit mode active}}
    \If{\(\Psi^{w}=1\)}
    \small{\Comment{Target WP uncovered}}
        \State Set \(\mathrm{op}_\text{next}^i = 1\);  
    \EndIf
\Else
\small{\Comment{Covering mode active}}
    \If{\(b_\text{W}^{w,i}=0, \forall w \in \mathbb{N}_\text{WP}\)}
    \small{\Comment{Won't cover new WP}}
        \State Set \(\mathrm{op}_\text{next}^i = 1\);  
    \EndIf
\EndIf
\EndFunction
\end{algorithmic}
\end{algorithm}

\begin{algorithm}[H]
\caption{Update Target Distance Dynamics Function}\label{algo_3}
\begin{algorithmic}[1]
\Function{update\_target\_distance\_dynamics}{}
\For{every UAS}  
    \State \(\mathrm{op}_\text{next}^i\) = determine\_next\_op(); 
    \small{\Comment{Algorithm \ref{algo_2}}}
    \If{\(\mathrm{op}^i = 0 \ \land \ \mathrm{op}_\text{next}^i = 1\)}
        \State Determine best fitting target waypoint \(w\);
        \State Calculate target distance cost coefficient \(W_\text{t}^{i,w}\);
        \State Add target distance dynamics for waypoint \(w\);
    \ElsIf{\(\mathrm{op}^i = 0 \ \land \ \mathrm{op}_\text{next}^i = 2\)}
        \State Calculate target distance cost coefficient \(W_\text{r}^{i}\);
        \State Add target distance dynamics for \(\mathbf{p}_\text{init}^i\);
    \ElsIf{\(\mathrm{op}^i = 1 \ \land \ \mathrm{op}_\text{next}^i = 0\)}
        \State Remove target distance dynamics;
    \ElsIf{\(\mathrm{op}^i = 1 \ \land \ \mathrm{op}_\text{next}^i = 1\)}
        \State Calculate  target distance cost coefficient \(W_\text{t}^{i,w}\);
        \State Update target distance dynamics for waypoint \(w\);
    \ElsIf{\(\mathrm{op}^i = 1 \ \land \ \mathrm{op}_\text{next}^i = 2\)}
        \State Calculate target distance cost coefficient \(W_\text{r}^{i}\);
        \State Add target distance dynamics for \(\mathbf{p}_\text{init}^i\);
    \ElsIf{\(\mathrm{op}^i = 2 \ \land \ \mathrm{op}_\text{next}^i = 2\)}
        \State Calculate target distance cost coefficient \(W_\text{r}^{i}\);
        \State Update target distance dynamics for \(\mathbf{p}_\text{init}^i\);
    \ElsIf{\(\mathrm{op}^i = 2 \ \land \ \mathrm{op}_\text{next}^i = 3\)}
        \State Set UAS as landed;
        \State Remove target distance dynamics;
    \EndIf
    \State Change \(\mathrm{op}^i = \mathrm{op}_\text{next}^i\);  
\EndFor
\If{target distance dynamics are changed}
    \State Update OCP;
\EndIf
\EndFunction
\end{algorithmic}
\end{algorithm}

\section{Simulation}
\label{cha:Simulation}
A C++ program is developed to set up a \ac{uas} swarm path planning simulation. It includes simulation scenarios, several \ac{uav} models, the \ac{mpc} and supporting functions to adapt the \ac{ocp} depending on the current situation. Furthermore, it includes output functions to save the simulation results, which are displayed in Matlab, see \citep{Mathworks(2023)}, with plots of various simulation values and an animation of the flight. The Gurobi Optimizer v11.0.1, see \citep{Gurobi(2023)}, is used to solve the \ac{milp} \ac{ocp}.

\begin{figure}[H]
    \centering
    \includegraphics[width=0.5\textwidth]{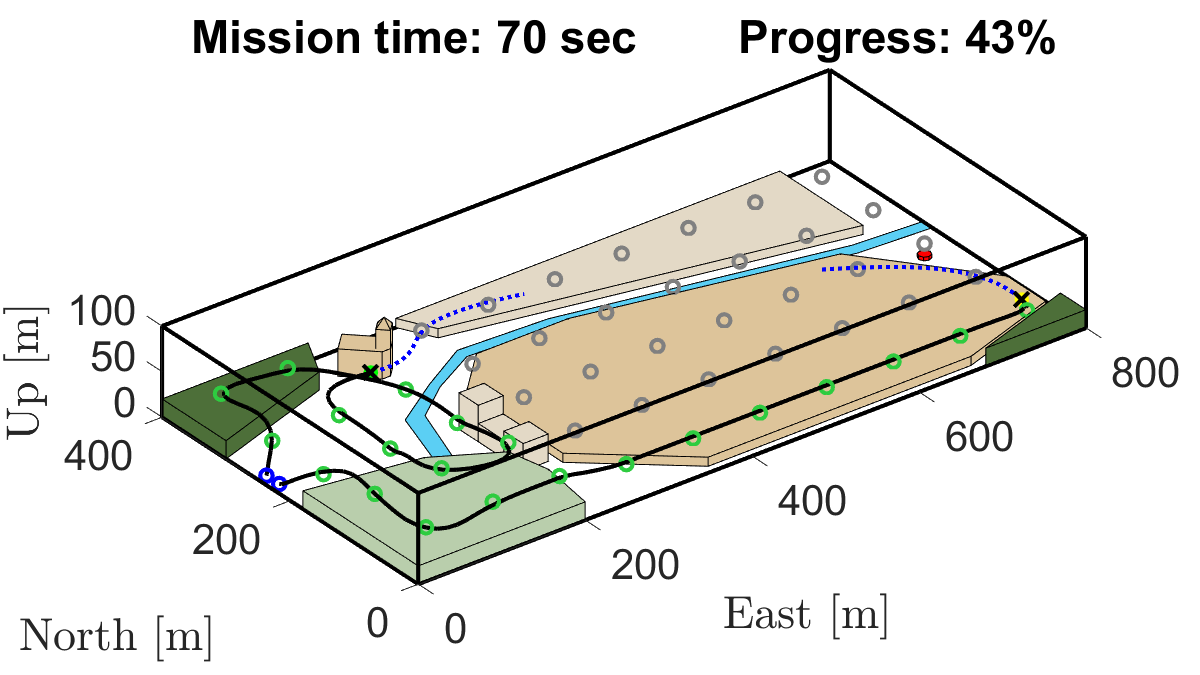}
    \vspace{-0.75cm}
    \caption{Snapshot of the simulation at time step 70}
    \label{fig:Sim_1}
\end{figure}

\begin{figure}[H]
    \centering
    \includegraphics[width=0.5\textwidth]{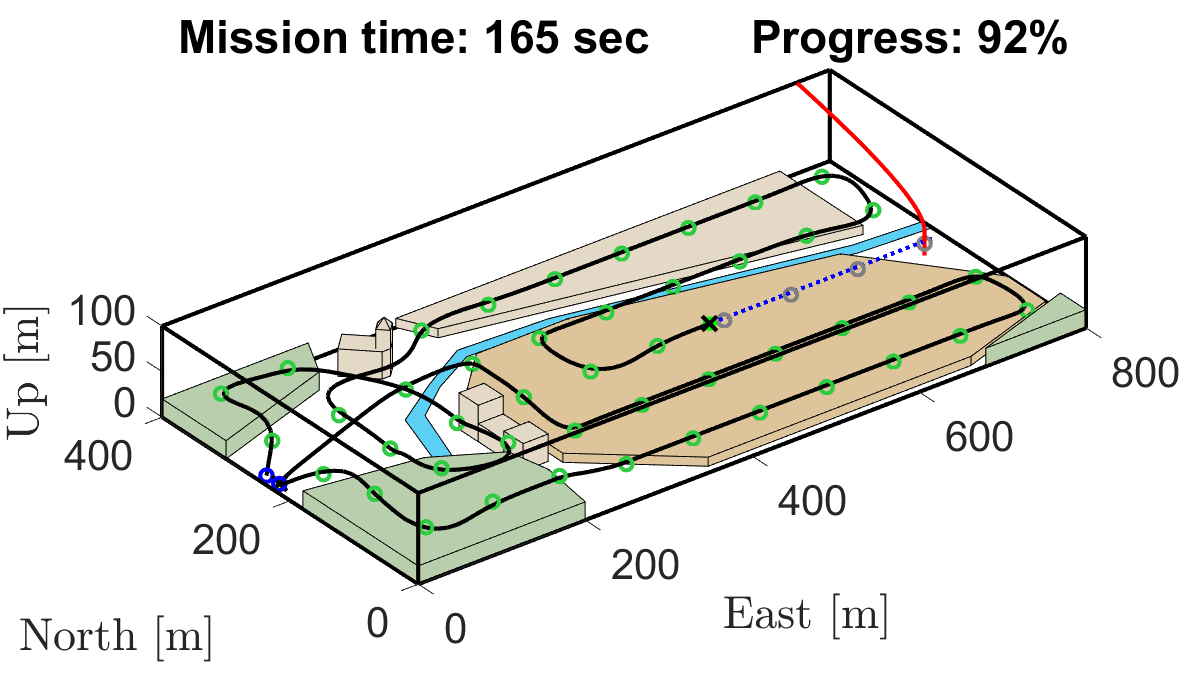}
    \vspace{-0.75cm}
    \caption{Snapshot of the simulation at time step 165}
    \label{fig:Sim_2}
\end{figure}

In the following, we present the results of a simulation using the simulation parameters, shown in Tab.~\ref{tab:0} and two "Holybro S500 V2" quadcopter models \citep{Holybro(2022)}, whose model parameters are listed in \cite{Gasche2024_ECM}. The simulation scenario, which is shown in Fig.~\ref{fig:Sim_1} and \ref{fig:Sim_2}, is defined as follows: A small village (brown), consisting of two residential areas, 3 high-rise buildings and a church, is located along a river (blue) and surrounded by forests (green). Due to heavy rainfall, the area is flooded. The rescue team arrive on site from the west, employing two \acp{uas}, which are equipped with (thermal) cameras to provide the rescue team with information about the current situation. The search area is enclosed by the geo fence (black rectangle). Meanwhile, an emergency helicopter (red cylinder) in the east is preparing to take-off. During operations, the \acp{uas} (black crosses) draw their past trajectories as black lines, while the predicted future trajectories are illustrated as blue dots. For this prediction, a prediction horizon of 18 with a sampling time of 1 second allows to predict the behavior of the \acp{uas} for 18 seconds into the future. The \ac{uas} start at the blue circles in the west. While \ac{uas} 1 is fully charged at the beginning, \ac{uas} 2 is already over 55\% discharged. The current state of charge of the \ac{uas} are shown as colored circles around the \acp{uas} (green: charged, yellow: medium charge, red: discharged).
The snapshot in Fig.~\ref{fig:Sim_1} shows the animated simulation results at time step 70. At the beginning of the simulation, both \acp{uas} start covering the search area. Due to the coverage costs, the forest area is covered first. Meanwhile, obstacles and other \acp{uas} are avoided due to the obstacle avoidance constraints and energy-saving maneuvers, such as low accelerations and straight or smoothly curved trajectories, are preferred. The effects of the target distance dynamics can be seen for \ac{uas} 2, which is currently in the east. Its depths of discharge has exceeded the defined threshold and the distance to its initial position is now penalized. The costs increase depending on the remaining charge so that \ac{uas} 2 continues the area coverage while heading for its initial position until the costs outweigh in the objective function. Then the area coverage is aborted and \ac{uas} 2 returns directly.
At time step \(165\), shown in Fig.~\ref{fig:Sim_2}, \ac{uas} 2 is already returned. \ac{uas} 1 continues the coverage until it reaches the last waypoint, which was previously blocked by the helicopter. Afterwards it also returns to conclude the mission. 
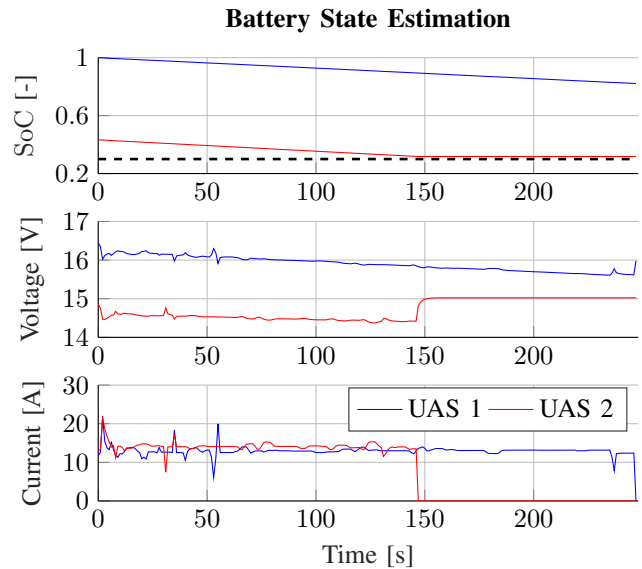
\begin{figure}[H]
    \centering
    \input{Tikz_Simulation_Plot_ECM_short}
    \vspace{-0.25cm}
    \caption{Battery state estimation for \ac{uas} 1 (blue and \ac{uas} 2 (red)}
    \label{fig:Sim_3}
\end{figure}
The corresponding battery state estimations are illustrated in Fig.~\ref{fig:Sim_3}. Here the state of charge, the battery voltage and the battery current of \ac{uas} 1 and \ac{uas} 2 are shown in blue and red, respectively. It can be seen that \ac{uas} 2 is drawing a higher current to compensate for its lower charge and therefore lower battery voltage. When climbing, braking, or accelerating the current draw, and thereby the power consumption is increased. Meanwhile, it decreases when descending due to the vehicle dynamics and constraints. Furthermore,  the power consumption during horizontal flight at maximum velocity is only slightly higher, compared to steady hover flight, due to the thrust correction. This effect is also observable in reality. 
\section{Discussion}
\label{cha:Discussion}
This section discusses the achievements of this work, starting with the \ac{uav} and energy consumption models, followed by the further developments made to the \ac{ppa}. 

\subsection{Multicopter and Energy Consumption Models}
In Section \ref{cha:UAV}, a discrete-time \ac{lpv} model for multicopter that incorporates an \ac{ecm} is introduced. This model relies on assumptions about the multicopter's shape, aerodynamic properties, and rotor dynamics, as described in \cite{Gasche2024_ECM}. This includes the simplification assuming that the air drag is independent of the multicopter’s orientation. Similarly, the aerodynamic force and torque parameters, \(k_\text{F}\) and \(k_\text{M}\), are considered constant, although these parameters actually depend on factors such as motor speed, air inflow velocity, and atmospheric conditions. These assumptions should be mitigated in future work in order to reduce the resulting model uncertainties. 

Furthermore, the \acp{ecm} shows a self-amplifying effect, increasing the charge estimation error over time, which is present in the simulation because the predicted next state of the last \ac{mpc} iteration is used as "measurements" instead of actual measurements. In reality, the state is measured between the iterations, which compensates for the self-amplifying estimation error. In the simulation, we compensate for this effect by implementing a safety buffer \(\mathrm{DoD}_\text{max}\). In real-world applications, this safety buffer should remain for unexpected circumstances, such as a new landing site being chosen at the last minute due to unexpected events.

Environmental disturbances, such as wind, varying air pressure, and temperature changes, also pose challenges. These factors are not modeled here but have a significant impact on the performance of the \ac{uav} and its power train. Wind, for instance, changes the aerodynamic forces and control effectiveness, affecting the accuracy of the \ac{ecm}. Similarly, temperature variations can influence battery performance.

\subsection{Path Planning Algorithm}
The \ac{ppa} introduced in Section \ref{cha:PPA} considers the \acp{uas} energy consumption to optimize flight efficiency and to return \acp{uas}, low on energy, to their initial positions. Moreover, it improves the path planning in already covered areas and near to obstacles. 

The inclusion of energy-aware dynamics and additional model dimensions increases the complexity of the \ac{ocp}, particularly in terms of variables and constraints. However, the highest impact on the number of optimization variables and constraints and, thereby, on the optimization time has the number of waypoints \(N_\text{WP}\). Reducing the reliance on a high number of waypoints should be a priority in future work, as it poses a significant challenge to real-time optimization. Furthermore, improving the accuracy of path planning requires increasing parameters such as the prediction horizon \(N\),  the approximation degrees \(H_\text{P}\), \(H_\text{WP}\), \(H_\text{O}\), and the discretization degree \(N_\text{dis}\). However, this increases the optimization time as well. This trade-off between accuracy and computational efficiency must be managed based on specific application needs. Depending on how these parameters are tuned, solving the \ac{ocp} in real-time may not always be feasible.

To enhance the robustness of the \ac{ppa} and improve the safety of planned paths, it will be necessary to consider uncertainties and disturbances. A suitable approach for this is tube \ac{mpc}, which maintains the system's trajectory within a predefined tube around a nominal trajectory. Here, the nominal system is optimized, while uncertainties are managed, ensuring that the actual trajectory remains within the bounds of the tube, even in the presence of significant disturbances.

Future developments should also account for the mission-specific requirements of the heterogeneous \ac{uas} swarm. Meaning that different \acp{uav} could be assigned different tasks based on energy levels, sensor capabilities, or operational roles. 

\section{Conclusion}
\label{cha:Conclusion}
The goal of this work was to integrate multicopter \ac{uav} models, incorporating their energy consumption, into a \ac{ppa} for a heterogeneous \ac{uas} swarm. The developed approach enables energy-efficient path planning within a dynamic \ac{3D} environment, ensuring that \acp{uas} return to their initial positions for recharging when their energy levels are low.
The proposed moving-horizon \ac{ppa} employs a combination of \ac{mpc} and \ac{milp}. It is able to plan paths for a \ac{uas} swarm to cover a defined search area while considering the \acp{uas}' physical capabilities, as well as obstacle and collision avoidance, while preferring low energy-consuming maneuvers. Furthermore, new dynamics are included to improve path planning in already covered areas and to return \acp{uas}, low on energy, to their initial positions. Upon arrival, the \acp{uav} deactivate themselves to avoid damage to their batteries. 

This allows fully autonomous guidance of a \ac{uas} swarm for search and rescue, surveillance, or monitoring missions. The \ac{uas} swarm can provide additional information about the current situation from a bird's eye view, without the need for manned aircraft in cluttered environments, which usually pose a danger to aircraft pilots.

In conclusion, the proposed \ac{ppa} optimizes the behavior of the \acp{uas} with regard to multiple objectives, including energy efficiency, safety, and mission success. This ensures effective and safe operations within complex and dynamic environments.

\appendix

\section{Simulation Parameters}
\label{cha:AppendixA}
\vspace{-0.25cm}
\begin{table}[H]
\caption{Simulation parameters}
\begin{center}
\begin{tabular}{|c|c|c|c|c|}
\hline
\(N=18\) & \(N_\text{UAS} = 2\) & \(N_\text{WP} = 49\) & \multicolumn{2}{c|}{\(\mathbf{p}_\text{0}^1 = ( 250,10,-1)^\top\)} \\ \hline
\(\Delta t = 1\text{s}\) & \(N_\text{O} = 11\) & \(H_\text{O} = 8\) & \multicolumn{2}{c|}{\(\mathbf{p}_\text{0}^2 = (230,10,-1)^\top\)} \\ \hline
\(N_\text{dis}=4\) & \(N_\text{MO} = 1\) & \(H_\text{P} = 8\) &  \(\mathrm{DoD}_\text{0}^1 = 0\) & \(\mathrm{DoD}_\text{0}^2 = 0.56\) \\ \hline
\multicolumn{2}{|c|}{\(N_\text{step,max} = 500\)} & \(H_\text{WP} = 4\) & \multicolumn{2}{c|}{\(\mathrm{DoD}_\text{max} = 0.75\)} \\ \hline
\end{tabular}
\end{center}
\label{tab:0}
\end{table} 
\begin{remark_new}
    The parameters \(H_\text{P},H_\text{O}\) and \(H_\text{WP}\) are the approximation degrees \(H\) for the formulations in Section \ref{sec:MILP_Round}, which are used for the physical constraints, the obstacle and collision avoidance constraints as well as the waypoint coverage constraints in Section \ref{sec:PPA}. 
\end{remark_new}
\clearpage

\ifCLASSOPTIONcaptionsoff
  \newpage
\fi

\small
\bibliographystyle{plainnat}
\bibliography{paper}

\normalsize
\end{document}

%% file: acronyms.tex
\DeclareAcronym{pso}{
short = PSO ,
long = particle swarm optimization,
short-plural = s ,
long-plural = s ,
}

\DeclareAcronym{aco}{
short = ACO ,
long = ant colony optimization,
short-plural = s ,
long-plural = s ,
}

\DeclareAcronym{prm}{
short = PRM ,
long = probabilistic roadmaps algorithm,
short-plural = s ,
long-plural = s ,
}

\DeclareAcronym{rrt}{
short = RRT ,
long = rapidly-exploring random trees algorithm,
short-plural = s ,
long-plural = s ,
}

\DeclareAcronym{3D}{
short = 3D ,
long = three-dimensional,
short-plural = s ,
long-plural = s ,
}
\DeclareAcronym{vtol}{
short = VTOL ,
long = vertical take-off and landing,
short-plural = s ,
long-plural = s ,
}

\DeclareAcronym{2D}{
short = 2D ,
long = two-dimensional,
short-plural = s ,
long-plural = s ,
}

\DeclareAcronym{ppa}{
short = PPA ,
long = path planning algorithm ,
short-plural = s ,
long-plural = s ,
}

\DeclareAcronym{uas}{
short = UAS ,
long = unmanned aircraft system,
short-plural = s ,
long-plural = s ,
}

\DeclareAcronym{uav}{
short = UAV ,
long = unmanned aerial vehicle,
short-plural = s ,
long-plural = s ,
}

\DeclareAcronym{ecm}{
short = ECM ,
long = energy consumption model,
short-plural = s ,
long-plural = s ,
}

\DeclareAcronym{mpc}{
short = MPC ,
long = model predictive control,
short-plural = s ,
long-plural = s ,
}

\DeclareAcronym{milp}{
short = MILP ,
long = mixed integer linear programming,
short-plural = s ,
long-plural = s ,
}

\DeclareAcronym{mip}{
short = MIP ,
long = mixed integer programming,
short-plural = s ,
long-plural = s ,
}

\DeclareAcronym{lpv}{
short = LPV ,
long = linear parameter-varying,
short-plural = s ,
long-plural = s ,
}

\DeclareAcronym{npv}{
short = NPV ,
long = nonlinear parameter-varying,
short-plural = s ,
long-plural = s ,
}

\DeclareAcronym{ocp}{
short = OCP ,
long = optimal control problem,
short-plural = s ,
long-plural = s ,
}

\DeclareAcronym{bldc}{
short = BLDC,
long = brushless direct current,
short-plural = s ,
long-plural = s ,
}

\DeclareAcronym{dc}{
short = DC,
long = direct current,
short-plural = s ,
long-plural = s ,
}

\DeclareAcronym{esc}{
short = ESC,
long = electric speed controller,
short-plural = s ,
long-plural = s ,
}

\DeclareAcronym{esc-bldc}{
short = ESC-BLDC,
long = ESC-BLDC, 
short-plural = s ,
long-plural = s ,
}
\DeclareAcronym{bldc-esc}{
short = BLDC-ESC,
long = BLDC-ESC, 
short-plural = s ,
long-plural = s ,
}

\DeclareAcronym{lib}{
short = LIB,
long = lithium-ion battery,
short-plural = s ,
long-plural = lithium-ion batteries ,
}

\DeclareAcronym{libc}{
short = LIBC,
long = lithium-ion battery cell,
short-plural = s ,
long-plural = s ,
}

\DeclareAcronym{pwm}{
short = PWM,
long = pulse width modulation,
short-plural = s ,
long-plural = s ,
}

\DeclareAcronym{nps}{
short = NPS,
long = number of polygon sides and affine linear functions,
short-plural = s ,
long-plural = s ,
}

%% file: Tikz_Quadcopter_Frames.tex
\tikzset{every picture/.style={line width=0.75pt}} %

\begin{tikzpicture}[x=0.6pt,y=0.6pt,yscale=-1,xscale=1]

\draw [color={rgb, 255:red, 239; green, 12; blue, 16 }  ,draw opacity=1 ]   (265.16,67.7) -- (265.16,134.9) ;
\draw [shift={(265.16,137.9)}, rotate = 270] [fill={rgb, 255:red, 239; green, 12; blue, 16 }  ,fill opacity=1 ][line width=0.08]  [draw opacity=0] (8.93,-4.29) -- (0,0) -- (8.93,4.29) -- cycle    ;
\draw [line width=3]    (265.16,51.22) -- (265.16,76.77) ;
\draw [line width=3]    (186.74,70.61) -- (343.59,70.61) ;
\draw  [draw opacity=0][fill={rgb, 255:red, 155; green, 155; blue, 155 }  ,fill opacity=1 ] (255.36,59.52) -- (274.97,59.52) -- (288.04,63.83) -- (288.04,77.39) -- (274.97,81.7) -- (255.36,81.7) -- (242.29,77.39) -- (242.29,63.83) -- cycle ;
\draw  [fill={rgb, 255:red, 184; green, 233; blue, 134 }  ,fill opacity=1 ] (257,53.97) .. controls (257,52.95) and (260.65,52.12) .. (265.16,52.12) .. controls (269.68,52.12) and (273.33,52.95) .. (273.33,53.97) .. controls (273.33,54.99) and (269.68,55.82) .. (265.16,55.82) .. controls (260.65,55.82) and (257,54.99) .. (257,53.97) -- cycle ;
\draw  [draw opacity=0][fill={rgb, 255:red, 184; green, 233; blue, 134 }  ,fill opacity=1 ] (257,44.73) -- (273.33,44.73) -- (273.33,53.97) -- (257,53.97) -- cycle ;
\draw  [fill={rgb, 255:red, 184; green, 233; blue, 134 }  ,fill opacity=1 ] (257,44.73) .. controls (257,43.71) and (260.65,42.88) .. (265.16,42.88) .. controls (269.68,42.88) and (273.33,43.71) .. (273.33,44.73) .. controls (273.33,45.75) and (269.68,46.58) .. (265.16,46.58) .. controls (260.65,46.58) and (257,45.75) .. (257,44.73) -- cycle ;
\draw    (257,44.73) -- (257,53.97) ;
\draw    (273.33,44.73) -- (273.33,53.97) ;
\draw  [fill={rgb, 255:red, 0; green, 0; blue, 0 }  ,fill opacity=1 ] (262.55,44.73) .. controls (262.55,44.22) and (263.72,43.8) .. (265.16,43.8) .. controls (266.61,43.8) and (267.78,44.22) .. (267.78,44.73) .. controls (267.78,45.24) and (266.61,45.65) .. (265.16,45.65) .. controls (263.72,45.65) and (262.55,45.24) .. (262.55,44.73) -- cycle ;
\draw  [fill={rgb, 255:red, 0; green, 0; blue, 0 }  ,fill opacity=1 ] (262.55,42.57) .. controls (262.55,42.06) and (263.72,41.65) .. (265.16,41.65) .. controls (266.61,41.65) and (267.78,42.06) .. (267.78,42.57) .. controls (267.78,43.08) and (266.61,43.5) .. (265.16,43.5) .. controls (263.72,43.5) and (262.55,43.08) .. (262.55,42.57) -- cycle ;
\draw  [fill={rgb, 255:red, 74; green, 74; blue, 74 }  ,fill opacity=1 ] (266.4,42.57) -- (304.38,42.57) -- (304.38,42.57) -- (304.38,43.8) -- (303.14,45.04) -- (265.16,45.04) -- (265.16,45.04) -- (265.16,43.8) -- cycle ;
\draw  [fill={rgb, 255:red, 74; green, 74; blue, 74 }  ,fill opacity=1 ] (227.19,42.42) -- (265.16,42.42) -- (265.16,42.42) -- (265.16,43.65) -- (263.93,44.88) -- (225.95,44.88) -- (225.95,44.88) -- (225.95,43.65) -- cycle ;
\draw  [fill={rgb, 255:red, 0; green, 0; blue, 0 }  ,fill opacity=1 ] (262.55,42.57) -- (267.78,42.57) -- (267.78,44.73) -- (262.55,44.73) -- cycle ;
\draw    (255.36,71.84) -- (274.97,71.84) ;
\draw   (255.36,71.84) -- (274.97,71.84) -- (274.97,81.7) -- (255.36,81.7) -- cycle ;
\draw    (274.97,81.7) -- (288.04,77.39) ;
\draw    (274.97,71.84) -- (288.04,67.53) ;
\draw    (288.04,77.39) -- (288.04,67.53) ;
\draw    (242.29,67.53) -- (255.36,71.84) ;
\draw    (242.29,67.53) -- (242.29,77.39) ;
\draw    (255.36,81.7) -- (242.29,77.39) ;
\draw    (242.29,67.53) -- (242.29,63.83) ;
\draw    (288.04,67.53) -- (288.04,63.83) ;
\draw    (242.29,63.83) -- (255.36,59.52) ;
\draw    (274.97,59.52) -- (255.36,59.52) ;
\draw    (288.04,63.83) -- (286.39,63.29) -- (274.97,59.52) ;
\draw  [fill={rgb, 255:red, 158; green, 202; blue, 255 }  ,fill opacity=1 ] (178.57,74.93) .. controls (178.57,73.9) and (182.23,73.08) .. (186.74,73.08) .. controls (191.25,73.08) and (194.91,73.9) .. (194.91,74.93) .. controls (194.91,75.95) and (191.25,76.77) .. (186.74,76.77) .. controls (182.23,76.77) and (178.57,75.95) .. (178.57,74.93) -- cycle ;
\draw  [draw opacity=0][fill={rgb, 255:red, 158; green, 202; blue, 255 }  ,fill opacity=1 ] (178.57,65.68) -- (194.91,65.68) -- (194.91,74.93) -- (178.57,74.93) -- cycle ;
\draw  [fill={rgb, 255:red, 158; green, 202; blue, 255 }  ,fill opacity=1 ] (178.57,65.68) .. controls (178.57,64.66) and (182.23,63.83) .. (186.74,63.83) .. controls (191.25,63.83) and (194.91,64.66) .. (194.91,65.68) .. controls (194.91,66.7) and (191.25,67.53) .. (186.74,67.53) .. controls (182.23,67.53) and (178.57,66.7) .. (178.57,65.68) -- cycle ;
\draw    (178.57,65.68) -- (178.57,74.93) ;
\draw    (194.91,65.68) -- (194.91,74.93) ;
\draw  [fill={rgb, 255:red, 0; green, 0; blue, 0 }  ,fill opacity=1 ] (184.13,65.68) .. controls (184.13,65.17) and (185.3,64.76) .. (186.74,64.76) .. controls (188.18,64.76) and (189.36,65.17) .. (189.36,65.68) .. controls (189.36,66.19) and (188.18,66.61) .. (186.74,66.61) .. controls (185.3,66.61) and (184.13,66.19) .. (184.13,65.68) -- cycle ;
\draw  [fill={rgb, 255:red, 0; green, 0; blue, 0 }  ,fill opacity=1 ] (184.13,63.52) .. controls (184.13,63.01) and (185.3,62.6) .. (186.74,62.6) .. controls (188.18,62.6) and (189.36,63.01) .. (189.36,63.52) .. controls (189.36,64.04) and (188.18,64.45) .. (186.74,64.45) .. controls (185.3,64.45) and (184.13,64.04) .. (184.13,63.52) -- cycle ;
\draw  [fill={rgb, 255:red, 74; green, 74; blue, 74 }  ,fill opacity=1 ] (187.97,63.52) -- (225.95,63.52) -- (225.95,63.52) -- (225.95,64.76) -- (224.72,65.99) -- (186.74,65.99) -- (186.74,65.99) -- (186.74,64.76) -- cycle ;
\draw  [fill={rgb, 255:red, 74; green, 74; blue, 74 }  ,fill opacity=1 ] (148.76,63.37) -- (186.74,63.37) -- (186.74,63.37) -- (186.74,64.6) -- (185.51,65.84) -- (147.53,65.84) -- (147.53,65.84) -- (147.53,64.6) -- cycle ;
\draw  [fill={rgb, 255:red, 0; green, 0; blue, 0 }  ,fill opacity=1 ] (184.13,63.52) -- (189.36,63.52) -- (189.36,65.68) -- (184.13,65.68) -- cycle ;

\draw [line width=3]    (265.16,76.77) -- (265.16,102.32) ;
\draw  [fill={rgb, 255:red, 184; green, 233; blue, 134 }  ,fill opacity=1 ] (257,106.36) .. controls (257,105.33) and (260.65,104.51) .. (265.16,104.51) .. controls (269.68,104.51) and (273.33,105.33) .. (273.33,106.36) .. controls (273.33,107.38) and (269.68,108.2) .. (265.16,108.2) .. controls (260.65,108.2) and (257,107.38) .. (257,106.36) -- cycle ;
\draw  [draw opacity=0][fill={rgb, 255:red, 184; green, 233; blue, 134 }  ,fill opacity=1 ] (257,97.11) -- (273.33,97.11) -- (273.33,106.36) -- (257,106.36) -- cycle ;
\draw  [fill={rgb, 255:red, 184; green, 233; blue, 134 }  ,fill opacity=1 ] (257,97.11) .. controls (257,96.09) and (260.65,95.26) .. (265.16,95.26) .. controls (269.68,95.26) and (273.33,96.09) .. (273.33,97.11) .. controls (273.33,98.13) and (269.68,98.96) .. (265.16,98.96) .. controls (260.65,98.96) and (257,98.13) .. (257,97.11) -- cycle ;
\draw    (257,97.11) -- (257,106.36) ;
\draw    (273.33,97.11) -- (273.33,106.36) ;
\draw  [fill={rgb, 255:red, 0; green, 0; blue, 0 }  ,fill opacity=1 ] (262.55,97.11) .. controls (262.55,96.6) and (263.72,96.19) .. (265.16,96.19) .. controls (266.61,96.19) and (267.78,96.6) .. (267.78,97.11) .. controls (267.78,97.62) and (266.61,98.04) .. (265.16,98.04) .. controls (263.72,98.04) and (262.55,97.62) .. (262.55,97.11) -- cycle ;
\draw  [fill={rgb, 255:red, 0; green, 0; blue, 0 }  ,fill opacity=1 ] (262.55,94.95) .. controls (262.55,94.44) and (263.72,94.03) .. (265.16,94.03) .. controls (266.61,94.03) and (267.78,94.44) .. (267.78,94.95) .. controls (267.78,95.46) and (266.61,95.88) .. (265.16,95.88) .. controls (263.72,95.88) and (262.55,95.46) .. (262.55,94.95) -- cycle ;
\draw  [fill={rgb, 255:red, 74; green, 74; blue, 74 }  ,fill opacity=1 ] (266.4,94.95) -- (304.38,94.95) -- (304.38,94.95) -- (304.38,96.19) -- (303.14,97.42) -- (265.16,97.42) -- (265.16,97.42) -- (265.16,96.19) -- cycle ;
\draw  [fill={rgb, 255:red, 74; green, 74; blue, 74 }  ,fill opacity=1 ] (227.19,94.8) -- (265.16,94.8) -- (265.16,94.8) -- (265.16,96.03) -- (263.93,97.27) -- (225.95,97.27) -- (225.95,97.27) -- (225.95,96.03) -- cycle ;
\draw  [fill={rgb, 255:red, 0; green, 0; blue, 0 }  ,fill opacity=1 ] (262.55,94.95) -- (267.78,94.95) -- (267.78,97.11) -- (262.55,97.11) -- cycle ;
\draw  [fill={rgb, 255:red, 0; green, 0; blue, 0 }  ,fill opacity=1 ] (263.86,66.46) .. controls (263.86,65.78) and (264.44,65.23) .. (265.16,65.23) .. controls (265.89,65.23) and (266.47,65.78) .. (266.47,66.46) .. controls (266.47,67.15) and (265.89,67.7) .. (265.16,67.7) .. controls (264.44,67.7) and (263.86,67.15) .. (263.86,66.46) -- cycle ;
\draw  [color={rgb, 255:red, 0; green, 0; blue, 0 }  ,draw opacity=1 ][fill={rgb, 255:red, 239; green, 12; blue, 16 }  ,fill opacity=1 ] (273.44,65.74) -- (267.37,63.56) -- (277.67,62.84) -- cycle ;
\draw [color={rgb, 255:red, 239; green, 12; blue, 16 }  ,draw opacity=1 ]   (265.16,66.46) -- (341.08,81.92) ;
\draw [shift={(344.02,82.52)}, rotate = 191.51] [fill={rgb, 255:red, 239; green, 12; blue, 16 }  ,fill opacity=1 ][line width=0.08]  [draw opacity=0] (8.93,-4.29) -- (0,0) -- (8.93,4.29) -- cycle    ;
\draw [color={rgb, 255:red, 239; green, 12; blue, 16 }  ,draw opacity=1 ]   (265.16,66.46) -- (340.29,42.82) ;
\draw [shift={(343.15,41.92)}, rotate = 162.53] [fill={rgb, 255:red, 239; green, 12; blue, 16 }  ,fill opacity=1 ][line width=0.08]  [draw opacity=0] (8.93,-4.29) -- (0,0) -- (8.93,4.29) -- cycle    ;
\draw [color={rgb, 255:red, 0; green, 0; blue, 0 }  ,draw opacity=1 ][line width=0.75]    (42,103.42) -- (42,160.42) ;
\draw [shift={(42,163.42)}, rotate = 270] [fill={rgb, 255:red, 0; green, 0; blue, 0 }  ,fill opacity=1 ][line width=0.08]  [draw opacity=0] (8.93,-4.29) -- (0,0) -- (8.93,4.29) -- cycle    ;
\draw [color={rgb, 255:red, 0; green, 0; blue, 0 }  ,draw opacity=1 ][line width=0.75]    (42,103.42) -- (99,103.42) ;
\draw [shift={(102,103.42)}, rotate = 180] [fill={rgb, 255:red, 0; green, 0; blue, 0 }  ,fill opacity=1 ][line width=0.08]  [draw opacity=0] (8.93,-4.29) -- (0,0) -- (8.93,4.29) -- cycle    ;
\draw [color={rgb, 255:red, 0; green, 0; blue, 0 }  ,draw opacity=1 ][line width=0.75]    (42,103.42) -- (74.85,71.51) ;
\draw [shift={(77,69.42)}, rotate = 135.83] [fill={rgb, 255:red, 0; green, 0; blue, 0 }  ,fill opacity=1 ][line width=0.08]  [draw opacity=0] (8.93,-4.29) -- (0,0) -- (8.93,4.29) -- cycle    ;
\draw  [fill={rgb, 255:red, 0; green, 0; blue, 0 }  ,fill opacity=1 ] (40,103.42) .. controls (40,102.32) and (40.9,101.42) .. (42,101.42) .. controls (43.1,101.42) and (44,102.32) .. (44,103.42) .. controls (44,104.53) and (43.1,105.42) .. (42,105.42) .. controls (40.9,105.42) and (40,104.53) .. (40,103.42) -- cycle ; \draw   (40,103.42) -- (44,103.42) ; \draw   (42,101.42) -- (42,105.42) ;
\draw  [dash pattern={on 0.84pt off 2.51pt}]  (42,103.42) -- (265.16,66.46) ;
\draw  [draw opacity=0] (248.94,120.49) .. controls (246.81,119.75) and (245.56,118.86) .. (245.56,117.9) .. controls (245.56,115.35) and (254.34,113.28) .. (265.16,113.28) .. controls (275.99,113.28) and (284.77,115.35) .. (284.77,117.9) .. controls (284.77,120.45) and (275.99,122.51) .. (265.16,122.51) .. controls (263.43,122.51) and (261.74,122.46) .. (260.13,122.36) -- (265.16,117.9) -- cycle ; \draw  [color={rgb, 255:red, 21; green, 10; blue, 255 }  ,draw opacity=1 ] (248.94,120.49) .. controls (246.81,119.75) and (245.56,118.86) .. (245.56,117.9) .. controls (245.56,115.35) and (254.34,113.28) .. (265.16,113.28) .. controls (275.99,113.28) and (284.77,115.35) .. (284.77,117.9) .. controls (284.77,120.45) and (275.99,122.51) .. (265.16,122.51) .. controls (263.43,122.51) and (261.74,122.46) .. (260.13,122.36) ;  
\draw [color={rgb, 255:red, 21; green, 10; blue, 255 }  ,draw opacity=1 ]   (260.13,122.36) -- (254.62,121.43) ;
\draw [shift={(251.66,120.94)}, rotate = 9.53] [fill={rgb, 255:red, 21; green, 10; blue, 255 }  ,fill opacity=1 ][line width=0.08]  [draw opacity=0] (6.25,-3) -- (0,0) -- (6.25,3) -- cycle    ;
\draw  [draw opacity=0] (323.38,60.5) .. controls (323.48,63.31) and (322.99,65.17) .. (321.85,65.53) .. controls (319.28,66.34) and (314.41,59.16) .. (310.97,49.5) .. controls (307.52,39.83) and (306.81,31.34) .. (309.38,30.53) .. controls (311.95,29.72) and (316.82,36.9) .. (320.26,46.57) .. controls (320.81,48.11) and (321.29,49.62) .. (321.7,51.08) -- (315.62,48.03) -- cycle ; \draw  [color={rgb, 255:red, 21; green, 10; blue, 255 }  ,draw opacity=1 ] (323.38,60.5) .. controls (323.48,63.31) and (322.99,65.17) .. (321.85,65.53) .. controls (319.28,66.34) and (314.41,59.16) .. (310.97,49.5) .. controls (307.52,39.83) and (306.81,31.34) .. (309.38,30.53) .. controls (311.95,29.72) and (316.82,36.9) .. (320.26,46.57) .. controls (320.81,48.11) and (321.29,49.62) .. (321.7,51.08) ;  
\draw [color={rgb, 255:red, 21; green, 10; blue, 255 }  ,draw opacity=1 ]   (321.64,50.88) -- (322.57,55.05) ;
\draw [shift={(323.23,57.98)}, rotate = 257.44] [fill={rgb, 255:red, 21; green, 10; blue, 255 }  ,fill opacity=1 ][line width=0.08]  [draw opacity=0] (6.25,-3) -- (0,0) -- (6.25,3) -- cycle    ;
\draw [color={rgb, 255:red, 21; green, 10; blue, 255 }  ,draw opacity=1 ]   (297.61,77.79) -- (297.42,81.97) ;
\draw [shift={(297.29,84.96)}, rotate = 272.61] [fill={rgb, 255:red, 21; green, 10; blue, 255 }  ,fill opacity=1 ][line width=0.08]  [draw opacity=0] (6.25,-3) -- (0,0) -- (6.25,3) -- cycle    ;
\draw [color={rgb, 255:red, 239; green, 12; blue, 16 }  ,draw opacity=1 ]   (265.16,67.45) -- (265.16,89.55) -- (265.16,117.15) ;
\draw [color={rgb, 255:red, 239; green, 12; blue, 16 }  ,draw opacity=1 ]   (307.54,53.11) -- (316.79,50.23) ;
\draw  [draw opacity=0] (0,0) -- (400,0) -- (400,190) -- (0,190) -- cycle ;
\draw  [fill={rgb, 255:red, 158; green, 202; blue, 255 }  ,fill opacity=1 ] (335.42,74.93) .. controls (335.42,73.9) and (339.08,73.08) .. (343.59,73.08) .. controls (348.1,73.08) and (351.76,73.9) .. (351.76,74.93) .. controls (351.76,75.95) and (348.1,76.77) .. (343.59,76.77) .. controls (339.08,76.77) and (335.42,75.95) .. (335.42,74.93) -- cycle ;
\draw  [draw opacity=0][fill={rgb, 255:red, 158; green, 202; blue, 255 }  ,fill opacity=1 ] (335.42,65.68) -- (351.76,65.68) -- (351.76,74.93) -- (335.42,74.93) -- cycle ;
\draw  [fill={rgb, 255:red, 158; green, 202; blue, 255 }  ,fill opacity=1 ] (335.42,65.68) .. controls (335.42,64.66) and (339.08,63.83) .. (343.59,63.83) .. controls (348.1,63.83) and (351.76,64.66) .. (351.76,65.68) .. controls (351.76,66.7) and (348.1,67.53) .. (343.59,67.53) .. controls (339.08,67.53) and (335.42,66.7) .. (335.42,65.68) -- cycle ;
\draw    (335.42,65.68) -- (335.42,74.93) ;
\draw    (351.76,65.68) -- (351.76,74.93) ;
\draw  [fill={rgb, 255:red, 0; green, 0; blue, 0 }  ,fill opacity=1 ] (340.97,65.68) .. controls (340.97,65.17) and (342.14,64.76) .. (343.59,64.76) .. controls (345.03,64.76) and (346.2,65.17) .. (346.2,65.68) .. controls (346.2,66.19) and (345.03,66.61) .. (343.59,66.61) .. controls (342.14,66.61) and (340.97,66.19) .. (340.97,65.68) -- cycle ;
\draw  [fill={rgb, 255:red, 0; green, 0; blue, 0 }  ,fill opacity=1 ] (340.97,63.52) .. controls (340.97,63.01) and (342.14,62.6) .. (343.59,62.6) .. controls (345.03,62.6) and (346.2,63.01) .. (346.2,63.52) .. controls (346.2,64.04) and (345.03,64.45) .. (343.59,64.45) .. controls (342.14,64.45) and (340.97,64.04) .. (340.97,63.52) -- cycle ;
\draw  [fill={rgb, 255:red, 74; green, 74; blue, 74 }  ,fill opacity=1 ] (344.82,63.52) -- (382.8,63.52) -- (382.8,63.52) -- (382.8,64.76) -- (381.57,65.99) -- (343.59,65.99) -- (343.59,65.99) -- (343.59,64.76) -- cycle ;
\draw  [fill={rgb, 255:red, 74; green, 74; blue, 74 }  ,fill opacity=1 ] (305.61,63.37) -- (343.59,63.37) -- (343.59,63.37) -- (343.59,64.6) -- (342.36,65.84) -- (304.38,65.84) -- (304.38,65.84) -- (304.38,64.6) -- cycle ;
\draw  [fill={rgb, 255:red, 0; green, 0; blue, 0 }  ,fill opacity=1 ] (340.97,63.52) -- (346.2,63.52) -- (346.2,65.68) -- (340.97,65.68) -- cycle ;

\draw  [draw opacity=0] (297.58,78.02) .. controls (297.78,76.57) and (298.04,75.05) .. (298.37,73.51) .. controls (300.45,63.5) and (304.29,55.79) .. (306.95,56.28) .. controls (309.6,56.77) and (310.07,65.28) .. (307.99,75.28) .. controls (305.91,85.28) and (302.06,93) .. (299.41,92.51) .. controls (298.26,92.3) and (297.52,90.6) .. (297.22,87.93) -- (303.18,74.39) -- cycle ; \draw  [color={rgb, 255:red, 21; green, 10; blue, 255 }  ,draw opacity=1 ] (297.58,78.02) .. controls (297.78,76.57) and (298.04,75.05) .. (298.37,73.51) .. controls (300.45,63.5) and (304.29,55.79) .. (306.95,56.28) .. controls (309.6,56.77) and (310.07,65.28) .. (307.99,75.28) .. controls (305.91,85.28) and (302.06,93) .. (299.41,92.51) .. controls (298.26,92.3) and (297.52,90.6) .. (297.22,87.93) ;  
\draw [color={rgb, 255:red, 239; green, 12; blue, 16 }  ,draw opacity=1 ]   (304.59,74.49) -- (310,75.58) ;
\draw  [draw opacity=0][fill={rgb, 255:red, 255; green, 255; blue, 255 }  ,fill opacity=0.5 ] (238.25,61.12) .. controls (238.25,59.13) and (239.86,57.52) .. (241.85,57.52) -- (259.65,57.52) .. controls (261.64,57.52) and (263.25,59.13) .. (263.25,61.12) -- (263.25,71.92) .. controls (263.25,73.9) and (261.64,75.52) .. (259.65,75.52) -- (241.85,75.52) .. controls (239.86,75.52) and (238.25,73.9) .. (238.25,71.92) -- cycle ;
\draw  [color={rgb, 255:red, 239; green, 12; blue, 16 }  ,draw opacity=1 ][fill={rgb, 255:red, 239; green, 12; blue, 16 }  ,fill opacity=1 ] (262.55,66.46) .. controls (262.55,65.1) and (263.72,64) .. (265.16,64) .. controls (266.61,64) and (267.78,65.1) .. (267.78,66.46) .. controls (267.78,67.82) and (266.61,68.92) .. (265.16,68.92) .. controls (263.72,68.92) and (262.55,67.82) .. (262.55,66.46) -- cycle ; \draw  [color={rgb, 255:red, 239; green, 12; blue, 16 }  ,draw opacity=1 ] (262.55,66.46) -- (267.78,66.46) ; \draw  [color={rgb, 255:red, 239; green, 12; blue, 16 }  ,draw opacity=1 ] (265.16,64) -- (265.16,68.92) ;
\draw [color={rgb, 255:red, 21; green, 10; blue, 255 }  ,draw opacity=1 ][line width=1.5]    (265.16,66.46) -- (264.68,15.31) ;
\draw [shift={(264.64,11.31)}, rotate = 89.46] [fill={rgb, 255:red, 21; green, 10; blue, 255 }  ,fill opacity=1 ][line width=0.08]  [draw opacity=0] (6.97,-3.35) -- (0,0) -- (6.97,3.35) -- cycle    ;

\draw (79,66.02) node [anchor=south west] [inner sep=0.75pt]  [color={rgb, 255:red, 0; green, 0; blue, 0 }  ,opacity=1 ]  {$\text{x}^{\text{I}}$};
\draw (104,103.42) node [anchor=west] [inner sep=0.75pt]  [color={rgb, 255:red, 0; green, 0; blue, 0 }  ,opacity=1 ]  {$\text{y}^{\text{I}}$};
\draw (42,166.82) node [anchor=north] [inner sep=0.75pt]  [color={rgb, 255:red, 0; green, 0; blue, 0 }  ,opacity=1 ]  {$\text{z}^{\text{I}}$};
\draw (40,103.42) node [anchor=east] [inner sep=0.75pt]  [color={rgb, 255:red, 0; green, 0; blue, 0 }  ,opacity=1 ]  {$O^{\text{I}}$};
\draw (250.75,66.52) node  [color={rgb, 255:red, 239; green, 12; blue, 16 }  ,opacity=1 ]  {$O^{\text{B}}$};
\draw (345.15,41.92) node [anchor=west] [inner sep=0.75pt]  [color={rgb, 255:red, 239; green, 12; blue, 16 }  ,opacity=1 ]  {$\text{x}^{\text{B}}$};
\draw (346.02,82.52) node [anchor=west] [inner sep=0.75pt]  [color={rgb, 255:red, 239; green, 12; blue, 16 }  ,opacity=1 ]  {$\text{y}^{\text{B}}$};
\draw (265.16,141.3) node [anchor=north] [inner sep=0.75pt]  [color={rgb, 255:red, 239; green, 12; blue, 16 }  ,opacity=1 ]  {$\text{z}^{\text{B}}$};
\draw (240.2,120.38) node [anchor=east] [inner sep=0.75pt]    {$\psi ,\textcolor[rgb]{0.08,0.04,1}{\tau }\textcolor[rgb]{0.08,0.04,1}{_{\text{z}}}$};
\draw (342.02,85.92) node [anchor=north east] [inner sep=0.75pt]    {$\theta ,\textcolor[rgb]{0.08,0.04,1}{\tau }\textcolor[rgb]{0.08,0.04,1}{_\text{y}}$};
\draw (310.18,28.31) node [anchor=south] [inner sep=0.75pt]    {$\phi ,\textcolor[rgb]{0.08,0.04,1}{\tau }\textcolor[rgb]{0.08,0.04,1}{_\text{x}}$};
\draw (260.45,21) node [anchor=east] [inner sep=0.75pt]  [color={rgb, 255:red, 21; green, 10; blue, 255 }  ,opacity=1 ]  {$\mathnormal{T}\approx\mathnormal{L}$};

\end{tikzpicture}

%% file: Tikz_Energy_Consumption_0_short.tex
\tikzset{every picture/.style={line width=0.75pt}} %

\begin{tikzpicture}[x=0.6pt,y=0.6pt,yscale=-1,xscale=1]

\draw [color={rgb, 255:red, 239; green, 12; blue, 16 }  ,draw opacity=1 ]   (100,30) -- (180,30) ;
\draw  [fill={rgb, 255:red, 128; green, 128; blue, 128 }  ,fill opacity=1 ] (360,37) -- (370,37) -- (370,43) -- (360,43) -- cycle ;
\draw  [fill={rgb, 255:red, 74; green, 74; blue, 74 }  ,fill opacity=1 ] (369.33,69.98) -- (369.33,43.98) -- (373.34,39.98) -- (377.34,39.98) -- (377.34,39.98) -- (377.34,65.97) -- (373.34,69.98) -- (369.33,69.98) -- cycle ;
\draw  [fill={rgb, 255:red, 74; green, 74; blue, 74 }  ,fill opacity=1 ][line width=0.75]  (377.33,14) -- (377.33,40) -- (377.33,40) -- (373.33,40) -- (369.33,36) -- (369.33,10) -- (369.33,10) -- (373.33,10) -- cycle ;
\draw  [fill={rgb, 255:red, 0; green, 0; blue, 0 }  ,fill opacity=1 ] (375.33,34.98) .. controls (378.09,34.98) and (380.33,37.22) .. (380.33,39.98) -- (380.33,39.98) .. controls (380.33,42.74) and (378.09,44.98) .. (375.33,44.98) -- (368.33,44.98) .. controls (368.33,44.98) and (368.33,44.98) .. (368.33,44.98) -- (368.33,34.98) .. controls (368.33,34.98) and (368.33,34.98) .. (368.33,34.98) -- cycle ;
\draw  [draw opacity=0] (0,0) -- (400,0) -- (400,150) -- (0,150) -- cycle ;
\draw   (95,50) .. controls (95,48.62) and (96.12,47.5) .. (97.5,47.5) .. controls (98.88,47.5) and (100,48.62) .. (100,50) .. controls (100,51.38) and (98.88,52.5) .. (97.5,52.5) .. controls (96.12,52.5) and (95,51.38) .. (95,50) -- cycle ;
\draw [color={rgb, 255:red, 239; green, 12; blue, 16 }  ,draw opacity=1 ] [dash pattern={on 4.5pt off 4.5pt}]  (130,70) -- (190,70) ;
\draw  [line width=0.75]  (20,28) .. controls (20,23.58) and (23.58,20) .. (28,20) -- (72,20) .. controls (76.42,20) and (80,23.58) .. (80,28) -- (80,52) .. controls (80,56.42) and (76.42,60) .. (72,60) -- (28,60) .. controls (23.58,60) and (20,56.42) .. (20,52) -- cycle ;
\draw   (95,30) .. controls (95,28.62) and (96.12,27.5) .. (97.5,27.5) .. controls (98.88,27.5) and (100,28.62) .. (100,30) .. controls (100,31.38) and (98.88,32.5) .. (97.5,32.5) .. controls (96.12,32.5) and (95,31.38) .. (95,30) -- cycle ;
\draw  [fill={rgb, 255:red, 128; green, 128; blue, 128 }  ,fill opacity=1 ] (360,117) -- (370,117) -- (370,123) -- (360,123) -- cycle ;
\draw  [fill={rgb, 255:red, 74; green, 74; blue, 74 }  ,fill opacity=1 ] (369.33,149.98) -- (369.33,123.98) -- (373.34,119.98) -- (377.34,119.98) -- (377.34,119.98) -- (377.34,145.97) -- (373.34,149.98) -- (369.33,149.98) -- cycle ;
\draw  [fill={rgb, 255:red, 74; green, 74; blue, 74 }  ,fill opacity=1 ][line width=0.75]  (377.33,94) -- (377.33,120) -- (377.33,120) -- (373.33,120) -- (369.33,116) -- (369.33,90) -- (369.33,90) -- (373.33,90) -- cycle ;
\draw  [fill={rgb, 255:red, 0; green, 0; blue, 0 }  ,fill opacity=1 ] (375.33,114.98) .. controls (378.09,114.98) and (380.33,117.22) .. (380.33,119.98) -- (380.33,119.98) .. controls (380.33,122.74) and (378.09,124.98) .. (375.33,124.98) -- (368.33,124.98) .. controls (368.33,124.98) and (368.33,124.98) .. (368.33,124.98) -- (368.33,114.98) .. controls (368.33,114.98) and (368.33,114.98) .. (368.33,114.98) -- cycle ;
\draw [line width=2.25]  [dash pattern={on 2.53pt off 3.02pt}]  (220,80) -- (242,80) ;
\draw   (180,30) .. controls (180,28.62) and (181.12,27.5) .. (182.5,27.5) .. controls (183.88,27.5) and (185,28.62) .. (185,30) .. controls (185,31.38) and (183.88,32.5) .. (182.5,32.5) .. controls (181.12,32.5) and (180,31.38) .. (180,30) -- cycle ;
\draw   (180,50) .. controls (180,48.62) and (181.12,47.5) .. (182.5,47.5) .. controls (183.88,47.5) and (185,48.62) .. (185,50) .. controls (185,51.38) and (183.88,52.5) .. (182.5,52.5) .. controls (181.12,52.5) and (180,51.38) .. (180,50) -- cycle ;
\draw   (275,50) .. controls (275,48.62) and (276.12,47.5) .. (277.5,47.5) .. controls (278.88,47.5) and (280,48.62) .. (280,50) .. controls (280,51.38) and (278.88,52.5) .. (277.5,52.5) .. controls (276.12,52.5) and (275,51.38) .. (275,50) -- cycle ;
\draw   (275,30) .. controls (275,28.62) and (276.12,27.5) .. (277.5,27.5) .. controls (278.88,27.5) and (280,28.62) .. (280,30) .. controls (280,31.38) and (278.88,32.5) .. (277.5,32.5) .. controls (276.12,32.5) and (275,31.38) .. (275,30) -- cycle ;
\draw   (275,110) .. controls (275,108.62) and (276.12,107.5) .. (277.5,107.5) .. controls (278.88,107.5) and (280,108.62) .. (280,110) .. controls (280,111.38) and (278.88,112.5) .. (277.5,112.5) .. controls (276.12,112.5) and (275,111.38) .. (275,110) -- cycle ;
\draw   (275,130) .. controls (275,128.62) and (276.12,127.5) .. (277.5,127.5) .. controls (278.88,127.5) and (280,128.62) .. (280,130) .. controls (280,131.38) and (278.88,132.5) .. (277.5,132.5) .. controls (276.12,132.5) and (275,131.38) .. (275,130) -- cycle ;
\draw   (180,130) .. controls (180,128.62) and (181.12,127.5) .. (182.5,127.5) .. controls (183.88,127.5) and (185,128.62) .. (185,130) .. controls (185,131.38) and (183.88,132.5) .. (182.5,132.5) .. controls (181.12,132.5) and (180,131.38) .. (180,130) -- cycle ;
\draw  [color={rgb, 255:red, 0; green, 0; blue, 0 }  ,draw opacity=1 ] (275,120) .. controls (275,118.62) and (276.12,117.5) .. (277.5,117.5) .. controls (278.88,117.5) and (280,118.62) .. (280,120) .. controls (280,121.38) and (278.88,122.5) .. (277.5,122.5) .. controls (276.12,122.5) and (275,121.38) .. (275,120) -- cycle ;
\draw  [color={rgb, 255:red, 0; green, 0; blue, 0 }  ,draw opacity=1 ] (275,40) .. controls (275,38.62) and (276.12,37.5) .. (277.5,37.5) .. controls (278.88,37.5) and (280,38.62) .. (280,40) .. controls (280,41.38) and (278.88,42.5) .. (277.5,42.5) .. controls (276.12,42.5) and (275,41.38) .. (275,40) -- cycle ;
\draw    (100,50) -- (180,50) ;
\draw    (280,30) -- (295,30) ;
\draw    (280,50) -- (295,50) ;
\draw    (280,110) -- (295,110) ;
\draw    (280,130) -- (295,130) ;
\draw    (260,30) -- (275,30) ;
\draw    (260,50) -- (275,50) ;
\draw    (260,110) -- (275,110) ;
\draw    (260,130) -- (275,130) ;
\draw    (185,130) -- (200,130) ;
\draw    (185,110) -- (200,110) ;
\draw    (185,50) -- (200,50) ;
\draw    (185,30) -- (200,30) ;
\draw    (80,50) -- (95,50) ;
\draw    (80,30) -- (95,30) ;
\draw [color={rgb, 255:red, 0; green, 0; blue, 0 }  ,draw opacity=1 ]   (280,40) -- (295,40) ;
\draw [color={rgb, 255:red, 0; green, 0; blue, 0 }  ,draw opacity=1 ]   (260,40) -- (275,40) ;
\draw [color={rgb, 255:red, 0; green, 0; blue, 0 }  ,draw opacity=1 ]   (260,120) -- (275,120) ;
\draw [color={rgb, 255:red, 0; green, 0; blue, 0 }  ,draw opacity=1 ]   (280,120) -- (295,120) ;
\draw  [line width=0.75]  (200,108) .. controls (200,103.58) and (203.58,100) .. (208,100) -- (252,100) .. controls (256.42,100) and (260,103.58) .. (260,108) -- (260,132) .. controls (260,136.42) and (256.42,140) .. (252,140) -- (208,140) .. controls (203.58,140) and (200,136.42) .. (200,132) -- cycle ;
\draw  [line width=0.75]  (295,108.02) .. controls (295,103.6) and (298.58,100.02) .. (303,100.02) -- (352,100.02) .. controls (356.42,100.02) and (360,103.6) .. (360,108.02) -- (360,132.02) .. controls (360,136.43) and (356.42,140.02) .. (352,140.02) -- (303,140.02) .. controls (298.58,140.02) and (295,136.43) .. (295,132.02) -- cycle ;
\draw  [line width=0.75]  (200,28) .. controls (200,23.58) and (203.58,20) .. (208,20) -- (252,20) .. controls (256.42,20) and (260,23.58) .. (260,28) -- (260,52) .. controls (260,56.42) and (256.42,60) .. (252,60) -- (208,60) .. controls (203.58,60) and (200,56.42) .. (200,52) -- cycle ;
\draw  [line width=0.75]  (295,28.02) .. controls (295,23.6) and (298.58,20.02) .. (303,20.02) -- (352,20.02) .. controls (356.42,20.02) and (360,23.6) .. (360,28.02) -- (360,52.02) .. controls (360,56.43) and (356.42,60.02) .. (352,60.02) -- (303,60.02) .. controls (298.58,60.02) and (295,56.43) .. (295,52.02) -- cycle ;
\draw  [dash pattern={on 4.5pt off 4.5pt}]  (110,90) -- (190,90) ;
\draw    (110,130) -- (180,130) ;
\draw [color={rgb, 255:red, 239; green, 12; blue, 16 }  ,draw opacity=1 ]   (130,110) -- (180,110) ;
\draw   (180,110) .. controls (180,108.62) and (181.12,107.5) .. (182.5,107.5) .. controls (183.88,107.5) and (185,108.62) .. (185,110) .. controls (185,111.38) and (183.88,112.5) .. (182.5,112.5) .. controls (181.12,112.5) and (180,111.38) .. (180,110) -- cycle ;
\draw [color={rgb, 255:red, 239; green, 12; blue, 16 }  ,draw opacity=1 ]   (130,110) -- (130,30) ;
\draw    (110,130) -- (110,50) ;

\draw (327.5,40.02) node  [font=\normalsize]  {$\text{BLDC}$};
\draw (230,40) node  [font=\normalsize]  {$\text{ESC}$};
\draw (50,40) node  [font=\normalsize]  {$\text{Battery}$};
\draw (230,120) node  [font=\normalsize]  {$\text{ESC}$};
\draw (327.5,120.02) node  [font=\normalsize]  {$\text{BLDC}$};

\end{tikzpicture}

%% file: Tikz_Control_Setup_short.tex
\tikzset{every picture/.style={line width=0.75pt}} %

\begin{tikzpicture}[x=0.6pt,y=0.6pt,yscale=-1,xscale=1]

\draw  [fill={rgb, 255:red, 209; green, 209; blue, 209 }  ,fill opacity=0.3 ] (135,15) -- (345,15) -- (345,104.79) -- (135,104.79) -- cycle ;
\draw  [fill={rgb, 255:red, 158; green, 202; blue, 255 }  ,fill opacity=0.4 ] (160,70) -- (320,70) -- (320,100) -- (160,100) -- cycle ;
\draw  [fill={rgb, 255:red, 128; green, 128; blue, 128 }  ,fill opacity=1 ] (292.83,76.36) -- (295.35,76.36) -- (295.35,77.3) -- (292.83,77.3) -- cycle ;
\draw  [fill={rgb, 255:red, 255; green, 255; blue, 255 }  ,fill opacity=1 ][line width=0.75]  (239.25,74.69) .. controls (239.25,73.9) and (239.89,73.25) .. (240.68,73.25) -- (255.41,73.25) .. controls (256.2,73.25) and (256.84,73.9) .. (256.84,74.69) -- (256.84,78.99) .. controls (256.84,79.79) and (256.2,80.43) .. (255.41,80.43) -- (240.68,80.43) .. controls (239.89,80.43) and (239.25,79.79) .. (239.25,78.99) -- cycle ;
\draw  [fill={rgb, 255:red, 255; green, 255; blue, 255 }  ,fill opacity=1 ][line width=0.75]  (272.23,74.51) .. controls (272.23,73.82) and (272.8,73.25) .. (273.49,73.25) -- (277.26,73.25) .. controls (277.95,73.25) and (278.51,73.82) .. (278.51,74.51) -- (278.51,79.17) .. controls (278.51,79.87) and (277.95,80.43) .. (277.26,80.43) -- (273.49,80.43) .. controls (272.8,80.43) and (272.23,79.87) .. (272.23,79.17) -- cycle ;
\draw  [fill={rgb, 255:red, 255; green, 255; blue, 255 }  ,fill opacity=1 ][line width=0.75]  (283.23,74.69) .. controls (283.23,73.9) and (283.87,73.25) .. (284.66,73.25) -- (290.59,73.25) .. controls (291.38,73.25) and (292.02,73.9) .. (292.02,74.69) -- (292.02,78.99) .. controls (292.02,79.79) and (291.38,80.43) .. (290.59,80.43) -- (284.66,80.43) .. controls (283.87,80.43) and (283.23,79.79) .. (283.23,78.99) -- cycle ;
\draw  [fill={rgb, 255:red, 74; green, 74; blue, 74 }  ,fill opacity=1 ] (295.35,81.49) -- (295.35,77.84) -- (296.37,76.83) -- (297.38,76.83) -- (297.38,76.83) -- (297.38,80.48) -- (296.37,81.49) -- (295.35,81.49) -- cycle ;
\draw  [fill={rgb, 255:red, 74; green, 74; blue, 74 }  ,fill opacity=1 ][line width=0.75]  (297.37,73.18) -- (297.37,76.83) -- (297.37,76.83) -- (296.36,76.83) -- (295.35,75.82) -- (295.35,72.17) -- (295.35,72.17) -- (296.36,72.17) -- cycle ;
\draw  [fill={rgb, 255:red, 0; green, 0; blue, 0 }  ,fill opacity=1 ][line width=0.75]  (297.36,76.05) .. controls (297.79,76.05) and (298.13,76.4) .. (298.13,76.83) -- (298.13,76.83) .. controls (298.13,77.26) and (297.79,77.6) .. (297.36,77.6) -- (295.1,77.6) .. controls (295.1,77.6) and (295.1,77.6) .. (295.1,77.6) -- (295.1,76.05) .. controls (295.1,76.05) and (295.1,76.05) .. (295.1,76.05) -- cycle ;
\draw    (256.84,79.08) -- (272.23,79.08) ;
\draw    (256.84,74.6) -- (272.23,74.6) ;
\draw    (278.53,74.67) -- (283.23,74.69) ;
\draw    (278.74,76.79) -- (283.44,76.81) ;
\draw    (278.53,78.97) -- (283.23,78.99) ;
\draw  [fill={rgb, 255:red, 128; green, 128; blue, 128 }  ,fill opacity=1 ] (292.94,92.79) -- (295.47,92.79) -- (295.47,93.72) -- (292.94,93.72) -- cycle ;
\draw  [fill={rgb, 255:red, 255; green, 255; blue, 255 }  ,fill opacity=1 ][line width=0.75]  (272.35,90.93) .. controls (272.35,90.24) and (272.91,89.68) .. (273.6,89.68) -- (277.37,89.68) .. controls (278.07,89.68) and (278.63,90.24) .. (278.63,90.93) -- (278.63,95.6) .. controls (278.63,96.29) and (278.07,96.85) .. (277.37,96.85) -- (273.6,96.85) .. controls (272.91,96.85) and (272.35,96.29) .. (272.35,95.6) -- cycle ;
\draw  [fill={rgb, 255:red, 255; green, 255; blue, 255 }  ,fill opacity=1 ][line width=0.75]  (283.34,91.11) .. controls (283.34,90.32) and (283.99,89.68) .. (284.78,89.68) -- (290.71,89.68) .. controls (291.5,89.68) and (292.14,90.32) .. (292.14,91.11) -- (292.14,95.42) .. controls (292.14,96.21) and (291.5,96.85) .. (290.71,96.85) -- (284.78,96.85) .. controls (283.99,96.85) and (283.34,96.21) .. (283.34,95.42) -- cycle ;
\draw  [fill={rgb, 255:red, 74; green, 74; blue, 74 }  ,fill opacity=1 ] (295.47,97.91) -- (295.47,94.26) -- (296.48,93.25) -- (297.49,93.25) -- (297.49,93.25) -- (297.49,96.9) -- (296.48,97.91) -- (295.47,97.91) -- cycle ;
\draw  [fill={rgb, 255:red, 74; green, 74; blue, 74 }  ,fill opacity=1 ][line width=0.75]  (297.49,89.6) -- (297.49,93.26) -- (297.49,93.26) -- (296.48,93.26) -- (295.47,92.24) -- (295.47,88.59) -- (295.47,88.59) -- (296.48,88.59) -- cycle ;
\draw  [fill={rgb, 255:red, 0; green, 0; blue, 0 }  ,fill opacity=1 ][line width=0.75]  (297.47,92.47) .. controls (297.9,92.47) and (298.25,92.82) .. (298.25,93.25) -- (298.25,93.25) .. controls (298.25,93.68) and (297.9,94.03) .. (297.47,94.03) -- (295.22,94.03) .. controls (295.22,94.03) and (295.22,94.03) .. (295.22,94.03) -- (295.22,92.47) .. controls (295.22,92.47) and (295.22,92.47) .. (295.22,92.47) -- cycle ;
\draw    (278.65,91.09) -- (283.34,91.11) ;
\draw    (278.86,93.22) -- (283.56,93.24) ;
\draw    (278.65,95.4) -- (283.34,95.42) ;
\draw  [dash pattern={on 0.84pt off 2.51pt}]  (276.92,85.06) -- (287.91,85.05) ;
\draw    (263.27,79.02) -- (263.22,95.57) -- (272.23,95.57) ;
\draw    (267.4,74.74) -- (267.4,91.08) -- (272.23,91.08) ;

\draw  [draw opacity=0] (190,40.82) -- (240,40.82) -- (240,70.82) -- (190,70.82) -- cycle ;
\draw  [fill={rgb, 255:red, 184; green, 233; blue, 134 }  ,fill opacity=0.4 ] (160,35) -- (320,35) -- (320,65) -- (160,65) -- cycle ;
\draw    (270,37.86) -- (270,41.72) ;
\draw  [color={rgb, 255:red, 0; green, 0; blue, 0 }  ,draw opacity=1 ][fill={rgb, 255:red, 184; green, 233; blue, 134 }  ,fill opacity=1 ] (266.67,39.91) -- (273.33,39.91) -- (273.33,43.52) -- (266.67,43.52) -- cycle ;
\draw [line width=1.5]    (270,41.72) -- (270,51.69) ;
\draw [line width=1.5]    (238.07,49) -- (301.93,49) ;
\draw  [draw opacity=0][fill={rgb, 255:red, 155; green, 155; blue, 155 }  ,fill opacity=1 ] (266.01,44.67) -- (273.99,44.67) -- (279.31,46.35) -- (279.31,51.65) -- (273.99,53.33) -- (266.01,53.33) -- (260.69,51.65) -- (260.69,46.35) -- cycle ;
\draw    (266.01,49.48) -- (273.99,49.48) ;
\draw    (273.99,53.33) -- (279.31,51.65) ;
\draw    (273.99,49.48) -- (279.31,47.8) ;
\draw    (260.69,47.8) -- (266.01,49.48) ;
\draw    (266.01,53.33) -- (260.69,51.65) ;
\draw    (260.69,51.65) -- (260.69,46.35) ;
\draw    (260.69,46.35) -- (266.01,44.67) ;
\draw    (273.99,44.67) -- (266.01,44.67) ;
\draw    (279.31,46.35) -- (278.64,46.14) -- (273.99,44.67) ;
\draw  [fill={rgb, 255:red, 74; green, 74; blue, 74 }  ,fill opacity=1 ] (222.62,44.63) -- (254.03,44.63) -- (254.03,44.63) -- (254.03,45.15) -- (253.51,45.67) -- (222.1,45.67) -- (222.1,45.67) -- (222.1,45.15) -- cycle ;
\draw [line width=1.5]    (270,51.41) -- (270,61.38) ;
\draw    (238.07,45.15) -- (238.07,49) ;
\draw    (270,57.52) -- (270,61.38) ;
\draw    (301.93,45.15) -- (301.93,49) ;
\draw  [fill={rgb, 255:red, 74; green, 74; blue, 74 }  ,fill opacity=1 ] (254.55,57.01) -- (285.97,57.01) -- (285.97,57.01) -- (285.97,57.52) -- (285.45,58.04) -- (254.03,58.04) -- (254.03,58.04) -- (254.03,57.52) -- cycle ;
\draw  [fill={rgb, 255:red, 74; green, 74; blue, 74 }  ,fill opacity=1 ] (254.55,37.34) -- (285.97,37.34) -- (285.97,37.34) -- (285.97,37.86) -- (285.45,38.38) -- (254.03,38.38) -- (254.03,38.38) -- (254.03,37.86) -- cycle ;
\draw  [fill={rgb, 255:red, 74; green, 74; blue, 74 }  ,fill opacity=1 ] (286.49,44.63) -- (317.9,44.63) -- (317.9,44.63) -- (317.9,45.15) -- (317.38,45.67) -- (285.97,45.67) -- (285.97,45.67) -- (285.97,45.15) -- cycle ;
\draw  [color={rgb, 255:red, 0; green, 0; blue, 0 }  ,draw opacity=1 ][fill={rgb, 255:red, 158; green, 202; blue, 255 }  ,fill opacity=1 ] (234.87,47.18) -- (241.26,47.18) -- (241.26,50.82) -- (234.87,50.82) -- cycle ;
\draw  [color={rgb, 255:red, 0; green, 0; blue, 0 }  ,draw opacity=1 ][fill={rgb, 255:red, 158; green, 202; blue, 255 }  ,fill opacity=1 ] (298.74,47.18) -- (305.13,47.18) -- (305.13,50.82) -- (298.74,50.82) -- cycle ;
\draw  [color={rgb, 255:red, 0; green, 0; blue, 0 }  ,draw opacity=1 ][fill={rgb, 255:red, 184; green, 233; blue, 134 }  ,fill opacity=1 ] (266.67,59.57) -- (273.33,59.57) -- (273.33,63.18) -- (266.67,63.18) -- cycle ;
\draw    (266.01,53.33) -- (273.99,53.33) ;
\draw    (279.31,51.65) -- (279.31,46.35) ;
\draw [line width=0.75]    (0,30) -- (132,30) ;
\draw [shift={(135,30)}, rotate = 180] [fill={rgb, 255:red, 0; green, 0; blue, 0 }  ][line width=0.08]  [draw opacity=0] (5.36,-2.57) -- (0,0) -- (5.36,2.57) -- cycle    ;
\draw [line width=0.75]    (345,59.9) -- (397,59.88) ;
\draw [shift={(400,59.88)}, rotate = 179.98] [fill={rgb, 255:red, 0; green, 0; blue, 0 }  ][line width=0.08]  [draw opacity=0] (5.36,-2.57) -- (0,0) -- (5.36,2.57) -- cycle    ;
\draw [line width=0.75]    (75,95) -- (129,95) -- (132,95) ;
\draw [shift={(135,95)}, rotate = 180] [fill={rgb, 255:red, 0; green, 0; blue, 0 }  ][line width=0.08]  [draw opacity=0] (5.36,-2.57) -- (0,0) -- (5.36,2.57) -- cycle    ;
\draw   (35,75) -- (75,75) -- (75,105) -- (35,105) -- cycle ;
\draw [line width=0.75]    (0,95) -- (32,95) ;
\draw [shift={(35,95)}, rotate = 180] [fill={rgb, 255:red, 0; green, 0; blue, 0 }  ][line width=0.08]  [draw opacity=0] (5.36,-2.57) -- (0,0) -- (5.36,2.57) -- cycle    ;
\draw [line width=0.75]    (0,50) -- (132,50) ;
\draw [shift={(135,50)}, rotate = 180] [fill={rgb, 255:red, 0; green, 0; blue, 0 }  ][line width=0.08]  [draw opacity=0] (5.36,-2.57) -- (0,0) -- (5.36,2.57) -- cycle    ;
\draw [line width=0.75]    (55,50) -- (55,72) ;
\draw [shift={(55,75)}, rotate = 270] [fill={rgb, 255:red, 0; green, 0; blue, 0 }  ][line width=0.08]  [draw opacity=0] (5.36,-2.57) -- (0,0) -- (5.36,2.57) -- cycle    ;
\draw [line width=2.25]    (135,15) -- (135,105) ;
\draw [line width=2.25]    (345,15) -- (345,104.79) ;
\draw  [draw opacity=0] (150,15) -- (240,15) -- (240,25) -- (150,25) -- cycle ;
\draw  [draw opacity=0] (0,0) -- (400,0) -- (400,105.55) -- (0,105.55) -- cycle ;
\draw  [draw opacity=0] (150,50) -- (160,50) -- (160,85) -- (150,85) -- cycle ;

\draw (347,56.5) node [anchor=south west] [inner sep=0.75pt]  [font=\scriptsize]  {$\mathbf{x}( k+1)$};
\draw (77,86.6) node [anchor=south west] [inner sep=0.75pt]  [font=\scriptsize]  {$\Delta T( k)$};

\draw (55,90) node  [font=\scriptsize]  {$f_{T}( \cdot )$};

\draw (135,11.6) node [anchor=south] [inner sep=0.75pt]  [font=\scriptsize]  {$\mathbf{u}( k)$};
\draw (240,25) node  [font=\scriptsize]  {$\mathbf{x}( k+1) =\mathbf{A}_{\text{d}}\mathbf{x}( k) +\mathbf{B}_{\text{d}}\mathbf{u}( k) +\mathbf{E}_{\text{d}}$};
\draw (345,11.6) node [anchor=south] [inner sep=0.75pt]  [font=\scriptsize]  {$\mathbf{x}( k+1)$};
\draw (162,50) node [anchor=west] [inner sep=0.75pt]  [font=\footnotesize] [align=left] {Lin. UAV};
\draw (162,85) node [anchor=west] [inner sep=0.75pt]  [font=\footnotesize] [align=left] {LPV ECM};

\draw (1,26.6) node [anchor=south west] [inner sep=0.75pt]  [font=\scriptsize]  {$\boldsymbol{\tau }( k)$};
\draw (1,46.6) node [anchor=south west] [inner sep=0.75pt]  [font=\scriptsize]  {$L(k)$};

\draw (1,75.6) node [anchor=south west] [inner sep=0.75pt]  [font=\scriptsize]  {$ \mathbf{v}(k)$};
\draw (1,91.6) node [anchor=south west] [inner sep=0.75pt]  [font=\scriptsize]  {$ \boldsymbol{\Psi}(k)$};

\end{tikzpicture}

%% file: Tikz_Approx_Sphere.tex
\tikzset{every picture/.style={line width=0.75pt}} %

\begin{tikzpicture}[x=0.6pt,y=0.6pt,yscale=-1,xscale=1]

\draw  [draw opacity=0][fill={rgb, 255:red, 158; green, 202; blue, 255 }  ,fill opacity=0.4 ] (82.93,61.33) -- (222.33,61.33) -- (209.4,80) -- (70,80) -- cycle ;
\draw  [draw opacity=0][fill={rgb, 255:red, 158; green, 202; blue, 255 }  ,fill opacity=0.4 ] (250,10) -- (390,10) -- (390,150) -- (250,150) -- cycle ;
\draw [color={rgb, 255:red, 239; green, 12; blue, 16 }  ,draw opacity=0.4 ] [dash pattern={on 0.84pt off 2.51pt}]  (400,70) -- (310,160) ;
\draw [color={rgb, 255:red, 239; green, 12; blue, 16 }  ,draw opacity=0.4 ] [dash pattern={on 0.84pt off 2.51pt}]  (310,0) -- (400,90) ;
\draw [color={rgb, 255:red, 239; green, 12; blue, 16 }  ,draw opacity=0.4 ] [dash pattern={on 0.84pt off 2.51pt}]  (330,0) -- (240,90) ;
\draw [color={rgb, 255:red, 239; green, 12; blue, 16 }  ,draw opacity=0.4 ] [dash pattern={on 0.84pt off 2.51pt}]  (250,30) -- (390,30) ;
\draw [color={rgb, 255:red, 239; green, 12; blue, 16 }  ,draw opacity=0.4 ] [dash pattern={on 0.84pt off 2.51pt}]  (270,150) -- (270,10) ;
\draw [color={rgb, 255:red, 239; green, 12; blue, 16 }  ,draw opacity=0.4 ] [dash pattern={on 0.84pt off 2.51pt}]  (240,70) -- (330,160) ;
\draw [color={rgb, 255:red, 239; green, 12; blue, 16 }  ,draw opacity=0.4 ] [dash pattern={on 0.84pt off 2.51pt}]  (370,150) -- (370,10) ;
\draw [color={rgb, 255:red, 239; green, 12; blue, 16 }  ,draw opacity=0.4 ] [dash pattern={on 0.84pt off 2.51pt}]  (250,130) -- (390,130) ;
\draw [color={rgb, 255:red, 239; green, 12; blue, 16 }  ,draw opacity=0.4 ] [dash pattern={on 0.84pt off 2.51pt}]  (150,0) -- (60,90) ;
\draw [color={rgb, 255:red, 239; green, 12; blue, 16 }  ,draw opacity=0.4 ] [dash pattern={on 0.84pt off 2.51pt}]  (220,70) -- (130,160) ;
\draw [color={rgb, 255:red, 239; green, 12; blue, 16 }  ,draw opacity=0.4 ] [dash pattern={on 0.84pt off 2.51pt}]  (190,150) -- (190,10) ;
\draw [color={rgb, 255:red, 239; green, 12; blue, 16 }  ,draw opacity=0.4 ] [dash pattern={on 0.84pt off 2.51pt}]  (70,30) -- (210,30) ;
\draw  [fill={rgb, 255:red, 209; green, 209; blue, 209 }  ,fill opacity=0.49 ] (10,20) .. controls (10,14.48) and (14.48,10) .. (20,10) .. controls (25.52,10) and (30,14.48) .. (30,20) .. controls (30,25.52) and (25.52,30) .. (20,30) .. controls (14.48,30) and (10,25.52) .. (10,20) -- cycle ;
\draw    (70,80) -- (207,80) ;
\draw [shift={(210,80)}, rotate = 180] [fill={rgb, 255:red, 0; green, 0; blue, 0 }  ][line width=0.08]  [draw opacity=0] (6.25,-3) -- (0,0) -- (6.25,3) -- cycle    ;
\draw    (140,147) -- (140,10) ;
\draw [shift={(140,150)}, rotate = 270] [fill={rgb, 255:red, 0; green, 0; blue, 0 }  ][line width=0.08]  [draw opacity=0] (6.25,-3) -- (0,0) -- (6.25,3) -- cycle    ;
\draw  [draw opacity=0] (0,0) -- (400,0) -- (400,160) -- (0,160) -- cycle ;
\draw    (20.03,20) -- (37.35,11.34) ;
\draw [shift={(40.03,10)}, rotate = 153.43] [fill={rgb, 255:red, 0; green, 0; blue, 0 }  ][line width=0.08]  [draw opacity=0] (5.36,-2.57) -- (0,0) -- (5.36,2.57) -- cycle    ;
\draw    (20.03,20) -- (47.03,20) ;
\draw [shift={(50.03,20)}, rotate = 180.01] [fill={rgb, 255:red, 0; green, 0; blue, 0 }  ][line width=0.08]  [draw opacity=0] (5.36,-2.57) -- (0,0) -- (5.36,2.57) -- cycle    ;
\draw    (20.03,20) -- (20.03,47) ;
\draw [shift={(20.03,50)}, rotate = 270] [fill={rgb, 255:red, 0; green, 0; blue, 0 }  ][line width=0.08]  [draw opacity=0] (5.36,-2.57) -- (0,0) -- (5.36,2.57) -- cycle    ;
\draw  [fill={rgb, 255:red, 0; green, 0; blue, 0 }  ,fill opacity=1 ] (18.7,20) .. controls (18.7,19.26) and (19.3,18.67) .. (20.03,18.67) .. controls (20.77,18.67) and (21.36,19.26) .. (21.36,20) .. controls (21.36,20.74) and (20.77,21.33) .. (20.03,21.33) .. controls (19.3,21.33) and (18.7,20.74) .. (18.7,20) -- cycle ;
\draw [color={rgb, 255:red, 0; green, 0; blue, 0 }  ,draw opacity=1 ]   (140,80) -- (172.88,47.12) ;
\draw [shift={(175,45)}, rotate = 135] [fill={rgb, 255:red, 0; green, 0; blue, 0 }  ,fill opacity=1 ][line width=0.08]  [draw opacity=0] (6.25,-3) -- (0,0) -- (6.25,3) -- cycle    ;
\draw [color={rgb, 255:red, 0; green, 0; blue, 0 }  ,draw opacity=1 ]   (140,80) -- (140,127) ;
\draw [shift={(140,130)}, rotate = 270] [fill={rgb, 255:red, 0; green, 0; blue, 0 }  ,fill opacity=1 ][line width=0.08]  [draw opacity=0] (6.25,-3) -- (0,0) -- (6.25,3) -- cycle    ;
\draw [color={rgb, 255:red, 0; green, 0; blue, 0 }  ,draw opacity=1 ]   (140,80) -- (187,80) ;
\draw [shift={(190,80)}, rotate = 180] [fill={rgb, 255:red, 0; green, 0; blue, 0 }  ,fill opacity=1 ][line width=0.08]  [draw opacity=0] (6.25,-3) -- (0,0) -- (6.25,3) -- cycle    ;
\draw [color={rgb, 255:red, 0; green, 0; blue, 0 }  ,draw opacity=1 ]   (140,80) -- (140,33) ;
\draw [shift={(140,30)}, rotate = 90] [fill={rgb, 255:red, 0; green, 0; blue, 0 }  ,fill opacity=1 ][line width=0.08]  [draw opacity=0] (6.25,-3) -- (0,0) -- (6.25,3) -- cycle    ;
\draw [color={rgb, 255:red, 0; green, 0; blue, 0 }  ,draw opacity=1 ]   (140,80) -- (172.88,112.88) ;
\draw [shift={(175,115)}, rotate = 225] [fill={rgb, 255:red, 0; green, 0; blue, 0 }  ,fill opacity=1 ][line width=0.08]  [draw opacity=0] (6.25,-3) -- (0,0) -- (6.25,3) -- cycle    ;
\draw  [draw opacity=0] (105,45) -- (175,45) -- (175,115) -- (105,115) -- cycle ;
\draw  [color={rgb, 255:red, 0; green, 0; blue, 0 }  ,draw opacity=1 ][dash pattern={on 4.5pt off 4.5pt}] (270,80) .. controls (270,52.39) and (292.39,30) .. (320,30) .. controls (347.61,30) and (370,52.39) .. (370,80) .. controls (370,107.61) and (347.61,130) .. (320,130) .. controls (292.39,130) and (270,107.61) .. (270,80) -- cycle ;
\draw    (250,80) -- (387,80) ;
\draw [shift={(390,80)}, rotate = 180] [fill={rgb, 255:red, 0; green, 0; blue, 0 }  ][line width=0.08]  [draw opacity=0] (6.25,-3) -- (0,0) -- (6.25,3) -- cycle    ;
\draw    (320,150) -- (320,13) ;
\draw [shift={(320,10)}, rotate = 90] [fill={rgb, 255:red, 0; green, 0; blue, 0 }  ][line width=0.08]  [draw opacity=0] (6.25,-3) -- (0,0) -- (6.25,3) -- cycle    ;
\draw [color={rgb, 255:red, 0; green, 0; blue, 0 }  ,draw opacity=1 ]   (320,80) -- (352.88,47.12) ;
\draw [shift={(355,45)}, rotate = 135] [fill={rgb, 255:red, 0; green, 0; blue, 0 }  ,fill opacity=1 ][line width=0.08]  [draw opacity=0] (6.25,-3) -- (0,0) -- (6.25,3) -- cycle    ;
\draw [color={rgb, 255:red, 0; green, 0; blue, 0 }  ,draw opacity=1 ]   (320,80) -- (320,127) ;
\draw [shift={(320,130)}, rotate = 270] [fill={rgb, 255:red, 0; green, 0; blue, 0 }  ,fill opacity=1 ][line width=0.08]  [draw opacity=0] (6.25,-3) -- (0,0) -- (6.25,3) -- cycle    ;
\draw [color={rgb, 255:red, 0; green, 0; blue, 0 }  ,draw opacity=1 ]   (320,80) -- (367,80) ;
\draw [shift={(370,80)}, rotate = 180] [fill={rgb, 255:red, 0; green, 0; blue, 0 }  ,fill opacity=1 ][line width=0.08]  [draw opacity=0] (6.25,-3) -- (0,0) -- (6.25,3) -- cycle    ;
\draw [color={rgb, 255:red, 0; green, 0; blue, 0 }  ,draw opacity=1 ]   (320,80) -- (320,33) ;
\draw [shift={(320,30)}, rotate = 90] [fill={rgb, 255:red, 0; green, 0; blue, 0 }  ,fill opacity=1 ][line width=0.08]  [draw opacity=0] (6.25,-3) -- (0,0) -- (6.25,3) -- cycle    ;
\draw [color={rgb, 255:red, 0; green, 0; blue, 0 }  ,draw opacity=1 ]   (320,80) -- (352.88,112.88) ;
\draw [shift={(355,115)}, rotate = 225] [fill={rgb, 255:red, 0; green, 0; blue, 0 }  ,fill opacity=1 ][line width=0.08]  [draw opacity=0] (6.25,-3) -- (0,0) -- (6.25,3) -- cycle    ;
\draw [color={rgb, 255:red, 0; green, 0; blue, 0 }  ,draw opacity=1 ]   (320,80) -- (287.12,112.88) ;
\draw [shift={(285,115)}, rotate = 315] [fill={rgb, 255:red, 0; green, 0; blue, 0 }  ,fill opacity=1 ][line width=0.08]  [draw opacity=0] (6.25,-3) -- (0,0) -- (6.25,3) -- cycle    ;
\draw [color={rgb, 255:red, 0; green, 0; blue, 0 }  ,draw opacity=1 ]   (320,80) -- (287.12,47.12) ;
\draw [shift={(285,45)}, rotate = 45] [fill={rgb, 255:red, 0; green, 0; blue, 0 }  ,fill opacity=1 ][line width=0.08]  [draw opacity=0] (6.25,-3) -- (0,0) -- (6.25,3) -- cycle    ;
\draw [color={rgb, 255:red, 0; green, 0; blue, 0 }  ,draw opacity=1 ]   (320,80) -- (273,80) ;
\draw [shift={(270,80)}, rotate = 360] [fill={rgb, 255:red, 0; green, 0; blue, 0 }  ,fill opacity=1 ][line width=0.08]  [draw opacity=0] (6.25,-3) -- (0,0) -- (6.25,3) -- cycle    ;
\draw  [draw opacity=0] (140,50) .. controls (140,50) and (140,50) .. (140,50) .. controls (148.07,50) and (155.39,53.18) .. (160.78,58.37) -- (140,80) -- cycle ; \draw [color={rgb, 255:red, 21; green, 10; blue, 255 }  ,draw opacity=1 ]   (140,50) .. controls (140,50) and (140,50) .. (140,50) .. controls (147.02,50) and (153.47,52.41) .. (158.59,56.45) ; \draw [shift={(160.78,58.37)}, rotate = 220.24] [fill={rgb, 255:red, 21; green, 10; blue, 255 }  ,fill opacity=1 ][line width=0.08]  [draw opacity=0] (5.36,-2.57) -- (0,0) -- (5.36,2.57) -- (3.56,0) -- cycle    ; 
\draw  [draw opacity=0] (320,50) .. controls (328.07,50) and (335.39,53.18) .. (340.78,58.37) -- (320,80) -- cycle ; \draw [color={rgb, 255:red, 21; green, 10; blue, 255 }  ,draw opacity=1 ]   (320,50) .. controls (327.02,50) and (333.47,52.41) .. (338.59,56.45) ; \draw [shift={(340.78,58.37)}, rotate = 220.24] [fill={rgb, 255:red, 21; green, 10; blue, 255 }  ,fill opacity=1 ][line width=0.08]  [draw opacity=0] (5.36,-2.57) -- (0,0) -- (5.36,2.57) -- (3.56,0) -- cycle    ; 
\draw [color={rgb, 255:red, 239; green, 12; blue, 16 }  ,draw opacity=0.4 ] [dash pattern={on 0.84pt off 2.51pt}]  (90,150) -- (90,10) ;
\draw [color={rgb, 255:red, 239; green, 12; blue, 16 }  ,draw opacity=0.4 ] [dash pattern={on 0.84pt off 2.51pt}]  (70,130) -- (210,130) ;
\draw [color={rgb, 255:red, 239; green, 12; blue, 16 }  ,draw opacity=0.4 ] [dash pattern={on 0.84pt off 2.51pt}]  (130,0) -- (220,90) ;
\draw [color={rgb, 255:red, 239; green, 12; blue, 16 }  ,draw opacity=0.4 ] [dash pattern={on 0.84pt off 2.51pt}]  (60,70) -- (150,160) ;
\draw  [draw opacity=0][fill={rgb, 255:red, 158; green, 202; blue, 255 }  ,fill opacity=0.4 ] (70.6,80) -- (210,80) -- (197.07,98.67) -- (57.67,98.67) -- cycle ;
\draw  [color={rgb, 255:red, 0; green, 0; blue, 0 }  ,draw opacity=1 ][dash pattern={on 4.5pt off 4.5pt}] (90,80) .. controls (90,52.39) and (112.39,30) .. (140,30) .. controls (167.61,30) and (190,52.39) .. (190,80) .. controls (190,107.61) and (167.61,130) .. (140,130) .. controls (112.39,130) and (90,107.61) .. (90,80) -- cycle ;
\draw  [color={rgb, 255:red, 239; green, 12; blue, 16 }  ,draw opacity=1 ] (189.77,100.62) -- (160.62,129.77) -- (119.38,129.77) -- (90.23,100.62) -- (90.23,59.38) -- (119.38,30.23) -- (160.62,30.23) -- (189.77,59.38) -- cycle ;
\draw  [color={rgb, 255:red, 239; green, 12; blue, 16 }  ,draw opacity=1 ] (369.77,100.62) -- (340.62,129.77) -- (299.38,129.77) -- (270.23,100.62) -- (270.23,59.38) -- (299.38,30.23) -- (340.62,30.23) -- (369.77,59.38) -- cycle ;

\draw (42.03,10) node [anchor=west] [inner sep=0.75pt]  [font=\tiny] [align=left] {x};
\draw (50.03,23) node [anchor=north] [inner sep=0.75pt]  [font=\tiny] [align=left] {y};
\draw (22.03,47) node [anchor=south west] [inner sep=0.75pt]  [font=\tiny] [align=left] {z};
\draw (142,150) node [anchor=west] [inner sep=0.75pt]   [align=left] {z};
\draw (210,83) node [anchor=north] [inner sep=0.75pt]   [align=left] {y};
\draw (322,13) node [anchor=north west][inner sep=0.75pt]   [align=left] {x};
\draw (388,83) node [anchor=north east] [inner sep=0.75pt]   [align=left] {y};
\draw (148.33,34.73) node [anchor=north west][inner sep=0.75pt]  [font=\scriptsize,color={rgb, 255:red, 21; green, 10; blue, 255 }  ,opacity=1 ]  {$\beta _{j}$};
\draw (328.67,36.07) node [anchor=north west][inner sep=0.75pt]  [font=\scriptsize,color={rgb, 255:red, 21; green, 10; blue, 255 }  ,opacity=1 ]  {$\alpha _{i}$};

\end{tikzpicture}

%% file: Tikz_PolygonApprox.tex
\tikzset{every picture/.style={line width=0.75pt}} %

\begin{tikzpicture}[x=0.6pt,y=0.6pt,yscale=-1,xscale=1]

\draw  [color={rgb, 255:red, 239; green, 12; blue, 16 }  ,draw opacity=1 ][fill={rgb, 255:red, 155; green, 155; blue, 155 }  ,fill opacity=1 ] (50,30) -- (150,30) -- (150,130) -- (50,130) -- cycle ;
\draw  [color={rgb, 255:red, 239; green, 12; blue, 16 }  ,draw opacity=1 ][fill={rgb, 255:red, 155; green, 155; blue, 155 }  ,fill opacity=1 ] (349.89,59.34) -- (349.89,100.66) -- (320.66,129.89) -- (279.34,129.89) -- (250.11,100.66) -- (250.11,59.34) -- (279.34,30.11) -- (320.66,30.11) -- cycle ;
\draw  [color={rgb, 255:red, 0; green, 0; blue, 0 }  ,draw opacity=1 ][fill={rgb, 255:red, 255; green, 255; blue, 255 }  ,fill opacity=1 ] (50,80) .. controls (50,52.39) and (72.39,30) .. (100,30) .. controls (127.61,30) and (150,52.39) .. (150,80) .. controls (150,107.61) and (127.61,130) .. (100,130) .. controls (72.39,130) and (50,107.61) .. (50,80) -- cycle ;
\draw    (30,80) -- (167,80) ;
\draw [shift={(170,80)}, rotate = 180] [fill={rgb, 255:red, 0; green, 0; blue, 0 }  ][line width=0.08]  [draw opacity=0] (6.25,-3) -- (0,0) -- (6.25,3) -- cycle    ;
\draw    (100,85) -- (100,13) ;
\draw [shift={(100,10)}, rotate = 90] [fill={rgb, 255:red, 0; green, 0; blue, 0 }  ][line width=0.08]  [draw opacity=0] (6.25,-3) -- (0,0) -- (6.25,3) -- cycle    ;
\draw  [draw opacity=0] (0,0) -- (400,0) -- (400,149.95) -- (0,149.95) -- cycle ;
\draw    (100,150) -- (100,105) ;
\draw [draw opacity=0]   (100,85) -- (100,93) -- (100,105) ;
\draw  [draw opacity=0] (30,10) -- (170,10) -- (170,150) -- (30,150) -- cycle ;
\draw  [color={rgb, 255:red, 0; green, 0; blue, 0 }  ,draw opacity=1 ][fill={rgb, 255:red, 255; green, 255; blue, 255 }  ,fill opacity=1 ] (250,80) .. controls (250,52.39) and (272.39,30) .. (300,30) .. controls (327.61,30) and (350,52.39) .. (350,80) .. controls (350,107.61) and (327.61,130) .. (300,130) .. controls (272.39,130) and (250,107.61) .. (250,80) -- cycle ;
\draw    (230,80) -- (367,80) ;
\draw [shift={(370,80)}, rotate = 180] [fill={rgb, 255:red, 0; green, 0; blue, 0 }  ][line width=0.08]  [draw opacity=0] (6.25,-3) -- (0,0) -- (6.25,3) -- cycle    ;
\draw    (300,85) -- (300,13) ;
\draw [shift={(300,10)}, rotate = 90] [fill={rgb, 255:red, 0; green, 0; blue, 0 }  ][line width=0.08]  [draw opacity=0] (6.25,-3) -- (0,0) -- (6.25,3) -- cycle    ;
\draw    (300,150) -- (300,105) ;
\draw [draw opacity=0]   (300,85) -- (300,93) -- (300,105) ;
\draw  [draw opacity=0] (230,10) -- (370,10) -- (370,150) -- (230,150) -- cycle ;
\draw  [draw opacity=0] (100,80) -- (200,80) -- (200,130) -- (100,130) -- cycle ;

\draw (100,93) node  [font=\scriptsize]  {$||\mathbf{v} || \leq v_{\text{max}}$};
\draw (30,10) node  [font=\normalsize]  {$H =4$};
\draw (102,10) node [anchor=west] [inner sep=0.75pt]    {$v_{\text{x}}$};
\draw (170,83.4) node [anchor=north] [inner sep=0.75pt]    {$v_{\text{y}}$};
\draw (230,10) node  [font=\normalsize]  {$H =8$};
\draw (302,10) node [anchor=west] [inner sep=0.75pt]    {$v_{\text{x}}$};
\draw (370,83.4) node [anchor=north] [inner sep=0.75pt]    {$v_{\text{y}}$};
\draw (300,93) node  [font=\scriptsize]  {$||\mathbf{v} || \leq v_{\text{max}}$};
\draw (200,130) node  [font=\scriptsize,color={rgb, 255:red, 239; green, 12; blue, 16 }  ,opacity=1 ]  {$ \begin{array}{l}
\ \ \ \ \ v_{h} \leq v_{\text{max}} \ \\
\forall h\in \{1,...,H\}
\end{array}$};

\end{tikzpicture}

%% file: Tikz_PPA.tex
\tikzset{every picture/.style={line width=0.75pt}} %

\begin{tikzpicture}[x=0.6pt,y=0.6pt,yscale=-1,xscale=1]

\draw  [draw opacity=0][fill={rgb, 255:red, 230; green, 230; blue, 230 }  ,fill opacity=1 ] (66,85.04) .. controls (66,74.25) and (74.75,65.5) .. (85.54,65.5) .. controls (96.33,65.5) and (105.08,74.25) .. (105.08,85.04) .. controls (105.08,95.83) and (96.33,104.58) .. (85.54,104.58) .. controls (74.75,104.58) and (66,95.83) .. (66,85.04) -- cycle ;
\draw  [color={rgb, 255:red, 126; green, 211; blue, 33 }  ,draw opacity=1 ] (82.5,85.5) .. controls (82.5,84.12) and (83.62,83) .. (85,83) .. controls (86.38,83) and (87.5,84.12) .. (87.5,85.5) .. controls (87.5,86.88) and (86.38,88) .. (85,88) .. controls (83.62,88) and (82.5,86.88) .. (82.5,85.5) -- cycle ;
\draw  [draw opacity=0][fill={rgb, 255:red, 230; green, 230; blue, 230 }  ,fill opacity=1 ] (17.82,24.85) .. controls (17.82,14.06) and (26.57,5.31) .. (37.36,5.31) .. controls (48.16,5.31) and (56.91,14.06) .. (56.91,24.85) .. controls (56.91,35.65) and (48.16,44.4) .. (37.36,44.4) .. controls (26.57,44.4) and (17.82,35.65) .. (17.82,24.85) -- cycle ;
\draw [draw opacity=0]   (91.93,79.19) -- (78.18,91.88) ;
\draw [draw opacity=0]   (78.18,79.19) -- (78.72,79.69) -- (91.93,91.88) ;
\draw [line width=1.5]    (80.76,89.5) -- (89.35,81.57) ;
\draw [line width=1.5]    (80.76,81.57) -- (89.35,89.5) ;
\draw  [fill={rgb, 255:red, 155; green, 155; blue, 155 }  ,fill opacity=1 ] (86.64,86.14) -- (85.71,87) -- (84.39,87) -- (83.46,86.14) -- (83.46,84.92) -- (84.39,84.07) -- (85.71,84.07) -- (86.64,84.92) -- cycle ;
\draw  [draw opacity=0] (78.18,79.19) -- (91.93,79.19) -- (91.93,91.88) -- (78.18,91.88) -- cycle ;
\draw  [fill={rgb, 255:red, 184; green, 233; blue, 134 }  ,fill opacity=1 ] (80.22,81.57) .. controls (80.22,81.29) and (80.46,81.07) .. (80.76,81.07) .. controls (81.05,81.07) and (81.29,81.29) .. (81.29,81.57) .. controls (81.29,81.84) and (81.05,82.06) .. (80.76,82.06) .. controls (80.46,82.06) and (80.22,81.84) .. (80.22,81.57) -- cycle ;
\draw  [fill={rgb, 255:red, 184; green, 233; blue, 134 }  ,fill opacity=1 ] (88.81,89.5) .. controls (88.81,89.22) and (89.05,89) .. (89.35,89) .. controls (89.65,89) and (89.89,89.22) .. (89.89,89.5) .. controls (89.89,89.77) and (89.65,89.99) .. (89.35,89.99) .. controls (89.05,89.99) and (88.81,89.77) .. (88.81,89.5) -- cycle ;
\draw  [fill={rgb, 255:red, 158; green, 202; blue, 255 }  ,fill opacity=1 ] (88.81,81.57) .. controls (88.81,81.29) and (89.05,81.07) .. (89.35,81.07) .. controls (89.65,81.07) and (89.89,81.29) .. (89.89,81.57) .. controls (89.89,81.84) and (89.65,82.06) .. (89.35,82.06) .. controls (89.05,82.06) and (88.81,81.84) .. (88.81,81.57) -- cycle ;
\draw  [fill={rgb, 255:red, 158; green, 202; blue, 255 }  ,fill opacity=1 ] (80.22,89.5) .. controls (80.22,89.22) and (80.46,89) .. (80.76,89) .. controls (81.05,89) and (81.29,89.22) .. (81.29,89.5) .. controls (81.29,89.77) and (81.05,89.99) .. (80.76,89.99) .. controls (80.46,89.99) and (80.22,89.77) .. (80.22,89.5) -- cycle ;
\draw  [fill={rgb, 255:red, 74; green, 74; blue, 74 }  ,fill opacity=1 ] (89.35,89.5) -- (90.89,87.8) -- (91.14,87.78) -- (91.27,87.88) -- (91.27,87.88) -- (89.72,89.58) -- (89.48,89.6) -- (89.35,89.5) -- cycle ;
\draw  [fill={rgb, 255:red, 74; green, 74; blue, 74 }  ,fill opacity=1 ] (89.35,89.5) -- (87.8,91.2) -- (87.56,91.22) -- (87.43,91.11) -- (87.43,91.11) -- (88.97,89.42) -- (89.22,89.39) -- (89.35,89.5) -- cycle ;
\draw  [fill={rgb, 255:red, 74; green, 74; blue, 74 }  ,fill opacity=1 ] (80.76,81.57) -- (82.3,79.87) -- (82.54,79.85) -- (82.68,79.95) -- (82.68,79.95) -- (81.13,81.65) -- (80.89,81.67) -- (80.76,81.57) -- cycle ;
\draw  [fill={rgb, 255:red, 74; green, 74; blue, 74 }  ,fill opacity=1 ] (80.76,81.57) -- (79.21,83.27) -- (78.97,83.29) -- (78.83,83.18) -- (78.83,83.18) -- (80.38,81.49) -- (80.62,81.46) -- (80.76,81.57) -- cycle ;
\draw  [fill={rgb, 255:red, 74; green, 74; blue, 74 }  ,fill opacity=1 ] (79.05,87.93) -- (80.76,89.5) -- (80.76,89.5) -- (80.63,89.61) -- (80.39,89.61) -- (78.69,88.04) -- (78.69,88.04) -- (78.81,87.93) -- cycle ;
\draw  [fill={rgb, 255:red, 74; green, 74; blue, 74 }  ,fill opacity=1 ] (81.12,89.38) -- (82.82,90.95) -- (82.82,90.95) -- (82.7,91.07) -- (82.46,91.07) -- (80.76,89.5) -- (80.76,89.5) -- (80.88,89.38) -- cycle ;
\draw  [fill={rgb, 255:red, 74; green, 74; blue, 74 }  ,fill opacity=1 ] (87.65,80) -- (89.35,81.57) -- (89.35,81.57) -- (89.23,81.68) -- (88.98,81.68) -- (87.28,80.11) -- (87.28,80.11) -- (87.4,80) -- cycle ;
\draw  [fill={rgb, 255:red, 74; green, 74; blue, 74 }  ,fill opacity=1 ] (89.71,81.45) -- (91.41,83.02) -- (91.41,83.02) -- (91.29,83.14) -- (91.05,83.14) -- (89.35,81.57) -- (89.35,81.57) -- (89.47,81.45) -- cycle ;
\draw  [fill={rgb, 255:red, 0; green, 0; blue, 0 }  ,fill opacity=1 ] (80.58,89.5) .. controls (80.58,89.41) and (80.66,89.34) .. (80.76,89.34) .. controls (80.85,89.34) and (80.93,89.41) .. (80.93,89.5) .. controls (80.93,89.58) and (80.85,89.65) .. (80.76,89.65) .. controls (80.66,89.65) and (80.58,89.58) .. (80.58,89.5) -- cycle ;
\draw  [fill={rgb, 255:red, 0; green, 0; blue, 0 }  ,fill opacity=1 ] (80.58,81.57) .. controls (80.58,81.48) and (80.66,81.41) .. (80.76,81.41) .. controls (80.85,81.41) and (80.93,81.48) .. (80.93,81.57) .. controls (80.93,81.65) and (80.85,81.72) .. (80.76,81.72) .. controls (80.66,81.72) and (80.58,81.65) .. (80.58,81.57) -- cycle ;
\draw  [fill={rgb, 255:red, 0; green, 0; blue, 0 }  ,fill opacity=1 ] (89.18,89.5) .. controls (89.18,89.41) and (89.25,89.34) .. (89.35,89.34) .. controls (89.44,89.34) and (89.52,89.41) .. (89.52,89.5) .. controls (89.52,89.58) and (89.44,89.65) .. (89.35,89.65) .. controls (89.25,89.65) and (89.18,89.58) .. (89.18,89.5) -- cycle ;
\draw  [fill={rgb, 255:red, 0; green, 0; blue, 0 }  ,fill opacity=1 ] (89.18,81.57) .. controls (89.18,81.48) and (89.25,81.41) .. (89.35,81.41) .. controls (89.44,81.41) and (89.52,81.48) .. (89.52,81.57) .. controls (89.52,81.65) and (89.44,81.72) .. (89.35,81.72) .. controls (89.25,81.72) and (89.18,81.65) .. (89.18,81.57) -- cycle ;
\draw  [draw opacity=0][fill={rgb, 255:red, 230; green, 230; blue, 230 }  ,fill opacity=1 ] (5,5.5) -- (37.33,5.5) -- (37.33,32.56) -- (5,32.56) -- cycle ;
\draw  [draw opacity=0][fill={rgb, 255:red, 230; green, 230; blue, 230 }  ,fill opacity=1 ] (5,5.5) .. controls (5,5.5) and (5,5.5) .. (5,5.5) -- (20.33,5.5) .. controls (28.8,5.5) and (35.67,12.36) .. (35.67,20.83) -- (35.67,44.23) .. controls (35.67,44.23) and (35.67,44.23) .. (35.67,44.23) -- (20.33,44.23) .. controls (11.86,44.23) and (5,37.36) .. (5,28.9) -- cycle ;
\draw   (5,5) -- (165,5) -- (165,105) -- (5,105) -- cycle ;
\draw  [color={rgb, 255:red, 126; green, 211; blue, 33 }  ,draw opacity=1 ] (22.5,25.5) .. controls (22.5,24.12) and (23.62,23) .. (25,23) .. controls (26.38,23) and (27.5,24.12) .. (27.5,25.5) .. controls (27.5,26.88) and (26.38,28) .. (25,28) .. controls (23.62,28) and (22.5,26.88) .. (22.5,25.5) -- cycle ;
\draw   (22.5,85.5) .. controls (22.5,84.12) and (23.62,83) .. (25,83) .. controls (26.38,83) and (27.5,84.12) .. (27.5,85.5) .. controls (27.5,86.88) and (26.38,88) .. (25,88) .. controls (23.62,88) and (22.5,86.88) .. (22.5,85.5) -- cycle ;
\draw   (22.5,55.5) .. controls (22.5,54.12) and (23.62,53) .. (25,53) .. controls (26.38,53) and (27.5,54.12) .. (27.5,55.5) .. controls (27.5,56.88) and (26.38,58) .. (25,58) .. controls (23.62,58) and (22.5,56.88) .. (22.5,55.5) -- cycle ;
\draw   (52.5,25.5) .. controls (52.5,24.12) and (53.62,23) .. (55,23) .. controls (56.38,23) and (57.5,24.12) .. (57.5,25.5) .. controls (57.5,26.88) and (56.38,28) .. (55,28) .. controls (53.62,28) and (52.5,26.88) .. (52.5,25.5) -- cycle ;
\draw   (52.5,85.5) .. controls (52.5,84.12) and (53.62,83) .. (55,83) .. controls (56.38,83) and (57.5,84.12) .. (57.5,85.5) .. controls (57.5,86.88) and (56.38,88) .. (55,88) .. controls (53.62,88) and (52.5,86.88) .. (52.5,85.5) -- cycle ;
\draw   (52.5,55.5) .. controls (52.5,54.12) and (53.62,53) .. (55,53) .. controls (56.38,53) and (57.5,54.12) .. (57.5,55.5) .. controls (57.5,56.88) and (56.38,58) .. (55,58) .. controls (53.62,58) and (52.5,56.88) .. (52.5,55.5) -- cycle ;
\draw   (82.5,25.5) .. controls (82.5,24.12) and (83.62,23) .. (85,23) .. controls (86.38,23) and (87.5,24.12) .. (87.5,25.5) .. controls (87.5,26.88) and (86.38,28) .. (85,28) .. controls (83.62,28) and (82.5,26.88) .. (82.5,25.5) -- cycle ;
\draw   (82.5,55.5) .. controls (82.5,54.12) and (83.62,53) .. (85,53) .. controls (86.38,53) and (87.5,54.12) .. (87.5,55.5) .. controls (87.5,56.88) and (86.38,58) .. (85,58) .. controls (83.62,58) and (82.5,56.88) .. (82.5,55.5) -- cycle ;
\draw   (112.5,25.5) .. controls (112.5,24.12) and (113.62,23) .. (115,23) .. controls (116.38,23) and (117.5,24.12) .. (117.5,25.5) .. controls (117.5,26.88) and (116.38,28) .. (115,28) .. controls (113.62,28) and (112.5,26.88) .. (112.5,25.5) -- cycle ;
\draw   (112.5,85.5) .. controls (112.5,84.12) and (113.62,83) .. (115,83) .. controls (116.38,83) and (117.5,84.12) .. (117.5,85.5) .. controls (117.5,86.88) and (116.38,88) .. (115,88) .. controls (113.62,88) and (112.5,86.88) .. (112.5,85.5) -- cycle ;
\draw   (112.5,55.5) .. controls (112.5,54.12) and (113.62,53) .. (115,53) .. controls (116.38,53) and (117.5,54.12) .. (117.5,55.5) .. controls (117.5,56.88) and (116.38,58) .. (115,58) .. controls (113.62,58) and (112.5,56.88) .. (112.5,55.5) -- cycle ;
\draw   (142.5,25.5) .. controls (142.5,24.12) and (143.62,23) .. (145,23) .. controls (146.38,23) and (147.5,24.12) .. (147.5,25.5) .. controls (147.5,26.88) and (146.38,28) .. (145,28) .. controls (143.62,28) and (142.5,26.88) .. (142.5,25.5) -- cycle ;
\draw   (142.5,85.5) .. controls (142.5,84.12) and (143.62,83) .. (145,83) .. controls (146.38,83) and (147.5,84.12) .. (147.5,85.5) .. controls (147.5,86.88) and (146.38,88) .. (145,88) .. controls (143.62,88) and (142.5,86.88) .. (142.5,85.5) -- cycle ;
\draw   (142.5,55.5) .. controls (142.5,54.12) and (143.62,53) .. (145,53) .. controls (146.38,53) and (147.5,54.12) .. (147.5,55.5) .. controls (147.5,56.88) and (146.38,58) .. (145,58) .. controls (143.62,58) and (142.5,56.88) .. (142.5,55.5) -- cycle ;
\draw  [fill={rgb, 255:red, 155; green, 155; blue, 155 }  ,fill opacity=1 ] (37.1,81.4) .. controls (39.1,80.6) and (43.5,86.6) .. (43.1,87.4) .. controls (42.7,88.2) and (47.1,94.6) .. (43.9,97.4) .. controls (40.7,100.2) and (38.3,98.6) .. (34.3,94.6) .. controls (30.3,90.6) and (35.1,82.2) .. (37.1,81.4) -- cycle ;
\draw  [fill={rgb, 255:red, 155; green, 155; blue, 155 }  ,fill opacity=1 ] (146.03,31.74) .. controls (148.03,30.94) and (152.43,36.94) .. (152.03,37.74) .. controls (151.63,38.54) and (156.03,44.94) .. (152.83,47.74) .. controls (149.63,50.54) and (141.23,50.33) .. (137.23,46.33) .. controls (133.23,42.33) and (144.03,32.54) .. (146.03,31.74) -- cycle ;
\draw [color={rgb, 255:red, 0; green, 0; blue, 0 }  ,draw opacity=0.6 ] [dash pattern={on 0.84pt off 2.51pt}]  (43.5,0) -- (153.5,110) ;
\draw  [color={rgb, 255:red, 0; green, 0; blue, 0 }  ,draw opacity=1 ][fill={rgb, 255:red, 239; green, 12; blue, 16 }  ,fill opacity=1 ] (92.3,57.4) .. controls (92.3,52.87) and (95.97,49.2) .. (100.5,49.2) .. controls (105.03,49.2) and (108.7,52.87) .. (108.7,57.4) .. controls (108.7,61.93) and (105.03,65.6) .. (100.5,65.6) .. controls (95.97,65.6) and (92.3,61.93) .. (92.3,57.4) -- cycle ;
\draw [color={rgb, 255:red, 0; green, 0; blue, 0 }  ,draw opacity=0.6 ] [dash pattern={on 0.84pt off 2.51pt}]  (96.5,0) -- (0,87) ;
\draw  [color={rgb, 255:red, 0; green, 0; blue, 0 }  ,draw opacity=1 ][fill={rgb, 255:red, 239; green, 12; blue, 16 }  ,fill opacity=1 ] (32.9,52.8) .. controls (32.9,49.93) and (35.23,47.6) .. (38.1,47.6) .. controls (40.97,47.6) and (43.3,49.93) .. (43.3,52.8) .. controls (43.3,55.67) and (40.97,58) .. (38.1,58) .. controls (35.23,58) and (32.9,55.67) .. (32.9,52.8) -- cycle ;
\draw [color={rgb, 255:red, 0; green, 0; blue, 0 }  ,draw opacity=1 ]   (38.1,52.8) -- (45.71,45.65) ;
\draw [shift={(47.9,43.6)}, rotate = 136.81] [fill={rgb, 255:red, 0; green, 0; blue, 0 }  ,fill opacity=1 ][line width=0.08]  [draw opacity=0] (3.57,-1.72) -- (0,0) -- (3.57,1.72) -- cycle    ;
\draw [color={rgb, 255:red, 0; green, 0; blue, 0 }  ,draw opacity=1 ]   (100.5,57.4) -- (116.18,73.08) ;
\draw [shift={(118.3,75.2)}, rotate = 225] [fill={rgb, 255:red, 0; green, 0; blue, 0 }  ,fill opacity=1 ][line width=0.08]  [draw opacity=0] (3.57,-1.72) -- (0,0) -- (3.57,1.72) -- cycle    ;
\draw [draw opacity=0]   (43.75,19) -- (30,31.69) ;
\draw [draw opacity=0]   (30,19) -- (30.55,19.51) -- (43.75,31.69) ;
\draw [line width=1.5]    (32.58,29.31) -- (41.17,21.38) ;
\draw [line width=1.5]    (32.58,21.38) -- (41.17,29.31) ;
\draw  [fill={rgb, 255:red, 155; green, 155; blue, 155 }  ,fill opacity=1 ] (38.46,25.95) -- (37.53,26.81) -- (36.22,26.81) -- (35.29,25.95) -- (35.29,24.74) -- (36.22,23.88) -- (37.53,23.88) -- (38.46,24.74) -- cycle ;
\draw  [draw opacity=0] (30,19) -- (43.75,19) -- (43.75,31.69) -- (30,31.69) -- cycle ;
\draw  [fill={rgb, 255:red, 184; green, 233; blue, 134 }  ,fill opacity=1 ] (32.04,21.38) .. controls (32.04,21.11) and (32.28,20.88) .. (32.58,20.88) .. controls (32.87,20.88) and (33.12,21.11) .. (33.12,21.38) .. controls (33.12,21.65) and (32.87,21.87) .. (32.58,21.87) .. controls (32.28,21.87) and (32.04,21.65) .. (32.04,21.38) -- cycle ;
\draw  [fill={rgb, 255:red, 184; green, 233; blue, 134 }  ,fill opacity=1 ] (40.63,29.31) .. controls (40.63,29.03) and (40.88,28.81) .. (41.17,28.81) .. controls (41.47,28.81) and (41.71,29.03) .. (41.71,29.31) .. controls (41.71,29.58) and (41.47,29.8) .. (41.17,29.8) .. controls (40.88,29.8) and (40.63,29.58) .. (40.63,29.31) -- cycle ;
\draw  [fill={rgb, 255:red, 158; green, 202; blue, 255 }  ,fill opacity=1 ] (40.63,21.38) .. controls (40.63,21.11) and (40.88,20.88) .. (41.17,20.88) .. controls (41.47,20.88) and (41.71,21.11) .. (41.71,21.38) .. controls (41.71,21.65) and (41.47,21.87) .. (41.17,21.87) .. controls (40.88,21.87) and (40.63,21.65) .. (40.63,21.38) -- cycle ;
\draw  [fill={rgb, 255:red, 158; green, 202; blue, 255 }  ,fill opacity=1 ] (32.04,29.31) .. controls (32.04,29.03) and (32.28,28.81) .. (32.58,28.81) .. controls (32.87,28.81) and (33.12,29.03) .. (33.12,29.31) .. controls (33.12,29.58) and (32.87,29.8) .. (32.58,29.8) .. controls (32.28,29.8) and (32.04,29.58) .. (32.04,29.31) -- cycle ;
\draw  [fill={rgb, 255:red, 74; green, 74; blue, 74 }  ,fill opacity=1 ] (41.17,29.31) -- (42.72,27.61) -- (42.96,27.59) -- (43.09,27.69) -- (43.09,27.69) -- (41.55,29.39) -- (41.3,29.41) -- (41.17,29.31) -- cycle ;
\draw  [fill={rgb, 255:red, 74; green, 74; blue, 74 }  ,fill opacity=1 ] (41.17,29.31) -- (39.63,31.01) -- (39.38,31.03) -- (39.25,30.93) -- (39.25,30.93) -- (40.8,29.23) -- (41.04,29.21) -- (41.17,29.31) -- cycle ;
\draw  [fill={rgb, 255:red, 74; green, 74; blue, 74 }  ,fill opacity=1 ] (32.58,21.38) -- (34.12,19.68) -- (34.37,19.66) -- (34.5,19.76) -- (34.5,19.76) -- (32.95,21.46) -- (32.71,21.48) -- (32.58,21.38) -- cycle ;
\draw  [fill={rgb, 255:red, 74; green, 74; blue, 74 }  ,fill opacity=1 ] (32.58,21.38) -- (31.03,23.08) -- (30.79,23.1) -- (30.66,23) -- (30.66,23) -- (32.2,21.3) -- (32.45,21.28) -- (32.58,21.38) -- cycle ;
\draw  [fill={rgb, 255:red, 74; green, 74; blue, 74 }  ,fill opacity=1 ] (30.88,27.74) -- (32.58,29.31) -- (32.58,29.31) -- (32.46,29.42) -- (32.21,29.42) -- (30.51,27.85) -- (30.51,27.85) -- (30.63,27.74) -- cycle ;
\draw  [fill={rgb, 255:red, 74; green, 74; blue, 74 }  ,fill opacity=1 ] (32.94,29.2) -- (34.64,30.77) -- (34.64,30.77) -- (34.52,30.88) -- (34.28,30.88) -- (32.58,29.31) -- (32.58,29.31) -- (32.7,29.2) -- cycle ;
\draw  [fill={rgb, 255:red, 74; green, 74; blue, 74 }  ,fill opacity=1 ] (39.47,19.81) -- (41.17,21.38) -- (41.17,21.38) -- (41.05,21.49) -- (40.81,21.49) -- (39.11,19.92) -- (39.11,19.92) -- (39.23,19.81) -- cycle ;
\draw  [fill={rgb, 255:red, 74; green, 74; blue, 74 }  ,fill opacity=1 ] (41.54,21.27) -- (43.24,22.84) -- (43.24,22.84) -- (43.12,22.95) -- (42.87,22.95) -- (41.17,21.38) -- (41.17,21.38) -- (41.29,21.27) -- cycle ;
\draw  [fill={rgb, 255:red, 0; green, 0; blue, 0 }  ,fill opacity=1 ] (32.41,29.31) .. controls (32.41,29.22) and (32.48,29.15) .. (32.58,29.15) .. controls (32.67,29.15) and (32.75,29.22) .. (32.75,29.31) .. controls (32.75,29.4) and (32.67,29.47) .. (32.58,29.47) .. controls (32.48,29.47) and (32.41,29.4) .. (32.41,29.31) -- cycle ;
\draw  [fill={rgb, 255:red, 0; green, 0; blue, 0 }  ,fill opacity=1 ] (32.41,21.38) .. controls (32.41,21.29) and (32.48,21.22) .. (32.58,21.22) .. controls (32.67,21.22) and (32.75,21.29) .. (32.75,21.38) .. controls (32.75,21.47) and (32.67,21.54) .. (32.58,21.54) .. controls (32.48,21.54) and (32.41,21.47) .. (32.41,21.38) -- cycle ;
\draw  [fill={rgb, 255:red, 0; green, 0; blue, 0 }  ,fill opacity=1 ] (41,29.31) .. controls (41,29.22) and (41.08,29.15) .. (41.17,29.15) .. controls (41.27,29.15) and (41.34,29.22) .. (41.34,29.31) .. controls (41.34,29.4) and (41.27,29.47) .. (41.17,29.47) .. controls (41.08,29.47) and (41,29.4) .. (41,29.31) -- cycle ;
\draw  [fill={rgb, 255:red, 0; green, 0; blue, 0 }  ,fill opacity=1 ] (41,21.38) .. controls (41,21.29) and (41.08,21.22) .. (41.17,21.22) .. controls (41.27,21.22) and (41.34,21.29) .. (41.34,21.38) .. controls (41.34,21.47) and (41.27,21.54) .. (41.17,21.54) .. controls (41.08,21.54) and (41,21.47) .. (41,21.38) -- cycle ;
\draw    (11,13.44) .. controls (17.39,21.2) and (21.75,24.37) .. (25.18,25.57) .. controls (29.26,27) and (32.01,25.65) .. (35.29,25.95) ;
\draw   (9.47,13.44) .. controls (9.47,12.59) and (10.15,11.91) .. (11,11.91) .. controls (11.85,11.91) and (12.53,12.59) .. (12.53,13.44) .. controls (12.53,14.28) and (11.85,14.97) .. (11,14.97) .. controls (10.15,14.97) and (9.47,14.28) .. (9.47,13.44) -- cycle ;
\draw  [fill={rgb, 255:red, 155; green, 155; blue, 155 }  ,fill opacity=1 ] (268,40.23) .. controls (272,37.73) and (266.3,27.47) .. (269.3,23.47) .. controls (272.3,19.47) and (289.22,16.95) .. (290.01,16.99) .. controls (290.8,17.03) and (313,27.83) .. (316.8,30.03) .. controls (320.6,32.23) and (306.42,43.39) .. (305.8,47.43) .. controls (305.18,51.47) and (318.4,66.63) .. (317,67.63) .. controls (315.6,68.63) and (304.8,77.83) .. (302.6,78.23) .. controls (300.4,78.63) and (285.6,67.75) .. (283.8,67.19) .. controls (282,66.63) and (274.5,65.72) .. (272.05,61.97) .. controls (269.6,58.23) and (261.4,43.03) .. (262.2,42.23) .. controls (263,41.43) and (264,42.73) .. (268,40.23) -- cycle ;
\draw  [color={rgb, 255:red, 21; green, 10; blue, 255 }  ,draw opacity=1 ][dash pattern={on 4.5pt off 4.5pt}] (255,0) -- (325,0) -- (325,110) -- (255,110) -- cycle ;
\draw  [draw opacity=0] (255,0) -- (290,0) -- (290,110) -- (255,110) -- cycle ;
\draw [color={rgb, 255:red, 239; green, 12; blue, 16 }  ,draw opacity=1 ]   (317.8,29.43) -- (318,67.43) ;
\draw [color={rgb, 255:red, 239; green, 12; blue, 16 }  ,draw opacity=1 ]   (318,67.43) -- (302.8,79.63) ;
\draw [color={rgb, 255:red, 239; green, 12; blue, 16 }  ,draw opacity=1 ]   (271.8,63.83) -- (302.8,79.63) ;
\draw [color={rgb, 255:red, 239; green, 12; blue, 16 }  ,draw opacity=1 ]   (261.2,42.63) -- (271.8,63.83) ;
\draw [color={rgb, 255:red, 239; green, 12; blue, 16 }  ,draw opacity=1 ]   (269.2,21.23) -- (261.2,42.63) ;
\draw [color={rgb, 255:red, 239; green, 12; blue, 16 }  ,draw opacity=1 ]   (290,15.83) -- (269.2,21.23) ;
\draw [color={rgb, 255:red, 239; green, 12; blue, 16 }  ,draw opacity=1 ]   (290,15.83) -- (317.8,29.43) ;
\draw  [draw opacity=0] (54.5,105) -- (114.5,105) -- (114.5,125) -- (54.5,125) -- cycle ;
\draw  [color={rgb, 255:red, 192; green, 10; blue, 255 }  ,draw opacity=1 ][dash pattern={on 4.5pt off 4.5pt}] (90.5,47.4) -- (110.5,47.4) -- (110.5,67.4) -- (90.5,67.4) -- cycle ;
\draw  [color={rgb, 255:red, 21; green, 10; blue, 255 }  ,draw opacity=1 ][dash pattern={on 4.5pt off 4.5pt}] (135.33,30.83) -- (155.33,30.83) -- (155.33,50.83) -- (135.33,50.83) -- cycle ;
\draw  [color={rgb, 255:red, 0; green, 114; blue, 67 }  ,draw opacity=1 ][dash pattern={on 4.5pt off 4.5pt}] (138.01,78.51) -- (151.99,78.51) -- (151.99,92.49) -- (138.01,92.49) -- cycle ;
\draw  [color={rgb, 255:red, 192; green, 10; blue, 255 }  ,draw opacity=1 ][dash pattern={on 4.5pt off 4.5pt}] (75.05,75.53) -- (95.05,75.53) -- (95.05,95.53) -- (75.05,95.53) -- cycle ;
\draw  [color={rgb, 255:red, 21; green, 10; blue, 255 }  ,draw opacity=1 ][dash pattern={on 4.5pt off 4.5pt}] (29.3,80) -- (49.3,80) -- (49.3,100) -- (29.3,100) -- cycle ;
\draw  [color={rgb, 255:red, 192; green, 10; blue, 255 }  ,draw opacity=1 ][dash pattern={on 4.5pt off 4.5pt}] (28.1,42.8) -- (48.1,42.8) -- (48.1,62.8) -- (28.1,62.8) -- cycle ;
\draw [draw opacity=0]   (390.35,24.72) -- (340.35,74.81) ;
\draw [draw opacity=0]   (340.35,24.72) -- (342.34,26.72) -- (390.35,74.81) ;
\draw [line width=1.5]    (349.72,65.42) -- (380.97,34.12) ;
\draw [line width=1.5]    (349.72,34.12) -- (380.97,65.42) ;
\draw  [fill={rgb, 255:red, 155; green, 155; blue, 155 }  ,fill opacity=1 ] (371.13,52.16) -- (367.74,55.55) -- (362.96,55.55) -- (359.57,52.16) -- (359.57,47.37) -- (362.96,43.98) -- (367.74,43.98) -- (371.13,47.37) -- cycle ;
\draw  [draw opacity=0] (340.35,24.72) -- (390.35,24.72) -- (390.35,74.81) -- (340.35,74.81) -- cycle ;
\draw  [fill={rgb, 255:red, 184; green, 233; blue, 134 }  ,fill opacity=1 ] (347.77,34.12) .. controls (347.77,33.03) and (348.65,32.16) .. (349.72,32.16) .. controls (350.8,32.16) and (351.68,33.03) .. (351.68,34.12) .. controls (351.68,35.2) and (350.8,36.07) .. (349.72,36.07) .. controls (348.65,36.07) and (347.77,35.2) .. (347.77,34.12) -- cycle ;
\draw  [fill={rgb, 255:red, 184; green, 233; blue, 134 }  ,fill opacity=1 ] (379.02,65.42) .. controls (379.02,64.34) and (379.9,63.46) .. (380.97,63.46) .. controls (382.05,63.46) and (382.93,64.34) .. (382.93,65.42) .. controls (382.93,66.5) and (382.05,67.37) .. (380.97,67.37) .. controls (379.9,67.37) and (379.02,66.5) .. (379.02,65.42) -- cycle ;
\draw  [fill={rgb, 255:red, 158; green, 202; blue, 255 }  ,fill opacity=1 ] (379.02,34.12) .. controls (379.02,33.03) and (379.9,32.16) .. (380.97,32.16) .. controls (382.05,32.16) and (382.93,33.03) .. (382.93,34.12) .. controls (382.93,35.2) and (382.05,36.07) .. (380.97,36.07) .. controls (379.9,36.07) and (379.02,35.2) .. (379.02,34.12) -- cycle ;
\draw  [fill={rgb, 255:red, 158; green, 202; blue, 255 }  ,fill opacity=1 ] (347.77,65.42) .. controls (347.77,64.34) and (348.65,63.46) .. (349.72,63.46) .. controls (350.8,63.46) and (351.68,64.34) .. (351.68,65.42) .. controls (351.68,66.5) and (350.8,67.37) .. (349.72,67.37) .. controls (348.65,67.37) and (347.77,66.5) .. (347.77,65.42) -- cycle ;
\draw  [fill={rgb, 255:red, 74; green, 74; blue, 74 }  ,fill opacity=1 ] (380.97,65.42) -- (386.6,58.7) -- (387.48,58.63) -- (387.96,59.03) -- (387.96,59.03) -- (382.33,65.74) -- (381.45,65.82) -- (380.97,65.42) -- cycle ;
\draw  [fill={rgb, 255:red, 74; green, 74; blue, 74 }  ,fill opacity=1 ] (380.97,65.42) -- (375.35,72.13) -- (374.47,72.21) -- (373.99,71.81) -- (373.99,71.81) -- (379.62,65.09) -- (380.5,65.02) -- (380.97,65.42) -- cycle ;
\draw  [fill={rgb, 255:red, 74; green, 74; blue, 74 }  ,fill opacity=1 ] (349.72,34.12) -- (355.35,27.4) -- (356.23,27.32) -- (356.71,27.73) -- (356.71,27.73) -- (351.08,34.44) -- (350.2,34.52) -- (349.72,34.12) -- cycle ;
\draw  [fill={rgb, 255:red, 74; green, 74; blue, 74 }  ,fill opacity=1 ] (349.72,34.12) -- (344.1,40.83) -- (343.22,40.91) -- (342.74,40.5) -- (342.74,40.5) -- (348.37,33.79) -- (349.25,33.71) -- (349.72,34.12) -- cycle ;
\draw  [fill={rgb, 255:red, 74; green, 74; blue, 74 }  ,fill opacity=1 ] (343.54,59.22) -- (349.72,65.42) -- (349.72,65.42) -- (349.28,65.86) -- (348.4,65.86) -- (342.21,59.66) -- (342.21,59.66) -- (342.65,59.22) -- cycle ;
\draw  [fill={rgb, 255:red, 74; green, 74; blue, 74 }  ,fill opacity=1 ] (351.05,64.97) -- (357.24,71.17) -- (357.24,71.17) -- (356.8,71.62) -- (355.91,71.62) -- (349.73,65.42) -- (349.73,65.42) -- (350.17,64.97) -- cycle ;
\draw  [fill={rgb, 255:red, 74; green, 74; blue, 74 }  ,fill opacity=1 ] (374.79,27.92) -- (380.97,34.12) -- (380.97,34.12) -- (380.53,34.56) -- (379.65,34.56) -- (373.46,28.36) -- (373.46,28.36) -- (373.9,27.92) -- cycle ;
\draw  [fill={rgb, 255:red, 74; green, 74; blue, 74 }  ,fill opacity=1 ] (382.3,33.67) -- (388.49,39.87) -- (388.49,39.87) -- (388.05,40.31) -- (387.16,40.31) -- (380.98,34.12) -- (380.98,34.12) -- (381.42,33.67) -- cycle ;
\draw  [fill={rgb, 255:red, 0; green, 0; blue, 0 }  ,fill opacity=1 ] (349.1,65.42) .. controls (349.1,65.07) and (349.38,64.79) .. (349.72,64.79) .. controls (350.07,64.79) and (350.35,65.07) .. (350.35,65.42) .. controls (350.35,65.76) and (350.07,66.04) .. (349.72,66.04) .. controls (349.38,66.04) and (349.1,65.76) .. (349.1,65.42) -- cycle ;
\draw  [fill={rgb, 255:red, 0; green, 0; blue, 0 }  ,fill opacity=1 ] (349.1,34.12) .. controls (349.1,33.77) and (349.38,33.49) .. (349.72,33.49) .. controls (350.07,33.49) and (350.35,33.77) .. (350.35,34.12) .. controls (350.35,34.46) and (350.07,34.74) .. (349.72,34.74) .. controls (349.38,34.74) and (349.1,34.46) .. (349.1,34.12) -- cycle ;
\draw  [fill={rgb, 255:red, 0; green, 0; blue, 0 }  ,fill opacity=1 ] (380.35,65.42) .. controls (380.35,65.07) and (380.63,64.79) .. (380.97,64.79) .. controls (381.32,64.79) and (381.6,65.07) .. (381.6,65.42) .. controls (381.6,65.76) and (381.32,66.04) .. (380.97,66.04) .. controls (380.63,66.04) and (380.35,65.76) .. (380.35,65.42) -- cycle ;
\draw  [fill={rgb, 255:red, 0; green, 0; blue, 0 }  ,fill opacity=1 ] (380.35,34.12) .. controls (380.35,33.77) and (380.63,33.49) .. (380.97,33.49) .. controls (381.32,33.49) and (381.6,33.77) .. (381.6,34.12) .. controls (381.6,34.46) and (381.32,34.74) .. (380.97,34.74) .. controls (380.63,34.74) and (380.35,34.46) .. (380.35,34.12) -- cycle ;

\draw  [color={rgb, 255:red, 239; green, 12; blue, 16 }  ,draw opacity=1 ][line width=0.75]  (392.73,61.2) -- (376.5,77.43) -- (353.54,77.43) -- (337.3,61.2) -- (337.3,38.24) -- (353.54,22) -- (376.5,22) -- (392.73,38.24) -- cycle ;

\draw   (364.85,49.77) .. controls (364.85,49.49) and (365.07,49.27) .. (365.35,49.27) .. controls (365.63,49.27) and (365.85,49.49) .. (365.85,49.77) .. controls (365.85,50.04) and (365.63,50.27) .. (365.35,50.27) .. controls (365.07,50.27) and (364.85,50.04) .. (364.85,49.77) -- cycle ;

\draw  [color={rgb, 255:red, 192; green, 10; blue, 255 }  ,draw opacity=1 ][dash pattern={on 4.5pt off 4.5pt}] (330,0) -- (400,0) -- (400,110) -- (330,110) -- cycle ;
\draw  [draw opacity=0] (330,0) -- (365,0) -- (365,110) -- (330,110) -- cycle ;
\draw  [color={rgb, 255:red, 0; green, 114; blue, 67 }  ,draw opacity=1 ][dash pattern={on 4.5pt off 4.5pt}] (180,0) -- (250,0) -- (250,110) -- (180,110) -- cycle ;
\draw  [draw opacity=0] (180,0) -- (215,0) -- (215,110) -- (180,110) -- cycle ;
\draw  [color={rgb, 255:red, 0; green, 0; blue, 0 }  ,draw opacity=1 ] (190,50) .. controls (190,36.19) and (201.19,25) .. (215,25) .. controls (228.81,25) and (240,36.19) .. (240,50) .. controls (240,63.81) and (228.81,75) .. (215,75) .. controls (201.19,75) and (190,63.81) .. (190,50) -- cycle ;
\draw  [color={rgb, 255:red, 239; green, 12; blue, 16 }  ,draw opacity=1 ] (240,50) -- (227.5,71.65) -- (202.5,71.65) -- (190,50) -- (202.5,28.35) -- (227.5,28.35) -- cycle ;
\draw  [draw opacity=0][dash pattern={on 4.5pt off 4.5pt}] (5,5.5) -- (85,5.5) -- (85,105.5) -- (5,105.5) -- cycle ;
\draw  [draw opacity=0] (0,0) -- (400,0) -- (400,120) -- (0,120) -- cycle ;
\draw [color={rgb, 255:red, 158; green, 202; blue, 255 }  ,draw opacity=1 ] [dash pattern={on 0.84pt off 2.51pt}]  (39.86,25.5) .. controls (50.97,25.62) and (76.43,25.5) .. (115,25.5) ;
\draw [color={rgb, 255:red, 158; green, 202; blue, 255 }  ,draw opacity=1 ] [dash pattern={on 0.84pt off 2.51pt}]  (85,83) .. controls (85.29,75.5) and (86.14,65.21) .. (85,55.5) .. controls (83.86,45.79) and (58.14,44.36) .. (55,55.5) .. controls (51.86,66.64) and (53.29,77.5) .. (55,85.5) ;
\draw   (214.5,50) .. controls (214.5,49.72) and (214.72,49.5) .. (215,49.5) .. controls (215.28,49.5) and (215.5,49.72) .. (215.5,50) .. controls (215.5,50.28) and (215.28,50.5) .. (215,50.5) .. controls (214.72,50.5) and (214.5,50.28) .. (214.5,50) -- cycle ;

\draw (163.95,7.64) node [anchor=north east] [inner sep=0.75pt]  [font=\footnotesize] [align=left] {Search Area};
\draw (85.54,107.58) node [anchor=north] [inner sep=0.75pt]  [font=\footnotesize] [align=left] {Geo Fence};
\draw (290,107) node [anchor=south] [inner sep=0.75pt]  [font=\footnotesize] [align=left] {\begin{minipage}[lt]{39pt}\setlength\topsep{0pt}
\begin{center}
Fixed\\Obstacles
\end{center}

\end{minipage}};
\draw (215,3.4) node [anchor=north] [inner sep=0.75pt]  [font=\footnotesize]  {$H_{\text{W}} =6$};
\draw (215,107) node [anchor=south] [inner sep=0.75pt]  [font=\footnotesize] [align=left] {Waypoints};
\draw (365,3.4) node [anchor=north] [inner sep=0.75pt]  [font=\footnotesize]  {$H_{\text{O}} =8$};
\draw (365,107) node [anchor=south] [inner sep=0.75pt]  [font=\footnotesize] [align=left] {\begin{minipage}[lt]{39pt}\setlength\topsep{0pt}
\begin{center}
Moving\\Obstacles
\end{center}

\end{minipage}};

\end{tikzpicture}

%% file: Tikz_CC.tex
\tikzset{every picture/.style={line width=0.75pt}} %

\begin{tikzpicture}[x=0.6pt,y=0.6pt,yscale=-1,xscale=1]

\draw  [draw opacity=0][fill={rgb, 255:red, 239; green, 12; blue, 16 }  ,fill opacity=0.4 ] (0,0) -- (399.34,0) -- (399.34,35.02) -- (0.06,35.02) -- cycle ;
\draw  [draw opacity=0][fill={rgb, 255:red, 239; green, 12; blue, 16 }  ,fill opacity=0.4 ] (0,35.02) -- (55.96,35.02) -- (80.5,93.41) -- (0.28,58.24) -- cycle ;
\draw  [draw opacity=0][fill={rgb, 255:red, 184; green, 233; blue, 134 }  ,fill opacity=0.4 ] (101.67,142.18) -- (123.94,192.6) -- (0,192.6) -- (0,154.95) -- cycle ;
\draw  [draw opacity=0][fill={rgb, 255:red, 184; green, 233; blue, 134 }  ,fill opacity=0.4 ] (101.67,142.18) -- (80.5,93.41) -- (172.82,132.62) -- cycle ;
\draw  [draw opacity=0][fill={rgb, 255:red, 239; green, 12; blue, 16 }  ,fill opacity=0.4 ] (55.96,35.02) -- (346.21,35.06) -- (337.3,112.8) -- (172.82,132.62) -- (80.5,93.41) -- cycle ;
\draw  [draw opacity=0][fill={rgb, 255:red, 239; green, 12; blue, 16 }  ,fill opacity=0.4 ] (346.21,35.06) -- (399.34,35.02) -- (400,104.76) -- (337.3,112.8) -- cycle ;
\draw  [draw opacity=0][fill={rgb, 255:red, 184; green, 233; blue, 134 }  ,fill opacity=0.4 ] (400,104.76) -- (399.34,192.6) -- (330.49,192.95) -- (337.3,112.8) -- cycle ;
\draw  [draw opacity=0][fill={rgb, 255:red, 184; green, 233; blue, 134 }  ,fill opacity=0.4 ] (337.3,112.8) -- (330.49,192.95) -- (310.78,192.95) -- (172.82,132.62) -- cycle ;
\draw  [draw opacity=0][fill={rgb, 255:red, 184; green, 233; blue, 134 }  ,fill opacity=0.4 ] (0,58.24) -- (80.5,93.41) -- (101.67,142.18) -- (0.06,154.95) -- cycle ;
\draw  [draw opacity=0][fill={rgb, 255:red, 158; green, 202; blue, 255 }  ,fill opacity=0.4 ] (101.67,142.18) -- (123.94,192.6) -- (310.78,192.95) -- (172.82,132.62) -- cycle ;
\draw  [fill={rgb, 255:red, 155; green, 155; blue, 155 }  ,fill opacity=1 ] (344.2,35.77) .. controls (344.2,34.61) and (338.02,110.35) .. (336.27,111.35) .. controls (334.51,112.35) and (182.55,130.45) .. (174.61,130.68) .. controls (166.67,130.91) and (96.92,98.13) .. (89.49,96.66) .. controls (82.07,95.19) and (58.03,36.68) .. (57.64,36.16) .. controls (57.24,35.63) and (344.2,36.94) .. (344.2,35.77) -- cycle ;
\draw  [dash pattern={on 0.84pt off 2.51pt}]  (40.5,0) -- (123.94,192.6) ;
\draw  [dash pattern={on 0.84pt off 2.51pt}]  (0.28,58.24) -- (310.78,192.95) ;
\draw  [dash pattern={on 0.84pt off 2.51pt}]  (400,104.76) -- (0.06,154.95) ;
\draw  [dash pattern={on 0.84pt off 2.51pt}]  (348.66,0) -- (330.49,192.95) ;
\draw  [draw opacity=0] (286.69,163.92) -- (332.76,163.92) -- (332.76,181.43) -- (286.69,181.43) -- cycle ;
\draw  [draw opacity=0] (332.76,163.92) -- (378.83,163.92) -- (378.83,181.43) -- (332.76,181.43) -- cycle ;
\draw  [dash pattern={on 0.84pt off 2.51pt}]  (0.06,35.02) -- (399.34,35.02) ;
\draw [color={rgb, 255:red, 0; green, 0; blue, 0 }  ,draw opacity=0.4 ][line width=1.5]    (342.37,135.38) -- (355.57,121.49) ;
\draw [color={rgb, 255:red, 0; green, 0; blue, 0 }  ,draw opacity=0.4 ][line width=1.5]    (342.37,121.49) -- (355.57,135.38) ;
\draw  [color={rgb, 255:red, 0; green, 0; blue, 0 }  ,draw opacity=0.4 ][fill={rgb, 255:red, 155; green, 155; blue, 155 }  ,fill opacity=1 ] (351.41,129.5) -- (349.98,131) -- (347.96,131) -- (346.53,129.5) -- (346.53,127.37) -- (347.96,125.87) -- (349.98,125.87) -- (351.41,127.37) -- cycle ;
\draw  [color={rgb, 255:red, 0; green, 0; blue, 0 }  ,draw opacity=0.4 ][fill={rgb, 255:red, 184; green, 233; blue, 134 }  ,fill opacity=1 ] (341.54,121.49) .. controls (341.54,121.01) and (341.91,120.62) .. (342.37,120.62) .. controls (342.83,120.62) and (343.19,121.01) .. (343.19,121.49) .. controls (343.19,121.97) and (342.83,122.36) .. (342.37,122.36) .. controls (341.91,122.36) and (341.54,121.97) .. (341.54,121.49) -- cycle ;
\draw  [color={rgb, 255:red, 0; green, 0; blue, 0 }  ,draw opacity=0.4 ][fill={rgb, 255:red, 184; green, 233; blue, 134 }  ,fill opacity=1 ] (354.74,135.38) .. controls (354.74,134.9) and (355.11,134.51) .. (355.57,134.51) .. controls (356.02,134.51) and (356.39,134.9) .. (356.39,135.38) .. controls (356.39,135.86) and (356.02,136.24) .. (355.57,136.24) .. controls (355.11,136.24) and (354.74,135.86) .. (354.74,135.38) -- cycle ;
\draw  [color={rgb, 255:red, 0; green, 0; blue, 0 }  ,draw opacity=0.4 ][fill={rgb, 255:red, 158; green, 202; blue, 255 }  ,fill opacity=1 ] (354.74,121.49) .. controls (354.74,121.01) and (355.11,120.62) .. (355.57,120.62) .. controls (356.02,120.62) and (356.39,121.01) .. (356.39,121.49) .. controls (356.39,121.97) and (356.02,122.36) .. (355.57,122.36) .. controls (355.11,122.36) and (354.74,121.97) .. (354.74,121.49) -- cycle ;
\draw  [color={rgb, 255:red, 0; green, 0; blue, 0 }  ,draw opacity=0.4 ][fill={rgb, 255:red, 158; green, 202; blue, 255 }  ,fill opacity=1 ] (341.54,135.38) .. controls (341.54,134.9) and (341.91,134.51) .. (342.37,134.51) .. controls (342.83,134.51) and (343.19,134.9) .. (343.19,135.38) .. controls (343.19,135.86) and (342.83,136.24) .. (342.37,136.24) .. controls (341.91,136.24) and (341.54,135.86) .. (341.54,135.38) -- cycle ;
\draw  [color={rgb, 255:red, 0; green, 0; blue, 0 }  ,draw opacity=0.4 ][fill={rgb, 255:red, 74; green, 74; blue, 74 }  ,fill opacity=1 ] (355.57,135.38) -- (357.94,132.4) -- (358.31,132.36) -- (358.52,132.54) -- (358.52,132.54) -- (356.14,135.52) -- (355.77,135.55) -- (355.57,135.38) -- cycle ;
\draw  [color={rgb, 255:red, 0; green, 0; blue, 0 }  ,draw opacity=0.4 ][fill={rgb, 255:red, 74; green, 74; blue, 74 }  ,fill opacity=1 ] (355.57,135.38) -- (353.19,138.36) -- (352.82,138.39) -- (352.62,138.21) -- (352.62,138.21) -- (355,135.23) -- (355.37,135.2) -- (355.57,135.38) -- cycle ;
\draw  [color={rgb, 255:red, 0; green, 0; blue, 0 }  ,draw opacity=0.4 ][fill={rgb, 255:red, 74; green, 74; blue, 74 }  ,fill opacity=1 ] (342.37,121.49) -- (344.75,118.51) -- (345.12,118.48) -- (345.32,118.66) -- (345.32,118.66) -- (342.94,121.64) -- (342.57,121.67) -- (342.37,121.49) -- cycle ;
\draw  [color={rgb, 255:red, 0; green, 0; blue, 0 }  ,draw opacity=0.4 ][fill={rgb, 255:red, 74; green, 74; blue, 74 }  ,fill opacity=1 ] (342.37,121.49) -- (339.99,124.47) -- (339.62,124.5) -- (339.42,124.33) -- (339.42,124.33) -- (341.8,121.35) -- (342.17,121.31) -- (342.37,121.49) -- cycle ;
\draw  [color={rgb, 255:red, 0; green, 0; blue, 0 }  ,draw opacity=0.4 ][fill={rgb, 255:red, 74; green, 74; blue, 74 }  ,fill opacity=1 ] (339.76,132.63) -- (342.37,135.38) -- (342.37,135.38) -- (342.18,135.57) -- (341.81,135.57) -- (339.2,132.82) -- (339.2,132.82) -- (339.38,132.63) -- cycle ;
\draw  [color={rgb, 255:red, 0; green, 0; blue, 0 }  ,draw opacity=0.4 ][fill={rgb, 255:red, 74; green, 74; blue, 74 }  ,fill opacity=1 ] (342.93,135.18) -- (345.54,137.93) -- (345.54,137.93) -- (345.36,138.12) -- (344.98,138.12) -- (342.37,135.38) -- (342.37,135.38) -- (342.56,135.18) -- cycle ;
\draw  [color={rgb, 255:red, 0; green, 0; blue, 0 }  ,draw opacity=0.4 ][fill={rgb, 255:red, 74; green, 74; blue, 74 }  ,fill opacity=1 ] (352.95,118.74) -- (355.57,121.49) -- (355.57,121.49) -- (355.38,121.69) -- (355.01,121.69) -- (352.39,118.94) -- (352.39,118.94) -- (352.58,118.74) -- cycle ;
\draw  [color={rgb, 255:red, 0; green, 0; blue, 0 }  ,draw opacity=0.4 ][fill={rgb, 255:red, 74; green, 74; blue, 74 }  ,fill opacity=1 ] (356.13,121.3) -- (358.74,124.04) -- (358.74,124.04) -- (358.55,124.24) -- (358.18,124.24) -- (355.57,121.49) -- (355.57,121.49) -- (355.75,121.3) -- cycle ;
\draw  [color={rgb, 255:red, 0; green, 0; blue, 0 }  ,draw opacity=0.4 ][fill={rgb, 255:red, 0; green, 0; blue, 0 }  ,fill opacity=1 ] (342.11,135.38) .. controls (342.11,135.22) and (342.22,135.1) .. (342.37,135.1) .. controls (342.52,135.1) and (342.63,135.22) .. (342.63,135.38) .. controls (342.63,135.53) and (342.52,135.65) .. (342.37,135.65) .. controls (342.22,135.65) and (342.11,135.53) .. (342.11,135.38) -- cycle ;
\draw  [color={rgb, 255:red, 0; green, 0; blue, 0 }  ,draw opacity=0.4 ][fill={rgb, 255:red, 0; green, 0; blue, 0 }  ,fill opacity=1 ] (342.11,121.49) .. controls (342.11,121.34) and (342.22,121.21) .. (342.37,121.21) .. controls (342.52,121.21) and (342.63,121.34) .. (342.63,121.49) .. controls (342.63,121.65) and (342.52,121.77) .. (342.37,121.77) .. controls (342.22,121.77) and (342.11,121.65) .. (342.11,121.49) -- cycle ;
\draw  [color={rgb, 255:red, 0; green, 0; blue, 0 }  ,draw opacity=0.4 ][fill={rgb, 255:red, 0; green, 0; blue, 0 }  ,fill opacity=1 ] (355.3,135.38) .. controls (355.3,135.22) and (355.42,135.1) .. (355.57,135.1) .. controls (355.71,135.1) and (355.83,135.22) .. (355.83,135.38) .. controls (355.83,135.53) and (355.71,135.65) .. (355.57,135.65) .. controls (355.42,135.65) and (355.3,135.53) .. (355.3,135.38) -- cycle ;
\draw  [color={rgb, 255:red, 0; green, 0; blue, 0 }  ,draw opacity=0.4 ][fill={rgb, 255:red, 0; green, 0; blue, 0 }  ,fill opacity=1 ] (355.3,121.49) .. controls (355.3,121.34) and (355.42,121.21) .. (355.57,121.21) .. controls (355.71,121.21) and (355.83,121.34) .. (355.83,121.49) .. controls (355.83,121.65) and (355.71,121.77) .. (355.57,121.77) .. controls (355.42,121.77) and (355.3,121.65) .. (355.3,121.49) -- cycle ;

\draw [color={rgb, 255:red, 0; green, 0; blue, 0 }  ,draw opacity=0.4 ][line width=1.5]    (348.9,70.38) -- (362.1,56.49) ;
\draw [color={rgb, 255:red, 0; green, 0; blue, 0 }  ,draw opacity=0.4 ][line width=1.5]    (348.9,56.49) -- (362.1,70.38) ;
\draw  [color={rgb, 255:red, 0; green, 0; blue, 0 }  ,draw opacity=0.4 ][fill={rgb, 255:red, 155; green, 155; blue, 155 }  ,fill opacity=1 ] (357.94,64.5) -- (356.51,66) -- (354.49,66) -- (353.06,64.5) -- (353.06,62.37) -- (354.49,60.87) -- (356.51,60.87) -- (357.94,62.37) -- cycle ;
\draw  [color={rgb, 255:red, 0; green, 0; blue, 0 }  ,draw opacity=0.4 ][fill={rgb, 255:red, 184; green, 233; blue, 134 }  ,fill opacity=1 ] (348.08,56.49) .. controls (348.08,56.01) and (348.45,55.62) .. (348.9,55.62) .. controls (349.36,55.62) and (349.73,56.01) .. (349.73,56.49) .. controls (349.73,56.97) and (349.36,57.36) .. (348.9,57.36) .. controls (348.45,57.36) and (348.08,56.97) .. (348.08,56.49) -- cycle ;
\draw  [color={rgb, 255:red, 0; green, 0; blue, 0 }  ,draw opacity=0.4 ][fill={rgb, 255:red, 184; green, 233; blue, 134 }  ,fill opacity=1 ] (361.28,70.38) .. controls (361.28,69.9) and (361.64,69.51) .. (362.1,69.51) .. controls (362.56,69.51) and (362.93,69.9) .. (362.93,70.38) .. controls (362.93,70.86) and (362.56,71.24) .. (362.1,71.24) .. controls (361.64,71.24) and (361.28,70.86) .. (361.28,70.38) -- cycle ;
\draw  [color={rgb, 255:red, 0; green, 0; blue, 0 }  ,draw opacity=0.4 ][fill={rgb, 255:red, 158; green, 202; blue, 255 }  ,fill opacity=1 ] (361.28,56.49) .. controls (361.28,56.01) and (361.64,55.62) .. (362.1,55.62) .. controls (362.56,55.62) and (362.93,56.01) .. (362.93,56.49) .. controls (362.93,56.97) and (362.56,57.36) .. (362.1,57.36) .. controls (361.64,57.36) and (361.28,56.97) .. (361.28,56.49) -- cycle ;
\draw  [color={rgb, 255:red, 0; green, 0; blue, 0 }  ,draw opacity=0.4 ][fill={rgb, 255:red, 158; green, 202; blue, 255 }  ,fill opacity=1 ] (348.08,70.38) .. controls (348.08,69.9) and (348.45,69.51) .. (348.9,69.51) .. controls (349.36,69.51) and (349.73,69.9) .. (349.73,70.38) .. controls (349.73,70.86) and (349.36,71.24) .. (348.9,71.24) .. controls (348.45,71.24) and (348.08,70.86) .. (348.08,70.38) -- cycle ;
\draw  [color={rgb, 255:red, 0; green, 0; blue, 0 }  ,draw opacity=0.4 ][fill={rgb, 255:red, 74; green, 74; blue, 74 }  ,fill opacity=1 ] (362.1,70.38) -- (364.48,67.4) -- (364.85,67.36) -- (365.05,67.54) -- (365.05,67.54) -- (362.67,70.52) -- (362.3,70.55) -- (362.1,70.38) -- cycle ;
\draw  [color={rgb, 255:red, 0; green, 0; blue, 0 }  ,draw opacity=0.4 ][fill={rgb, 255:red, 74; green, 74; blue, 74 }  ,fill opacity=1 ] (362.1,70.38) -- (359.72,73.36) -- (359.35,73.39) -- (359.15,73.21) -- (359.15,73.21) -- (361.53,70.23) -- (361.9,70.2) -- (362.1,70.38) -- cycle ;
\draw  [color={rgb, 255:red, 0; green, 0; blue, 0 }  ,draw opacity=0.4 ][fill={rgb, 255:red, 74; green, 74; blue, 74 }  ,fill opacity=1 ] (348.9,56.49) -- (351.28,53.51) -- (351.65,53.48) -- (351.85,53.66) -- (351.85,53.66) -- (349.47,56.64) -- (349.1,56.67) -- (348.9,56.49) -- cycle ;
\draw  [color={rgb, 255:red, 0; green, 0; blue, 0 }  ,draw opacity=0.4 ][fill={rgb, 255:red, 74; green, 74; blue, 74 }  ,fill opacity=1 ] (348.9,56.49) -- (346.53,59.47) -- (346.16,59.5) -- (345.95,59.33) -- (345.95,59.33) -- (348.33,56.35) -- (348.7,56.31) -- (348.9,56.49) -- cycle ;
\draw  [color={rgb, 255:red, 0; green, 0; blue, 0 }  ,draw opacity=0.4 ][fill={rgb, 255:red, 74; green, 74; blue, 74 }  ,fill opacity=1 ] (346.29,67.63) -- (348.9,70.38) -- (348.9,70.38) -- (348.72,70.57) -- (348.34,70.57) -- (345.73,67.82) -- (345.73,67.82) -- (345.92,67.63) -- cycle ;
\draw  [color={rgb, 255:red, 0; green, 0; blue, 0 }  ,draw opacity=0.4 ][fill={rgb, 255:red, 74; green, 74; blue, 74 }  ,fill opacity=1 ] (349.46,70.18) -- (352.08,72.93) -- (352.08,72.93) -- (351.89,73.12) -- (351.52,73.12) -- (348.9,70.38) -- (348.9,70.38) -- (349.09,70.18) -- cycle ;
\draw  [color={rgb, 255:red, 0; green, 0; blue, 0 }  ,draw opacity=0.4 ][fill={rgb, 255:red, 74; green, 74; blue, 74 }  ,fill opacity=1 ] (359.49,53.74) -- (362.1,56.49) -- (362.1,56.49) -- (361.91,56.69) -- (361.54,56.69) -- (358.93,53.94) -- (358.93,53.94) -- (359.11,53.74) -- cycle ;
\draw  [color={rgb, 255:red, 0; green, 0; blue, 0 }  ,draw opacity=0.4 ][fill={rgb, 255:red, 74; green, 74; blue, 74 }  ,fill opacity=1 ] (362.66,56.3) -- (365.27,59.04) -- (365.27,59.04) -- (365.09,59.24) -- (364.71,59.24) -- (362.1,56.49) -- (362.1,56.49) -- (362.29,56.3) -- cycle ;
\draw  [color={rgb, 255:red, 0; green, 0; blue, 0 }  ,draw opacity=0.4 ][fill={rgb, 255:red, 0; green, 0; blue, 0 }  ,fill opacity=1 ] (348.64,70.38) .. controls (348.64,70.22) and (348.76,70.1) .. (348.9,70.1) .. controls (349.05,70.1) and (349.17,70.22) .. (349.17,70.38) .. controls (349.17,70.53) and (349.05,70.65) .. (348.9,70.65) .. controls (348.76,70.65) and (348.64,70.53) .. (348.64,70.38) -- cycle ;
\draw  [color={rgb, 255:red, 0; green, 0; blue, 0 }  ,draw opacity=0.4 ][fill={rgb, 255:red, 0; green, 0; blue, 0 }  ,fill opacity=1 ] (348.64,56.49) .. controls (348.64,56.34) and (348.76,56.21) .. (348.9,56.21) .. controls (349.05,56.21) and (349.17,56.34) .. (349.17,56.49) .. controls (349.17,56.65) and (349.05,56.77) .. (348.9,56.77) .. controls (348.76,56.77) and (348.64,56.65) .. (348.64,56.49) -- cycle ;
\draw  [color={rgb, 255:red, 0; green, 0; blue, 0 }  ,draw opacity=0.4 ][fill={rgb, 255:red, 0; green, 0; blue, 0 }  ,fill opacity=1 ] (361.84,70.38) .. controls (361.84,70.22) and (361.95,70.1) .. (362.1,70.1) .. controls (362.25,70.1) and (362.36,70.22) .. (362.36,70.38) .. controls (362.36,70.53) and (362.25,70.65) .. (362.1,70.65) .. controls (361.95,70.65) and (361.84,70.53) .. (361.84,70.38) -- cycle ;
\draw  [color={rgb, 255:red, 0; green, 0; blue, 0 }  ,draw opacity=0.4 ][fill={rgb, 255:red, 0; green, 0; blue, 0 }  ,fill opacity=1 ] (361.84,56.49) .. controls (361.84,56.34) and (361.95,56.21) .. (362.1,56.21) .. controls (362.25,56.21) and (362.36,56.34) .. (362.36,56.49) .. controls (362.36,56.65) and (362.25,56.77) .. (362.1,56.77) .. controls (361.95,56.77) and (361.84,56.65) .. (361.84,56.49) -- cycle ;

\draw [color={rgb, 255:red, 0; green, 0; blue, 0 }  ,draw opacity=0.4 ][line width=1.5]    (201.1,20.31) -- (214.3,6.43) ;
\draw [color={rgb, 255:red, 0; green, 0; blue, 0 }  ,draw opacity=0.4 ][line width=1.5]    (201.1,6.43) -- (214.3,20.31) ;
\draw  [color={rgb, 255:red, 0; green, 0; blue, 0 }  ,draw opacity=0.4 ][fill={rgb, 255:red, 155; green, 155; blue, 155 }  ,fill opacity=1 ] (210.14,14.43) -- (208.71,15.93) -- (206.69,15.93) -- (205.26,14.43) -- (205.26,12.3) -- (206.69,10.8) -- (208.71,10.8) -- (210.14,12.3) -- cycle ;
\draw  [color={rgb, 255:red, 0; green, 0; blue, 0 }  ,draw opacity=0.4 ][fill={rgb, 255:red, 184; green, 233; blue, 134 }  ,fill opacity=1 ] (200.28,6.43) .. controls (200.28,5.95) and (200.65,5.56) .. (201.1,5.56) .. controls (201.56,5.56) and (201.93,5.95) .. (201.93,6.43) .. controls (201.93,6.9) and (201.56,7.29) .. (201.1,7.29) .. controls (200.65,7.29) and (200.28,6.9) .. (200.28,6.43) -- cycle ;
\draw  [color={rgb, 255:red, 0; green, 0; blue, 0 }  ,draw opacity=0.4 ][fill={rgb, 255:red, 184; green, 233; blue, 134 }  ,fill opacity=1 ] (213.48,20.31) .. controls (213.48,19.83) and (213.84,19.44) .. (214.3,19.44) .. controls (214.76,19.44) and (215.13,19.83) .. (215.13,20.31) .. controls (215.13,20.79) and (214.76,21.18) .. (214.3,21.18) .. controls (213.84,21.18) and (213.48,20.79) .. (213.48,20.31) -- cycle ;
\draw  [color={rgb, 255:red, 0; green, 0; blue, 0 }  ,draw opacity=0.4 ][fill={rgb, 255:red, 158; green, 202; blue, 255 }  ,fill opacity=1 ] (213.48,6.43) .. controls (213.48,5.95) and (213.84,5.56) .. (214.3,5.56) .. controls (214.76,5.56) and (215.13,5.95) .. (215.13,6.43) .. controls (215.13,6.9) and (214.76,7.29) .. (214.3,7.29) .. controls (213.84,7.29) and (213.48,6.9) .. (213.48,6.43) -- cycle ;
\draw  [color={rgb, 255:red, 0; green, 0; blue, 0 }  ,draw opacity=0.4 ][fill={rgb, 255:red, 158; green, 202; blue, 255 }  ,fill opacity=1 ] (200.28,20.31) .. controls (200.28,19.83) and (200.65,19.44) .. (201.1,19.44) .. controls (201.56,19.44) and (201.93,19.83) .. (201.93,20.31) .. controls (201.93,20.79) and (201.56,21.18) .. (201.1,21.18) .. controls (200.65,21.18) and (200.28,20.79) .. (200.28,20.31) -- cycle ;
\draw  [color={rgb, 255:red, 0; green, 0; blue, 0 }  ,draw opacity=0.4 ][fill={rgb, 255:red, 74; green, 74; blue, 74 }  ,fill opacity=1 ] (214.3,20.31) -- (216.68,17.33) -- (217.05,17.3) -- (217.25,17.48) -- (217.25,17.48) -- (214.87,20.46) -- (214.5,20.49) -- (214.3,20.31) -- cycle ;
\draw  [color={rgb, 255:red, 0; green, 0; blue, 0 }  ,draw opacity=0.4 ][fill={rgb, 255:red, 74; green, 74; blue, 74 }  ,fill opacity=1 ] (214.3,20.31) -- (211.92,23.29) -- (211.55,23.32) -- (211.35,23.14) -- (211.35,23.14) -- (213.73,20.16) -- (214.1,20.13) -- (214.3,20.31) -- cycle ;
\draw  [color={rgb, 255:red, 0; green, 0; blue, 0 }  ,draw opacity=0.4 ][fill={rgb, 255:red, 74; green, 74; blue, 74 }  ,fill opacity=1 ] (201.1,6.43) -- (203.48,3.45) -- (203.85,3.41) -- (204.05,3.59) -- (204.05,3.59) -- (201.67,6.57) -- (201.3,6.6) -- (201.1,6.43) -- cycle ;
\draw  [color={rgb, 255:red, 0; green, 0; blue, 0 }  ,draw opacity=0.4 ][fill={rgb, 255:red, 74; green, 74; blue, 74 }  ,fill opacity=1 ] (201.1,6.43) -- (198.73,9.41) -- (198.36,9.44) -- (198.15,9.26) -- (198.15,9.26) -- (200.53,6.28) -- (200.9,6.25) -- (201.1,6.43) -- cycle ;
\draw  [color={rgb, 255:red, 0; green, 0; blue, 0 }  ,draw opacity=0.4 ][fill={rgb, 255:red, 74; green, 74; blue, 74 }  ,fill opacity=1 ] (198.49,17.56) -- (201.1,20.31) -- (201.1,20.31) -- (200.92,20.51) -- (200.54,20.51) -- (197.93,17.76) -- (197.93,17.76) -- (198.12,17.56) -- cycle ;
\draw  [color={rgb, 255:red, 0; green, 0; blue, 0 }  ,draw opacity=0.4 ][fill={rgb, 255:red, 74; green, 74; blue, 74 }  ,fill opacity=1 ] (201.66,20.11) -- (204.28,22.86) -- (204.28,22.86) -- (204.09,23.06) -- (203.72,23.06) -- (201.1,20.31) -- (201.1,20.31) -- (201.29,20.11) -- cycle ;
\draw  [color={rgb, 255:red, 0; green, 0; blue, 0 }  ,draw opacity=0.4 ][fill={rgb, 255:red, 74; green, 74; blue, 74 }  ,fill opacity=1 ] (211.69,3.68) -- (214.3,6.43) -- (214.3,6.43) -- (214.11,6.62) -- (213.74,6.62) -- (211.13,3.87) -- (211.13,3.87) -- (211.31,3.68) -- cycle ;
\draw  [color={rgb, 255:red, 0; green, 0; blue, 0 }  ,draw opacity=0.4 ][fill={rgb, 255:red, 74; green, 74; blue, 74 }  ,fill opacity=1 ] (214.86,6.23) -- (217.47,8.98) -- (217.47,8.98) -- (217.29,9.17) -- (216.91,9.17) -- (214.3,6.43) -- (214.3,6.43) -- (214.49,6.23) -- cycle ;
\draw  [color={rgb, 255:red, 0; green, 0; blue, 0 }  ,draw opacity=0.4 ][fill={rgb, 255:red, 0; green, 0; blue, 0 }  ,fill opacity=1 ] (200.84,20.31) .. controls (200.84,20.16) and (200.96,20.03) .. (201.1,20.03) .. controls (201.25,20.03) and (201.37,20.16) .. (201.37,20.31) .. controls (201.37,20.46) and (201.25,20.59) .. (201.1,20.59) .. controls (200.96,20.59) and (200.84,20.46) .. (200.84,20.31) -- cycle ;
\draw  [color={rgb, 255:red, 0; green, 0; blue, 0 }  ,draw opacity=0.4 ][fill={rgb, 255:red, 0; green, 0; blue, 0 }  ,fill opacity=1 ] (200.84,6.43) .. controls (200.84,6.27) and (200.96,6.15) .. (201.1,6.15) .. controls (201.25,6.15) and (201.37,6.27) .. (201.37,6.43) .. controls (201.37,6.58) and (201.25,6.7) .. (201.1,6.7) .. controls (200.96,6.7) and (200.84,6.58) .. (200.84,6.43) -- cycle ;
\draw  [color={rgb, 255:red, 0; green, 0; blue, 0 }  ,draw opacity=0.4 ][fill={rgb, 255:red, 0; green, 0; blue, 0 }  ,fill opacity=1 ] (214.04,20.31) .. controls (214.04,20.16) and (214.15,20.03) .. (214.3,20.03) .. controls (214.45,20.03) and (214.56,20.16) .. (214.56,20.31) .. controls (214.56,20.46) and (214.45,20.59) .. (214.3,20.59) .. controls (214.15,20.59) and (214.04,20.46) .. (214.04,20.31) -- cycle ;
\draw  [color={rgb, 255:red, 0; green, 0; blue, 0 }  ,draw opacity=0.4 ][fill={rgb, 255:red, 0; green, 0; blue, 0 }  ,fill opacity=1 ] (214.04,6.43) .. controls (214.04,6.27) and (214.15,6.15) .. (214.3,6.15) .. controls (214.45,6.15) and (214.56,6.27) .. (214.56,6.43) .. controls (214.56,6.58) and (214.45,6.7) .. (214.3,6.7) .. controls (214.15,6.7) and (214.04,6.58) .. (214.04,6.43) -- cycle ;

\draw [color={rgb, 255:red, 0; green, 114; blue, 67 }  ,draw opacity=1 ]   (207.2,170.27) -- (347.42,128.86) ;
\draw [shift={(349.33,128.29)}, rotate = 163.55] [color={rgb, 255:red, 0; green, 114; blue, 67 }  ,draw opacity=1 ][line width=0.75]    (6.56,-1.97) .. controls (4.17,-0.84) and (1.99,-0.18) .. (0,0) .. controls (1.99,0.18) and (4.17,0.84) .. (6.56,1.97)   ;
\draw [color={rgb, 255:red, 239; green, 12; blue, 16 }  ,draw opacity=1 ]   (207.2,170.27) -- (207.7,15.37) ;
\draw [shift={(207.7,13.37)}, rotate = 90.18] [color={rgb, 255:red, 239; green, 12; blue, 16 }  ,draw opacity=1 ][line width=0.75]    (6.56,-1.97) .. controls (4.17,-0.84) and (1.99,-0.18) .. (0,0) .. controls (1.99,0.18) and (4.17,0.84) .. (6.56,1.97)   ;
\draw [color={rgb, 255:red, 239; green, 12; blue, 16 }  ,draw opacity=1 ]   (207.2,170.27) -- (353.88,64.6) ;
\draw [shift={(355.5,63.43)}, rotate = 144.23] [color={rgb, 255:red, 239; green, 12; blue, 16 }  ,draw opacity=1 ][line width=0.75]    (6.56,-2.94) .. controls (4.17,-1.38) and (1.99,-0.4) .. (0,0) .. controls (1.99,0.4) and (4.17,1.38) .. (6.56,2.94)   ;
\draw [draw opacity=0]   (217.76,159.16) -- (196.64,181.37) ;
\draw [draw opacity=0]   (196.64,159.16) -- (197.48,160.05) -- (217.76,181.37) ;
\draw [line width=1.5]    (200.6,177.21) -- (213.8,163.33) ;
\draw [line width=1.5]    (200.6,163.33) -- (213.8,177.21) ;
\draw  [fill={rgb, 255:red, 155; green, 155; blue, 155 }  ,fill opacity=1 ] (209.64,171.33) -- (208.21,172.83) -- (206.19,172.83) -- (204.76,171.33) -- (204.76,169.2) -- (206.19,167.7) -- (208.21,167.7) -- (209.64,169.2) -- cycle ;
\draw  [fill={rgb, 255:red, 184; green, 233; blue, 134 }  ,fill opacity=1 ] (199.78,163.33) .. controls (199.78,162.85) and (200.15,162.46) .. (200.6,162.46) .. controls (201.06,162.46) and (201.43,162.85) .. (201.43,163.33) .. controls (201.43,163.8) and (201.06,164.19) .. (200.6,164.19) .. controls (200.15,164.19) and (199.78,163.8) .. (199.78,163.33) -- cycle ;
\draw  [fill={rgb, 255:red, 184; green, 233; blue, 134 }  ,fill opacity=1 ] (212.98,177.21) .. controls (212.98,176.73) and (213.34,176.34) .. (213.8,176.34) .. controls (214.26,176.34) and (214.63,176.73) .. (214.63,177.21) .. controls (214.63,177.69) and (214.26,178.08) .. (213.8,178.08) .. controls (213.34,178.08) and (212.98,177.69) .. (212.98,177.21) -- cycle ;
\draw  [fill={rgb, 255:red, 158; green, 202; blue, 255 }  ,fill opacity=1 ] (212.98,163.33) .. controls (212.98,162.85) and (213.34,162.46) .. (213.8,162.46) .. controls (214.26,162.46) and (214.63,162.85) .. (214.63,163.33) .. controls (214.63,163.8) and (214.26,164.19) .. (213.8,164.19) .. controls (213.34,164.19) and (212.98,163.8) .. (212.98,163.33) -- cycle ;
\draw  [fill={rgb, 255:red, 158; green, 202; blue, 255 }  ,fill opacity=1 ] (199.78,177.21) .. controls (199.78,176.73) and (200.15,176.34) .. (200.6,176.34) .. controls (201.06,176.34) and (201.43,176.73) .. (201.43,177.21) .. controls (201.43,177.69) and (201.06,178.08) .. (200.6,178.08) .. controls (200.15,178.08) and (199.78,177.69) .. (199.78,177.21) -- cycle ;
\draw  [fill={rgb, 255:red, 74; green, 74; blue, 74 }  ,fill opacity=1 ] (213.8,177.21) -- (216.18,174.23) -- (216.55,174.2) -- (216.75,174.38) -- (216.75,174.38) -- (214.37,177.36) -- (214,177.39) -- (213.8,177.21) -- cycle ;
\draw  [fill={rgb, 255:red, 74; green, 74; blue, 74 }  ,fill opacity=1 ] (213.8,177.21) -- (211.42,180.19) -- (211.05,180.22) -- (210.85,180.04) -- (210.85,180.04) -- (213.23,177.06) -- (213.6,177.03) -- (213.8,177.21) -- cycle ;
\draw  [fill={rgb, 255:red, 74; green, 74; blue, 74 }  ,fill opacity=1 ] (200.6,163.33) -- (202.98,160.35) -- (203.35,160.31) -- (203.55,160.49) -- (203.55,160.49) -- (201.17,163.47) -- (200.8,163.5) -- (200.6,163.33) -- cycle ;
\draw  [fill={rgb, 255:red, 74; green, 74; blue, 74 }  ,fill opacity=1 ] (200.6,163.33) -- (198.23,166.31) -- (197.86,166.34) -- (197.65,166.16) -- (197.65,166.16) -- (200.03,163.18) -- (200.4,163.15) -- (200.6,163.33) -- cycle ;
\draw  [fill={rgb, 255:red, 74; green, 74; blue, 74 }  ,fill opacity=1 ] (197.99,174.46) -- (200.6,177.21) -- (200.6,177.21) -- (200.42,177.41) -- (200.04,177.41) -- (197.43,174.66) -- (197.43,174.66) -- (197.62,174.46) -- cycle ;
\draw  [fill={rgb, 255:red, 74; green, 74; blue, 74 }  ,fill opacity=1 ] (201.16,177.01) -- (203.78,179.76) -- (203.78,179.76) -- (203.59,179.96) -- (203.22,179.96) -- (200.6,177.21) -- (200.6,177.21) -- (200.79,177.01) -- cycle ;
\draw  [fill={rgb, 255:red, 74; green, 74; blue, 74 }  ,fill opacity=1 ] (211.19,160.58) -- (213.8,163.33) -- (213.8,163.33) -- (213.61,163.52) -- (213.24,163.52) -- (210.63,160.77) -- (210.63,160.77) -- (210.81,160.58) -- cycle ;
\draw  [fill={rgb, 255:red, 74; green, 74; blue, 74 }  ,fill opacity=1 ] (214.36,163.13) -- (216.97,165.88) -- (216.97,165.88) -- (216.79,166.07) -- (216.41,166.07) -- (213.8,163.33) -- (213.8,163.33) -- (213.99,163.13) -- cycle ;
\draw  [fill={rgb, 255:red, 0; green, 0; blue, 0 }  ,fill opacity=1 ] (200.34,177.21) .. controls (200.34,177.06) and (200.46,176.93) .. (200.6,176.93) .. controls (200.75,176.93) and (200.87,177.06) .. (200.87,177.21) .. controls (200.87,177.36) and (200.75,177.49) .. (200.6,177.49) .. controls (200.46,177.49) and (200.34,177.36) .. (200.34,177.21) -- cycle ;
\draw  [fill={rgb, 255:red, 0; green, 0; blue, 0 }  ,fill opacity=1 ] (200.34,163.33) .. controls (200.34,163.17) and (200.46,163.05) .. (200.6,163.05) .. controls (200.75,163.05) and (200.87,163.17) .. (200.87,163.33) .. controls (200.87,163.48) and (200.75,163.6) .. (200.6,163.6) .. controls (200.46,163.6) and (200.34,163.48) .. (200.34,163.33) -- cycle ;
\draw  [fill={rgb, 255:red, 0; green, 0; blue, 0 }  ,fill opacity=1 ] (213.54,177.21) .. controls (213.54,177.06) and (213.65,176.93) .. (213.8,176.93) .. controls (213.95,176.93) and (214.06,177.06) .. (214.06,177.21) .. controls (214.06,177.36) and (213.95,177.49) .. (213.8,177.49) .. controls (213.65,177.49) and (213.54,177.36) .. (213.54,177.21) -- cycle ;
\draw  [fill={rgb, 255:red, 0; green, 0; blue, 0 }  ,fill opacity=1 ] (213.54,163.33) .. controls (213.54,163.17) and (213.65,163.05) .. (213.8,163.05) .. controls (213.95,163.05) and (214.06,163.17) .. (214.06,163.33) .. controls (214.06,163.48) and (213.95,163.6) .. (213.8,163.6) .. controls (213.65,163.6) and (213.54,163.48) .. (213.54,163.33) -- cycle ;

\draw [color={rgb, 255:red, 0; green, 114; blue, 67 }  ,draw opacity=1 ] [dash pattern={on 4.5pt off 4.5pt}]  (348.97,128.43) -- (355.3,65.42) ;
\draw [shift={(355.5,63.43)}, rotate = 95.74] [color={rgb, 255:red, 0; green, 114; blue, 67 }  ,draw opacity=1 ][line width=0.75]    (6.56,-1.97) .. controls (4.17,-0.84) and (1.99,-0.18) .. (0,0) .. controls (1.99,0.18) and (4.17,0.84) .. (6.56,1.97)   ;

\draw (49.52,3.08) node [anchor=north west][inner sep=0.75pt]  [font=\footnotesize,rotate=-359.97]  {$b^{1} =1$};
\draw (333.45,187.6) node [anchor=north west][inner sep=0.75pt]  [font=\footnotesize,rotate=-276.6]  {$b^{4} =0$};
\draw (328.28,186.7) node [anchor=south west] [inner sep=0.75pt]  [font=\footnotesize,rotate=-276.63]  {$b^{4} =1$};
\draw (1.53,151.29) node [anchor=south west] [inner sep=0.75pt]  [font=\footnotesize,rotate=-351.4]  {$b^{3} =1$};
\draw (5.79,63.15) node [anchor=north west][inner sep=0.75pt]  [font=\footnotesize,rotate=-23.9]  {$b^{2} =0$};
\draw (7.18,59.19) node [anchor=south west] [inner sep=0.75pt]  [font=\footnotesize,rotate=-23.86]  {$b^{2} =1$};
\draw (2.54,158.01) node [anchor=north west][inner sep=0.75pt]  [font=\footnotesize,rotate=-351.38]  {$b^{3} =0$};
\draw (43.5,3.07) node [anchor=north east] [inner sep=0.75pt]  [font=\footnotesize]  {$b^{1} =0$};
\draw (397.34,38.42) node [anchor=north east] [inner sep=0.75pt]  [font=\footnotesize]  {$b^{5} =1$};
\draw (397.34,31.62) node [anchor=south east] [inner sep=0.75pt]  [font=\footnotesize]  {$b^{5} =0$};

\end{tikzpicture}

%% file: Tikz_Simulation_Plot_ECM_short.tex
\begin{tikzpicture}

\begin{axis}[%
width=2.826in,
height=0.607in,
at={(0.474in,2.039in)},
scale only axis,
xmin=0,
xmax=248,
ymin=0.2,
ymax=1,
ytick={0.2, 0.6,   1},
ylabel style={font=\color{white!15!black}},
ylabel={$\mathrm{SoC}$ [-]},
axis background/.style={fill=white},
title style={font=\bfseries},
title={Battery State Estimation},
axis x line*=bottom,
axis y line*=left,
xmajorgrids,
ymajorgrids
]
\addplot [color=blue, forget plot]
  table[row sep=crcr]{%
0	1\\
1	0.999346\\
2	0.998647\\
3	0.997429\\
4	0.996574\\
5	0.995808\\
6	0.99507\\
7	0.994226\\
8	0.993489\\
9	0.992832\\
10	0.992203\\
11	0.991523\\
12	0.990847\\
13	0.990141\\
14	0.98938\\
15	0.988619\\
16	0.987851\\
17	0.987079\\
18	0.986327\\
19	0.98561\\
20	0.984914\\
21	0.984305\\
22	0.983677\\
23	0.983074\\
24	0.982369\\
25	0.981663\\
26	0.980957\\
27	0.98025\\
28	0.979558\\
29	0.978791\\
30	0.978089\\
31	0.977387\\
32	0.976685\\
33	0.975983\\
34	0.975281\\
35	0.974585\\
36	0.973565\\
37	0.972872\\
38	0.972176\\
39	0.971481\\
40	0.970789\\
41	0.970205\\
42	0.969493\\
43	0.96875\\
44	0.967998\\
45	0.967236\\
46	0.966465\\
47	0.965727\\
48	0.965012\\
49	0.964311\\
50	0.96359\\
51	0.962859\\
52	0.962128\\
53	0.961498\\
54	0.961162\\
55	0.960547\\
56	0.959438\\
57	0.958741\\
58	0.958045\\
59	0.957349\\
60	0.956652\\
61	0.955956\\
62	0.955259\\
63	0.954562\\
64	0.95385\\
65	0.953133\\
66	0.952391\\
67	0.951615\\
68	0.950837\\
69	0.95006\\
70	0.949323\\
71	0.948633\\
72	0.947929\\
73	0.94721\\
74	0.94649\\
75	0.945771\\
76	0.94505\\
77	0.944347\\
78	0.943635\\
79	0.942908\\
80	0.942181\\
81	0.941454\\
82	0.940726\\
83	0.939997\\
84	0.939274\\
85	0.938553\\
86	0.937831\\
87	0.937109\\
88	0.936388\\
89	0.935668\\
90	0.934949\\
91	0.934232\\
92	0.933515\\
93	0.932798\\
94	0.93208\\
95	0.931363\\
96	0.930645\\
97	0.929927\\
98	0.92921\\
99	0.92849\\
100	0.927781\\
101	0.927086\\
102	0.92639\\
103	0.925694\\
104	0.924998\\
105	0.924298\\
106	0.923594\\
107	0.922888\\
108	0.922182\\
109	0.921476\\
110	0.920766\\
111	0.920057\\
112	0.919338\\
113	0.918596\\
114	0.917851\\
115	0.917105\\
116	0.916359\\
117	0.915613\\
118	0.91489\\
119	0.914181\\
120	0.913448\\
121	0.912693\\
122	0.911919\\
123	0.911146\\
124	0.910397\\
125	0.909668\\
126	0.90894\\
127	0.908208\\
128	0.907473\\
129	0.906738\\
130	0.906011\\
131	0.905289\\
132	0.904568\\
133	0.903846\\
134	0.903124\\
135	0.902402\\
136	0.90168\\
137	0.900958\\
138	0.900236\\
139	0.899514\\
140	0.898792\\
141	0.898069\\
142	0.897347\\
143	0.896625\\
144	0.895907\\
145	0.895217\\
146	0.894537\\
147	0.893817\\
148	0.893062\\
149	0.892295\\
150	0.891518\\
151	0.890761\\
152	0.890042\\
153	0.88932\\
154	0.888584\\
155	0.88784\\
156	0.887087\\
157	0.886333\\
158	0.885579\\
159	0.884825\\
160	0.884085\\
161	0.883352\\
162	0.882616\\
163	0.881877\\
164	0.881138\\
165	0.880406\\
166	0.87968\\
167	0.878955\\
168	0.878229\\
169	0.877504\\
170	0.876778\\
171	0.876052\\
172	0.875327\\
173	0.874601\\
174	0.873875\\
175	0.873149\\
176	0.872422\\
177	0.871696\\
178	0.87097\\
179	0.870259\\
180	0.869573\\
181	0.868892\\
182	0.868211\\
183	0.867529\\
184	0.866848\\
185	0.86616\\
186	0.86544\\
187	0.864711\\
188	0.863981\\
189	0.86325\\
190	0.86252\\
191	0.86179\\
192	0.861059\\
193	0.860328\\
194	0.859597\\
195	0.858867\\
196	0.858136\\
197	0.857404\\
198	0.856673\\
199	0.855942\\
200	0.85521\\
201	0.854479\\
202	0.853747\\
203	0.853017\\
204	0.852287\\
205	0.851558\\
206	0.850829\\
207	0.850099\\
208	0.849369\\
209	0.84864\\
210	0.84791\\
211	0.84718\\
212	0.84645\\
213	0.84572\\
214	0.844989\\
215	0.844259\\
216	0.843529\\
217	0.842798\\
218	0.842067\\
219	0.841336\\
220	0.840605\\
221	0.839886\\
222	0.839164\\
223	0.838443\\
224	0.83772\\
225	0.836998\\
226	0.836267\\
227	0.835537\\
228	0.834806\\
229	0.834076\\
230	0.833345\\
231	0.832614\\
232	0.831883\\
233	0.831151\\
234	0.83042\\
235	0.829689\\
236	0.828957\\
237	0.828333\\
238	0.827899\\
239	0.827217\\
240	0.826535\\
241	0.825848\\
242	0.825161\\
243	0.824473\\
244	0.823786\\
245	0.823098\\
246	0.82241\\
247	0.821722\\
};
\addplot [color=red, forget plot]
  table[row sep=crcr]{%
0	0.433\\
1	0.432278\\
2	0.431537\\
3	0.43035\\
4	0.429311\\
5	0.428379\\
6	0.427527\\
7	0.426753\\
8	0.425957\\
9	0.425336\\
10	0.424549\\
11	0.423762\\
12	0.422991\\
13	0.422268\\
14	0.421511\\
15	0.420754\\
16	0.419996\\
17	0.419232\\
18	0.418443\\
19	0.417641\\
20	0.416834\\
21	0.416013\\
22	0.415191\\
23	0.414369\\
24	0.413556\\
25	0.412764\\
26	0.411977\\
27	0.411187\\
28	0.410393\\
29	0.409599\\
30	0.408814\\
31	0.408034\\
32	0.40762\\
33	0.406842\\
34	0.406063\\
35	0.405284\\
36	0.404336\\
37	0.403555\\
38	0.402774\\
39	0.401993\\
40	0.401213\\
41	0.400431\\
42	0.399662\\
43	0.398909\\
44	0.398156\\
45	0.397404\\
46	0.396634\\
47	0.395846\\
48	0.395055\\
49	0.39426\\
50	0.393468\\
51	0.392683\\
52	0.391902\\
53	0.391121\\
54	0.390339\\
55	0.389558\\
56	0.388776\\
57	0.387994\\
58	0.387213\\
59	0.386431\\
60	0.385649\\
61	0.384867\\
62	0.384085\\
63	0.383303\\
64	0.382523\\
65	0.381744\\
66	0.380995\\
67	0.380258\\
68	0.379477\\
69	0.37866\\
70	0.377841\\
71	0.377018\\
72	0.376197\\
73	0.375409\\
74	0.374666\\
75	0.37392\\
76	0.373134\\
77	0.372294\\
78	0.371446\\
79	0.370598\\
80	0.369755\\
81	0.368941\\
82	0.368154\\
83	0.36737\\
84	0.366586\\
85	0.365802\\
86	0.365019\\
87	0.364235\\
88	0.363451\\
89	0.362668\\
90	0.361884\\
91	0.3611\\
92	0.360314\\
93	0.359542\\
94	0.358786\\
95	0.35803\\
96	0.357274\\
97	0.3565\\
98	0.355709\\
99	0.354917\\
100	0.354126\\
101	0.353334\\
102	0.352543\\
103	0.351755\\
104	0.350971\\
105	0.350186\\
106	0.349408\\
107	0.348661\\
108	0.347923\\
109	0.347174\\
110	0.346389\\
111	0.345588\\
112	0.344786\\
113	0.343984\\
114	0.343182\\
115	0.342383\\
116	0.341604\\
117	0.340839\\
118	0.340075\\
119	0.33931\\
120	0.338541\\
121	0.337771\\
122	0.337025\\
123	0.336286\\
124	0.335508\\
125	0.334675\\
126	0.333824\\
127	0.332974\\
128	0.332122\\
129	0.3313\\
130	0.330513\\
131	0.329742\\
132	0.329102\\
133	0.328427\\
134	0.327691\\
135	0.326926\\
136	0.326149\\
137	0.325381\\
138	0.324609\\
139	0.323831\\
140	0.323054\\
141	0.322277\\
142	0.3215\\
143	0.320741\\
144	0.32\\
145	0.319251\\
146	0.318502\\
147	0.317752\\
148	0.317752\\
149	0.317752\\
150	0.317752\\
151	0.317752\\
152	0.317752\\
153	0.317752\\
154	0.317752\\
155	0.317752\\
156	0.317752\\
157	0.317752\\
158	0.317752\\
159	0.317752\\
160	0.317752\\
161	0.317752\\
162	0.317752\\
163	0.317752\\
164	0.317752\\
165	0.317752\\
166	0.317752\\
167	0.317752\\
168	0.317752\\
169	0.317752\\
170	0.317752\\
171	0.317752\\
172	0.317752\\
173	0.317752\\
174	0.317752\\
175	0.317752\\
176	0.317752\\
177	0.317752\\
178	0.317752\\
179	0.317752\\
180	0.317752\\
181	0.317752\\
182	0.317752\\
183	0.317752\\
184	0.317752\\
185	0.317752\\
186	0.317752\\
187	0.317752\\
188	0.317752\\
189	0.317752\\
190	0.317752\\
191	0.317752\\
192	0.317752\\
193	0.317752\\
194	0.317752\\
195	0.317752\\
196	0.317752\\
197	0.317752\\
198	0.317752\\
199	0.317752\\
200	0.317752\\
201	0.317752\\
202	0.317752\\
203	0.317752\\
204	0.317752\\
205	0.317752\\
206	0.317752\\
207	0.317752\\
208	0.317752\\
209	0.317752\\
210	0.317752\\
211	0.317752\\
212	0.317752\\
213	0.317752\\
214	0.317752\\
215	0.317752\\
216	0.317752\\
217	0.317752\\
218	0.317752\\
219	0.317752\\
220	0.317752\\
221	0.317752\\
222	0.317752\\
223	0.317752\\
224	0.317752\\
225	0.317752\\
226	0.317752\\
227	0.317752\\
228	0.317752\\
229	0.317752\\
230	0.317752\\
231	0.317752\\
232	0.317752\\
233	0.317752\\
234	0.317752\\
235	0.317752\\
236	0.317752\\
237	0.317752\\
238	0.317752\\
239	0.317752\\
240	0.317752\\
241	0.317752\\
242	0.317752\\
243	0.317752\\
244	0.317752\\
245	0.317752\\
246	0.317752\\
247	0.317752\\
};
\addplot [color=black, dashed, line width=1.0pt, forget plot]
  table[row sep=crcr]{%
0	0.3\\
1	0.3\\
2	0.3\\
3	0.3\\
4	0.3\\
5	0.3\\
6	0.3\\
7	0.3\\
8	0.3\\
9	0.3\\
10	0.3\\
11	0.3\\
12	0.3\\
13	0.3\\
14	0.3\\
15	0.3\\
16	0.3\\
17	0.3\\
18	0.3\\
19	0.3\\
20	0.3\\
21	0.3\\
22	0.3\\
23	0.3\\
24	0.3\\
25	0.3\\
26	0.3\\
27	0.3\\
28	0.3\\
29	0.3\\
30	0.3\\
31	0.3\\
32	0.3\\
33	0.3\\
34	0.3\\
35	0.3\\
36	0.3\\
37	0.3\\
38	0.3\\
39	0.3\\
40	0.3\\
41	0.3\\
42	0.3\\
43	0.3\\
44	0.3\\
45	0.3\\
46	0.3\\
47	0.3\\
48	0.3\\
49	0.3\\
50	0.3\\
51	0.3\\
52	0.3\\
53	0.3\\
54	0.3\\
55	0.3\\
56	0.3\\
57	0.3\\
58	0.3\\
59	0.3\\
60	0.3\\
61	0.3\\
62	0.3\\
63	0.3\\
64	0.3\\
65	0.3\\
66	0.3\\
67	0.3\\
68	0.3\\
69	0.3\\
70	0.3\\
71	0.3\\
72	0.3\\
73	0.3\\
74	0.3\\
75	0.3\\
76	0.3\\
77	0.3\\
78	0.3\\
79	0.3\\
80	0.3\\
81	0.3\\
82	0.3\\
83	0.3\\
84	0.3\\
85	0.3\\
86	0.3\\
87	0.3\\
88	0.3\\
89	0.3\\
90	0.3\\
91	0.3\\
92	0.3\\
93	0.3\\
94	0.3\\
95	0.3\\
96	0.3\\
97	0.3\\
98	0.3\\
99	0.3\\
100	0.3\\
101	0.3\\
102	0.3\\
103	0.3\\
104	0.3\\
105	0.3\\
106	0.3\\
107	0.3\\
108	0.3\\
109	0.3\\
110	0.3\\
111	0.3\\
112	0.3\\
113	0.3\\
114	0.3\\
115	0.3\\
116	0.3\\
117	0.3\\
118	0.3\\
119	0.3\\
120	0.3\\
121	0.3\\
122	0.3\\
123	0.3\\
124	0.3\\
125	0.3\\
126	0.3\\
127	0.3\\
128	0.3\\
129	0.3\\
130	0.3\\
131	0.3\\
132	0.3\\
133	0.3\\
134	0.3\\
135	0.3\\
136	0.3\\
137	0.3\\
138	0.3\\
139	0.3\\
140	0.3\\
141	0.3\\
142	0.3\\
143	0.3\\
144	0.3\\
145	0.3\\
146	0.3\\
147	0.3\\
148	0.3\\
149	0.3\\
150	0.3\\
151	0.3\\
152	0.3\\
153	0.3\\
154	0.3\\
155	0.3\\
156	0.3\\
157	0.3\\
158	0.3\\
159	0.3\\
160	0.3\\
161	0.3\\
162	0.3\\
163	0.3\\
164	0.3\\
165	0.3\\
166	0.3\\
167	0.3\\
168	0.3\\
169	0.3\\
170	0.3\\
171	0.3\\
172	0.3\\
173	0.3\\
174	0.3\\
175	0.3\\
176	0.3\\
177	0.3\\
178	0.3\\
179	0.3\\
180	0.3\\
181	0.3\\
182	0.3\\
183	0.3\\
184	0.3\\
185	0.3\\
186	0.3\\
187	0.3\\
188	0.3\\
189	0.3\\
190	0.3\\
191	0.3\\
192	0.3\\
193	0.3\\
194	0.3\\
195	0.3\\
196	0.3\\
197	0.3\\
198	0.3\\
199	0.3\\
200	0.3\\
201	0.3\\
202	0.3\\
203	0.3\\
204	0.3\\
205	0.3\\
206	0.3\\
207	0.3\\
208	0.3\\
209	0.3\\
210	0.3\\
211	0.3\\
212	0.3\\
213	0.3\\
214	0.3\\
215	0.3\\
216	0.3\\
217	0.3\\
218	0.3\\
219	0.3\\
220	0.3\\
221	0.3\\
222	0.3\\
223	0.3\\
224	0.3\\
225	0.3\\
226	0.3\\
227	0.3\\
228	0.3\\
229	0.3\\
230	0.3\\
231	0.3\\
232	0.3\\
233	0.3\\
234	0.3\\
235	0.3\\
236	0.3\\
237	0.3\\
238	0.3\\
239	0.3\\
240	0.3\\
241	0.3\\
242	0.3\\
243	0.3\\
244	0.3\\
245	0.3\\
246	0.3\\
247	0.3\\
};
\end{axis}

\begin{axis}[%
width=2.826in,
height=0.607in,
at={(0.474in,1.181in)},
scale only axis,
xmin=0,
xmax=248,
ymin=14,
ymax=17,
ytick={14, 15, 16, 17},
ylabel style={font=\color{white!15!black}},
ylabel={Voltage [V]},
axis background/.style={fill=white},
axis x line*=bottom,
axis y line*=left,
xmajorgrids,
ymajorgrids
]
\addplot [color=blue, forget plot]
  table[row sep=crcr]{%
0	16.4509\\
1	16.3478\\
2	16.0188\\
3	16.1188\\
4	16.1557\\
5	16.1739\\
6	16.122\\
7	16.1669\\
8	16.215\\
9	16.2415\\
10	16.2234\\
11	16.223\\
12	16.2056\\
13	16.1705\\
14	16.1593\\
15	16.1482\\
16	16.1399\\
17	16.1457\\
18	16.1622\\
19	16.1758\\
20	16.225\\
21	16.2256\\
22	16.2421\\
23	16.1915\\
24	16.1801\\
25	16.1721\\
26	16.1667\\
27	16.1701\\
28	16.1284\\
29	16.1521\\
30	16.1526\\
31	16.1518\\
32	16.1502\\
33	16.1482\\
34	16.1489\\
35	15.975\\
36	16.1074\\
37	16.1209\\
38	16.1284\\
39	16.1334\\
40	16.1913\\
41	16.1361\\
42	16.1104\\
43	16.0953\\
44	16.0822\\
45	16.0702\\
46	16.082\\
47	16.0931\\
48	16.1022\\
49	16.093\\
50	16.0847\\
51	16.0807\\
52	16.1314\\
53	16.2976\\
54	16.1892\\
55	15.9136\\
56	16.0657\\
57	16.076\\
58	16.0808\\
59	16.0824\\
60	16.0822\\
61	16.0807\\
62	16.0789\\
63	16.069\\
64	16.0621\\
65	16.0448\\
66	16.0197\\
67	16.0105\\
68	16.004\\
69	16.0209\\
70	16.0468\\
71	16.045\\
72	16.0366\\
73	16.0336\\
74	16.0308\\
75	16.0278\\
76	16.0343\\
77	16.0291\\
78	16.0185\\
79	16.0142\\
80	16.0105\\
81	16.0073\\
82	16.0037\\
83	16.0037\\
84	16.003\\
85	16.0003\\
86	15.9978\\
87	15.9956\\
88	15.994\\
89	15.9928\\
90	15.9911\\
91	15.9889\\
92	15.9864\\
93	15.9839\\
94	15.9814\\
95	15.9788\\
96	15.9762\\
97	15.9738\\
98	15.9703\\
99	15.9732\\
100	15.9787\\
101	15.9785\\
102	15.9772\\
103	15.9755\\
104	15.9713\\
105	15.966\\
106	15.9622\\
107	15.9591\\
108	15.9562\\
109	15.9517\\
110	15.9488\\
111	15.9407\\
112	15.9248\\
113	15.9168\\
114	15.9117\\
115	15.9077\\
116	15.9042\\
117	15.9129\\
118	15.9209\\
119	15.9081\\
120	15.8928\\
121	15.8772\\
122	15.8703\\
123	15.8785\\
124	15.8883\\
125	15.8892\\
126	15.8868\\
127	15.883\\
128	15.8804\\
129	15.8821\\
130	15.8835\\
131	15.8821\\
132	15.8802\\
133	15.8779\\
134	15.8756\\
135	15.8731\\
136	15.8705\\
137	15.8679\\
138	15.8653\\
139	15.8627\\
140	15.8601\\
141	15.8574\\
142	15.8553\\
143	15.8546\\
144	15.8677\\
145	15.8743\\
146	15.8535\\
147	15.8296\\
148	15.815\\
149	15.8016\\
150	15.8063\\
151	15.8235\\
152	15.8244\\
153	15.8165\\
154	15.8096\\
155	15.8004\\
156	15.7959\\
157	15.7921\\
158	15.7887\\
159	15.7929\\
160	15.7961\\
161	15.7934\\
162	15.7895\\
163	15.7867\\
164	15.7881\\
165	15.7897\\
166	15.7885\\
167	15.7867\\
168	15.7845\\
169	15.7821\\
170	15.7796\\
171	15.777\\
172	15.7744\\
173	15.7718\\
174	15.7692\\
175	15.7665\\
176	15.7639\\
177	15.7612\\
178	15.7672\\
179	15.7798\\
180	15.7836\\
181	15.784\\
182	15.7831\\
183	15.7816\\
184	15.7761\\
185	15.7567\\
186	15.7445\\
187	15.7383\\
188	15.7336\\
189	15.7298\\
190	15.7265\\
191	15.7234\\
192	15.7205\\
193	15.7177\\
194	15.715\\
195	15.7123\\
196	15.7096\\
197	15.7069\\
198	15.7043\\
199	15.7016\\
200	15.6989\\
201	15.6963\\
202	15.6943\\
203	15.6925\\
204	15.6901\\
205	15.6876\\
206	15.685\\
207	15.6824\\
208	15.6798\\
209	15.6771\\
210	15.6745\\
211	15.6718\\
212	15.6691\\
213	15.6665\\
214	15.6638\\
215	15.6612\\
216	15.6585\\
217	15.6558\\
218	15.6532\\
219	15.6505\\
220	15.6538\\
221	15.6518\\
222	15.6498\\
223	15.6466\\
224	15.6445\\
225	15.6377\\
226	15.6341\\
227	15.6309\\
228	15.628\\
229	15.6251\\
230	15.6224\\
231	15.6197\\
232	15.617\\
233	15.6143\\
234	15.6116\\
235	15.6089\\
236	15.6634\\
237	15.7758\\
238	15.6711\\
239	15.6567\\
240	15.6443\\
241	15.6371\\
242	15.6319\\
243	15.6278\\
244	15.6244\\
245	15.6214\\
246	15.6186\\
247	15.983\\
};
\addplot [color=red, forget plot]
  table[row sep=crcr]{%
0	14.856\\
1	14.7593\\
2	14.4691\\
3	14.4639\\
4	14.4907\\
5	14.5277\\
6	14.5753\\
7	14.5756\\
8	14.6721\\
9	14.6064\\
10	14.5985\\
11	14.602\\
12	14.6259\\
13	14.6119\\
14	14.6096\\
15	14.6076\\
16	14.6025\\
17	14.5866\\
18	14.5745\\
19	14.567\\
20	14.5551\\
21	14.5503\\
22	14.5468\\
23	14.5491\\
24	14.5589\\
25	14.563\\
26	14.5618\\
27	14.559\\
28	14.5581\\
29	14.5611\\
30	14.5633\\
31	14.7583\\
32	14.6072\\
33	14.5871\\
34	14.575\\
35	14.4779\\
36	14.5417\\
37	14.5469\\
38	14.5491\\
39	14.5501\\
40	14.5487\\
41	14.5547\\
42	14.5638\\
43	14.5655\\
44	14.5657\\
45	14.556\\
46	14.5436\\
47	14.5373\\
48	14.5319\\
49	14.5302\\
50	14.5319\\
51	14.5328\\
52	14.532\\
53	14.5309\\
54	14.5296\\
55	14.5282\\
56	14.5267\\
57	14.5252\\
58	14.5237\\
59	14.5222\\
60	14.5207\\
61	14.5192\\
62	14.5176\\
63	14.517\\
64	14.5168\\
65	14.531\\
66	14.54\\
67	14.5187\\
68	14.4948\\
69	14.4857\\
70	14.4789\\
71	14.4752\\
72	14.4905\\
73	14.5162\\
74	14.5204\\
75	14.5015\\
76	14.4679\\
77	14.4548\\
78	14.4477\\
79	14.4455\\
80	14.4583\\
81	14.474\\
82	14.4783\\
83	14.4805\\
84	14.4808\\
85	14.4803\\
86	14.4794\\
87	14.4782\\
88	14.4769\\
89	14.4754\\
90	14.4742\\
91	14.4716\\
92	14.4769\\
93	14.4856\\
94	14.4871\\
95	14.4872\\
96	14.4774\\
97	14.4648\\
98	14.4601\\
99	14.4568\\
100	14.4543\\
101	14.4527\\
102	14.4529\\
103	14.4531\\
104	14.452\\
105	14.4547\\
106	14.4705\\
107	14.478\\
108	14.4741\\
109	14.4546\\
110	14.4401\\
111	14.4343\\
112	14.4303\\
113	14.4273\\
114	14.4268\\
115	14.436\\
116	14.4442\\
117	14.4457\\
118	14.4459\\
119	14.4432\\
120	14.4415\\
121	14.4527\\
122	14.4576\\
123	14.4381\\
124	14.4036\\
125	14.3845\\
126	14.3759\\
127	14.3702\\
128	14.3816\\
129	14.401\\
130	14.414\\
131	14.487\\
132	14.4852\\
133	14.458\\
134	14.4374\\
135	14.4238\\
136	14.4227\\
137	14.4176\\
138	14.4119\\
139	14.4089\\
140	14.407\\
141	14.4054\\
142	14.4128\\
143	14.4235\\
144	14.4208\\
145	14.4199\\
146	14.4191\\
147	14.8177\\
148	14.9066\\
149	14.9569\\
150	14.9855\\
151	15.0017\\
152	15.0109\\
153	15.0161\\
154	15.0191\\
155	15.0208\\
156	15.0217\\
157	15.0222\\
158	15.0225\\
159	15.0227\\
160	15.0228\\
161	15.0229\\
162	15.0229\\
163	15.0229\\
164	15.0229\\
165	15.0229\\
166	15.0229\\
167	15.0229\\
168	15.0229\\
169	15.0229\\
170	15.0229\\
171	15.0229\\
172	15.0229\\
173	15.0229\\
174	15.0229\\
175	15.0229\\
176	15.0229\\
177	15.0229\\
178	15.0229\\
179	15.0229\\
180	15.0229\\
181	15.0229\\
182	15.0229\\
183	15.0229\\
184	15.0229\\
185	15.0229\\
186	15.0229\\
187	15.0229\\
188	15.0229\\
189	15.0229\\
190	15.0229\\
191	15.0229\\
192	15.0229\\
193	15.0229\\
194	15.0229\\
195	15.0229\\
196	15.0229\\
197	15.0229\\
198	15.0229\\
199	15.0229\\
200	15.0229\\
201	15.0229\\
202	15.0229\\
203	15.0229\\
204	15.0229\\
205	15.0229\\
206	15.0229\\
207	15.0229\\
208	15.0229\\
209	15.0229\\
210	15.0229\\
211	15.0229\\
212	15.0229\\
213	15.0229\\
214	15.0229\\
215	15.0229\\
216	15.0229\\
217	15.0229\\
218	15.0229\\
219	15.0229\\
220	15.0229\\
221	15.0229\\
222	15.0229\\
223	15.0229\\
224	15.0229\\
225	15.0229\\
226	15.0229\\
227	15.0229\\
228	15.0229\\
229	15.0229\\
230	15.0229\\
231	15.0229\\
232	15.0229\\
233	15.0229\\
234	15.0229\\
235	15.0229\\
236	15.0229\\
237	15.0229\\
238	15.0229\\
239	15.0229\\
240	15.0229\\
241	15.0229\\
242	15.0229\\
243	15.0229\\
244	15.0229\\
245	15.0229\\
246	15.0229\\
247	15.0229\\
};
\end{axis}

\begin{axis}[%
width=2.826in,
height=0.607in,
at={(0.474in,0.324in)},
scale only axis,
xmin=0,
xmax=248,
xlabel style={font=\color{white!15!black}},
xlabel={Time [s]},
ymin=0,
ymax=30,
ytick={ 0, 10, 20, 30},
ylabel style={font=\color{white!15!black}},
ylabel={Current [A]},
axis background/.style={fill=white},
axis x line*=bottom,
axis y line*=left,
xmajorgrids,
ymajorgrids,
legend style={legend columns=2, legend cell align=left, align=left, draw=white!15!black}
]
\addplot [color=blue]
  table[row sep=crcr]{%
0	11.7807\\
1	12.569\\
2	21.9306\\
3	15.3925\\
4	13.7862\\
5	13.2892\\
6	15.1801\\
7	13.2757\\
8	11.823\\
9	11.3207\\
10	12.2327\\
11	12.1757\\
12	12.7019\\
13	13.6965\\
14	13.7047\\
15	13.8256\\
16	13.8966\\
17	13.5281\\
18	12.9153\\
19	12.5251\\
20	10.9567\\
21	11.3126\\
22	10.8566\\
23	12.6905\\
24	12.6979\\
25	12.7159\\
26	12.7169\\
27	12.4607\\
28	13.8107\\
29	12.6329\\
30	12.6325\\
31	12.6331\\
32	12.6343\\
33	12.6357\\
34	12.5385\\
35	18.3502\\
36	12.4867\\
37	12.527\\
38	12.5099\\
39	12.4433\\
40	10.5263\\
41	12.8144\\
42	13.3684\\
43	13.5382\\
44	13.7103\\
45	13.8814\\
46	13.2747\\
47	12.8821\\
48	12.6133\\
49	12.9729\\
50	13.1612\\
51	13.1641\\
52	11.3398\\
53	6.04727\\
54	11.058\\
55	19.9657\\
56	12.5437\\
57	12.5362\\
58	12.5328\\
59	12.5328\\
60	12.534\\
61	12.5476\\
62	12.5477\\
63	12.8086\\
64	12.9063\\
65	13.3542\\
66	13.984\\
67	13.9907\\
68	13.9954\\
69	13.2624\\
70	12.4248\\
71	12.658\\
72	12.9513\\
73	12.9534\\
74	12.9555\\
75	12.9666\\
76	12.656\\
77	12.8133\\
78	13.0885\\
79	13.0891\\
80	13.0922\\
81	13.0946\\
82	13.1186\\
83	13.017\\
84	12.9713\\
85	12.9966\\
86	12.9984\\
87	12.9897\\
88	12.9625\\
89	12.9258\\
90	12.9089\\
91	12.9105\\
92	12.9123\\
93	12.9141\\
94	12.916\\
95	12.9171\\
96	12.9218\\
97	12.9147\\
98	12.9482\\
99	12.7597\\
100	12.5252\\
101	12.5241\\
102	12.5248\\
103	12.526\\
104	12.5978\\
105	12.6845\\
106	12.7031\\
107	12.7071\\
108	12.7092\\
109	12.7709\\
110	12.7657\\
111	12.9483\\
112	13.3534\\
113	13.4165\\
114	13.4202\\
115	13.4232\\
116	13.4257\\
117	13.0275\\
118	12.7505\\
119	13.2023\\
120	13.5932\\
121	13.9209\\
122	13.9259\\
123	13.4793\\
124	13.1147\\
125	13.1087\\
126	13.1686\\
127	13.2344\\
128	13.2344\\
129	13.0883\\
130	12.9872\\
131	12.9882\\
132	12.9896\\
133	12.9912\\
134	12.9929\\
135	12.9947\\
136	12.9966\\
137	12.9985\\
138	13.0004\\
139	13.0023\\
140	13.0042\\
141	13.0061\\
142	12.9896\\
143	12.9301\\
144	12.4194\\
145	12.2348\\
146	12.9676\\
147	13.5882\\
148	13.7941\\
149	13.9993\\
150	13.6161\\
151	12.9527\\
152	12.9906\\
153	13.2484\\
154	13.3814\\
155	13.5656\\
156	13.568\\
157	13.5708\\
158	13.5732\\
159	13.3283\\
160	13.1824\\
161	13.2443\\
162	13.3087\\
163	13.3107\\
164	13.1725\\
165	13.0627\\
166	13.0562\\
167	13.0575\\
168	13.0591\\
169	13.0609\\
170	13.0627\\
171	13.0646\\
172	13.0664\\
173	13.0684\\
174	13.0703\\
175	13.0722\\
176	13.0741\\
177	13.076\\
178	12.7882\\
179	12.3417\\
180	12.2659\\
181	12.2656\\
182	12.2662\\
183	12.2674\\
184	12.3864\\
185	12.9443\\
186	13.1354\\
187	13.14\\
188	13.1434\\
189	13.1462\\
190	13.1486\\
191	13.1508\\
192	13.1529\\
193	13.1549\\
194	13.1569\\
195	13.1589\\
196	13.1608\\
197	13.1628\\
198	13.1647\\
199	13.1667\\
200	13.1686\\
201	13.1701\\
202	13.1479\\
203	13.126\\
204	13.1277\\
205	13.1296\\
206	13.1314\\
207	13.1333\\
208	13.1353\\
209	13.1372\\
210	13.1391\\
211	13.141\\
212	13.143\\
213	13.1449\\
214	13.1469\\
215	13.1488\\
216	13.1507\\
217	13.1527\\
218	13.1546\\
219	13.1565\\
220	12.9574\\
221	12.9829\\
222	12.9843\\
223	13.0146\\
224	13.0005\\
225	13.1464\\
226	13.1489\\
227	13.1513\\
228	13.1534\\
229	13.1555\\
230	13.1575\\
231	13.1595\\
232	13.1614\\
233	13.1634\\
234	13.1653\\
235	13.1673\\
236	11.2402\\
237	7.80127\\
238	12.2863\\
239	12.2653\\
240	12.3672\\
241	12.3724\\
242	12.3762\\
243	12.3791\\
244	12.3816\\
245	12.3838\\
246	12.3858\\
247	0\\
};
\addlegendentry{UAS 1}

\addplot [color=red]
  table[row sep=crcr]{%
0	12.9964\\
1	13.3306\\
2	21.3654\\
3	18.7146\\
4	16.7706\\
5	15.3427\\
6	13.9295\\
7	14.324\\
8	11.1866\\
9	14.1579\\
10	14.164\\
11	13.8759\\
12	13.0168\\
13	13.6299\\
14	13.6317\\
15	13.6333\\
16	13.7455\\
17	14.2038\\
18	14.4417\\
19	14.5236\\
20	14.7862\\
21	14.79\\
22	14.7926\\
23	14.6295\\
24	14.2629\\
25	14.1617\\
26	14.2266\\
27	14.2981\\
28	14.2789\\
29	14.1302\\
30	14.0413\\
31	7.45141\\
32	14.0073\\
33	14.0229\\
34	14.0323\\
35	17.0555\\
36	14.058\\
37	14.0533\\
38	14.0548\\
39	14.0431\\
40	14.0823\\
41	13.8461\\
42	13.5512\\
43	13.5467\\
44	13.5465\\
45	13.8588\\
46	14.1803\\
47	14.2397\\
48	14.3032\\
49	14.2554\\
50	14.1277\\
51	14.0649\\
52	14.0655\\
53	14.0663\\
54	14.0673\\
55	14.0684\\
56	14.0695\\
57	14.0707\\
58	14.0719\\
59	14.073\\
60	14.0742\\
61	14.0754\\
62	14.0766\\
63	14.0463\\
64	14.0096\\
65	13.4942\\
66	13.2641\\
67	14.0534\\
68	14.7035\\
69	14.7539\\
70	14.8026\\
71	14.7917\\
72	14.1802\\
73	13.3707\\
74	13.4241\\
75	14.1416\\
76	15.1351\\
77	15.2552\\
78	15.2607\\
79	15.1773\\
80	14.6485\\
81	14.1646\\
82	14.1276\\
83	14.1053\\
84	14.1051\\
85	14.1054\\
86	14.1061\\
87	14.107\\
88	14.1074\\
89	14.1117\\
90	14.1017\\
91	14.1419\\
92	13.9062\\
93	13.6117\\
94	13.6073\\
95	13.6072\\
96	13.9196\\
97	14.2458\\
98	14.2494\\
99	14.252\\
100	14.254\\
101	14.2368\\
102	14.1726\\
103	14.1265\\
104	14.1273\\
105	13.9957\\
106	13.4475\\
107	13.2887\\
108	13.4863\\
109	14.1193\\
110	14.4282\\
111	14.4332\\
112	14.4363\\
113	14.4386\\
114	14.377\\
115	14.015\\
116	13.767\\
117	13.7665\\
118	13.7689\\
119	13.8429\\
120	13.8537\\
121	13.4244\\
122	13.3044\\
123	13.9974\\
124	15.0023\\
125	15.3095\\
126	15.3162\\
127	15.3205\\
128	14.8085\\
129	14.1691\\
130	13.8626\\
131	11.5218\\
132	12.1602\\
133	13.2411\\
134	13.7691\\
135	13.99\\
136	13.8232\\
137	13.8979\\
138	13.9949\\
139	13.9972\\
140	13.9855\\
141	13.9767\\
142	13.6708\\
143	13.3272\\
144	13.4838\\
145	13.4949\\
146	13.4904\\
147	0\\
148	0\\
149	0\\
150	0\\
151	0\\
152	0\\
153	0\\
154	0\\
155	0\\
156	0\\
157	0\\
158	0\\
159	0\\
160	0\\
161	0\\
162	0\\
163	0\\
164	0\\
165	0\\
166	0\\
167	0\\
168	0\\
169	0\\
170	0\\
171	0\\
172	0\\
173	0\\
174	0\\
175	0\\
176	0\\
177	0\\
178	0\\
179	0\\
180	0\\
181	0\\
182	0\\
183	0\\
184	0\\
185	0\\
186	0\\
187	0\\
188	0\\
189	0\\
190	0\\
191	0\\
192	0\\
193	0\\
194	0\\
195	0\\
196	0\\
197	0\\
198	0\\
199	0\\
200	0\\
201	0\\
202	0\\
203	0\\
204	0\\
205	0\\
206	0\\
207	0\\
208	0\\
209	0\\
210	0\\
211	0\\
212	0\\
213	0\\
214	0\\
215	0\\
216	0\\
217	0\\
218	0\\
219	0\\
220	0\\
221	0\\
222	0\\
223	0\\
224	0\\
225	0\\
226	0\\
227	0\\
228	0\\
229	0\\
230	0\\
231	0\\
232	0\\
233	0\\
234	0\\
235	0\\
236	0\\
237	0\\
238	0\\
239	0\\
240	0\\
241	0\\
242	0\\
243	0\\
244	0\\
245	0\\
246	0\\
247	0\\
};
\addlegendentry{UAS 2}

\end{axis}
\end{tikzpicture}%

%% file: paper.bbl
\begin{thebibliography}{49}
\providecommand{\natexlab}[1]{#1}
\providecommand{\url}[1]{\texttt{#1}}
\expandafter\ifx\csname urlstyle\endcsname\relax
  \providecommand{\doi}[1]{doi: #1}\else
  \providecommand{\doi}{doi: \begingroup \urlstyle{rm}\Url}\fi

\bibitem[Aurambout et~al.(2019)Aurambout, Gkoumas, and Ciuffo]{Aurambout(2019)}
J.-P. Aurambout, K.~Gkoumas, and B.~Ciuffo.
\newblock Last mile delivery by drones: An estimation of viable market
  potential and access to citizens across european cities.
\newblock \emph{European Transport Research Review}, 11:\penalty0 30, 2019.

\bibitem[Di~Franco and Buttazzo(2015)]{DiFranco(2015)}
C.~Di~Franco and G.~Buttazzo.
\newblock Energy-aware coverage path planning of uavs.
\newblock In \emph{2015 IEEE International Conference on Autonomous Robot
  Systems and Competitions}, pages 111--117, 2015.

\bibitem[Dijkstra(1959)]{Dijkstra1959}
E.~Dijkstra.
\newblock A note on two problems in connection with graphs.
\newblock \emph{Numerische Mathematik}, 1, 1959.

\bibitem[Driessens and Pounds(2015)]{Driessens(2015)}
S.~Driessens and P.~Pounds.
\newblock The triangular quadrotor: A more efficient quadrotor configuration.
\newblock \emph{IEEE Transactions on Robotics}, 31\penalty0 (6):\penalty0
  1517--1526, 2015.

\bibitem[Elsayed and Findeisen(2023)]{elsayed2023generic}
B.~Elsayed and R.~Findeisen.
\newblock Generic motion primitives-based safe motion planner under uncertainty
  for autonomous navigation in cluttered environments.
\newblock In \emph{2023 XXIX International Conference on Information,
  Communication and Automation Technologies (ICAT)}, pages 1--6. IEEE, 2023.

\bibitem[Fouad et~al.(2017)Fouad, Rizoug, Bouhali, and Hamerlain]{Fouad(2017)}
Y.~Fouad, N.~Rizoug, O.~Bouhali, and M.~Hamerlain.
\newblock Optimization of energy consumption for quadrotor uav.
\newblock In \emph{International Micro Air Vehicle Conference and Flight
  Competition (IMAV)}, 2017.

\bibitem[Gasche et~al.(2025)Gasche, Kallies, Himmel, and
  Findeisen]{Gasche2024_ECM}
S.~Gasche, C.~Kallies, A.~Himmel, and R.~Findeisen.
\newblock A modular energy aware framework for multicopter modeling in control
  and planning applications.
\newblock \emph{arXiv.org}, 2025.
\newblock (preprint).

\bibitem[Grüne and Pannek(2017)]{Gruene(2017)}
L.~Grüne and J.~Pannek.
\newblock \emph{Nonlinear Model Predictive Control}.
\newblock Communications and Control Engineering. Springer, 2017.

\bibitem[{Gurobi Optimization, LLC}(2023)]{Gurobi(2023)}
{Gurobi Optimization, LLC}.
\newblock {Gurobi Optimizer Reference Manual}, 2023.
\newblock URL \url{https://www.gurobi.com}.

\bibitem[Hagag et~al.(2024)Hagag, Gasche, J{\"a}ger, and
  Kallies]{hagag2024energy}
N.~Hagag, S.~Gasche, F.~J{\"a}ger, and C.~Kallies.
\newblock {Energy Demand Analysis for eVTOLs in Cluttered and Dynamic
  Environments based on Adaptive Trajectory Prediction}.
\newblock In \emph{2024 Integrated Communications, Navigation and Surveillance
  Conference (ICNS)}, pages 1--15. IEEE, 2024.

\bibitem[Hart et~al.(1968)Hart, Nilsson, and Raphael]{Hart1968}
P.~Hart, N.~Nilsson, and B.~Raphael.
\newblock A formal basis for the heuristic determination of minimum cost paths.
\newblock \emph{IEEE Transactions on Systems Science and Cybernetics},
  4\penalty0 (2):\penalty0 100--107, 1968.

\bibitem[Hildmann and Kovacs(2019)]{Hildmann(2019)}
H.~Hildmann and E.~Kovacs.
\newblock Review: Using unmanned aerial vehicles (uavs) as mobile sensing
  platforms (msps) for disaster response, civil security and public safety.
\newblock \emph{Drones}, 3\penalty0 (3):\penalty0 59, 2019.

\bibitem[{Holybro}(15.09.2022)]{Holybro(2022)}
{Holybro}, 15.09.2022.
\newblock URL \url{http://www.holybro.com}.

\bibitem[Ibrahim(2020)]{Ibrahim(2020)}
M.~Ibrahim.
\newblock \emph{Real-time Moving-horizon Planning and Control of Aerial Systems
  Under Uncertainties}.
\newblock PhD thesis, Otto-von-Guericke-Universit\"at Magdeburg, 2020.

\bibitem[Kallies et~al.(2024)Kallies, Gasche, and
  Kar{\'a}sek]{kallies2024multi}
C.~Kallies, S.~Gasche, and R.~Kar{\'a}sek.
\newblock {Multi-Agent Cooperative Path Planning via Model Predictive Control}.
\newblock In \emph{2024 Integrated Communications, Navigation and Surveillance
  Conference (ICNS)}, pages 1--7. IEEE, 2024.

\bibitem[Karydis and Kumar(2017)]{Karydis(2017)}
K.~Karydis and V.~Kumar.
\newblock Energetics in robotic flight at small scales.
\newblock \emph{Interface Focus}, 7\penalty0 (1):\penalty0 20160088, 2017.

\bibitem[Kavraki et~al.(1996)Kavraki, Svestka, Latombe, and
  Overmars]{Kavraki1996}
L.~Kavraki, P.~Svestka, J.-C. Latombe, and M.~Overmars.
\newblock Probabilistic roadmaps for path planning in high-dimensional
  configuration spaces.
\newblock \emph{IEEE Transactions on Robotics and Automation}, 12\penalty0
  (4):\penalty0 566--580, 1996.

\bibitem[Kingston et~al.(2018)Kingston, Moll, and Kavraki]{kingston2018}
Z.~Kingston, M.~Moll, and L.~E. Kavraki.
\newblock Sampling-based methods for motion planning with constraints.
\newblock \emph{Annual review of control, robotics, and autonomous systems},
  1\penalty0 (1):\penalty0 159--185, 2018.

\bibitem[K{\"o}gel et~al.(2023)K{\"o}gel, Ibrahim, Kallies, and
  Findeisen]{kogel2023safe}
M.~K{\"o}gel, M.~Ibrahim, C.~Kallies, and R.~Findeisen.
\newblock {Safe Hierarchical Model Predictive Control and Planning for
  Autonomous Systems}.
\newblock \emph{International Journal of Robust and Nonlinear Control}, 2023.

\bibitem[Kong et~al.(2010)Kong, Kuhn, and Rustem]{kong2010cutting}
F.~W. Kong, D.~Kuhn, and B.~Rustem.
\newblock A cutting-plane method for mixed-logical semidefinite programs with
  an application to multi-vehicle robust path planning.
\newblock In \emph{49th IEEE Conference on Decision and Control (CDC)}, pages
  1360--1365. IEEE, 2010.

\bibitem[Korolkov et~al.(2018)Korolkov, Pustovalov1, Tikhomirov, Telminov, and
  Kurakov]{Korolkov(2018)}
V.~Korolkov, A.~Pustovalov1, A.~Tikhomirov, A.~Telminov, and S.~Kurakov.
\newblock Autonomous weather stations for unmanned aerial vehicles. preliminary
  results of measurements of meteorological profiles.
\newblock \emph{IOP Conference Series: Earth and Environmental Science},
  211:\penalty0 012069, 2018.

\bibitem[Kreciglowa et~al.(2017)Kreciglowa, Karydis, and
  Kumar]{Kreciglowa(2017)}
N.~Kreciglowa, K.~Karydis, and V.~Kumar.
\newblock Energy efficiency of trajectory generation methods for stop-and-go
  aerial robot navigation.
\newblock In \emph{2017 International Conference on Unmanned Aircraft Systems
  (ICUAS)}, pages 656--662, 2017.

\bibitem[Kruskal(1956)]{Kruskal1956}
J.~Kruskal.
\newblock On the shortest spanning subtree of a graph and the traveling
  salesman problem.
\newblock \emph{Proceedings of the American Mathematical Society}, 7\penalty0
  (1):\penalty0 48--50, 1956.

\bibitem[Kuffner and LaValle(2000)]{Kuffner2000}
J.~Kuffner and S.~LaValle.
\newblock Rrt-connect: An efficient approach to single-query path planning.
\newblock In \emph{Proceedings 2000 ICRA. Millennium Conference. IEEE
  International Conference on Robotics and Automation. Symposia Proceedings},
  volume~2, pages 995--1001, 2000.

\bibitem[Lee et~al.(2010)Lee, Baek, Oh, and Choi]{Lee2010}
T.-K. Lee, S.-H. Baek, S.-Y. Oh, and Y.-H. Choi.
\newblock Complete coverage algorithm-based on linked smooth spiral paths for
  mobile robots.
\newblock In \emph{2010 11th International Conference on Control Automation
  Robotics \& Vision}, pages 609--614, 2010.

\bibitem[Li et~al.(2022)Li, Jia, Gong, and Guo]{Li(2022)}
M.~Li, G.~Jia, S.~Gong, and R.~Guo.
\newblock Energy consumption model of bldc quadrotor uavs for mobile
  communication trajectory planning.
\newblock \emph{IEEE Wireless Communications Letters}, 2022.

\bibitem[Lu et~al.(2018)Lu, Chen, Zhai, Chen, and Zhao]{Lu(2018)}
H.~Lu, K.~Chen, X.~Zhai, B.~Chen, and Y.~Zhao.
\newblock Tradeoff between duration and energy optimization for speed control
  of quadrotor unmanned aerial vehicle.
\newblock In \emph{2018 IEEE Symposium on Product Compliance Engineering - Asia
  (ISPCE-CN)}, pages 1--5, 2018.

\bibitem[Luo and Yang(2008)]{Luo(2008)}
C.~Luo and Simon~X. Yang.
\newblock A bioinspired neural network for real-time concurrent map building
  and complete coverage robot navigation in unknown environments.
\newblock \emph{IEEE Transactions on Neural Networks}, 19\penalty0
  (7):\penalty0 1279--1298, 2008.

\bibitem[Mac et~al.(2016)Mac, Copot, Tran, and {De Keyser}]{Mac2016}
Thi~Thoa Mac, Cosmin Copot, Duc~Trung Tran, and Robin {De Keyser}.
\newblock Heuristic approaches in robot path planning: A survey.
\newblock \emph{Robotics and Autonomous Systems}, 86:\penalty0 13--28, 2016.

\bibitem[Malyuta et~al.(2022)Malyuta, Reynolds, Szmuk, Lew, Bonalli, and
  Açıkmeşe]{Malyuta2022}
D.~Malyuta, T.~P. Reynolds, M.~Szmuk, T.~Lew, M.~Bonalli, R.and~Pavone, and
  B.~Açıkmeşe.
\newblock Convex optimization for trajectory generation: A tutorial on
  generating dynamically feasible trajectories reliably and efficiently.
\newblock \emph{IEEE Control Systems Magazine}, 42\penalty0 (5):\penalty0
  40--113, 2022.

\bibitem[Mirzaei et~al.(2011)Mirzaei, Sharifi, Gordon, Rabbath, and
  Zhang]{Mirzaei(2011)}
M.~Mirzaei, F.~Sharifi, B.~W. Gordon, C.~A. Rabbath, and Y.~M. Zhang.
\newblock Cooperative multi-vehicle search and coverage problem in uncertain
  environments.
\newblock In \emph{2011 50th IEEE Conference on Decision and Control and
  European Control Conference}, pages 4140--4145, 2011.

\bibitem[Morbidi et~al.(2016)Morbidi, Cano, and Lara]{Morbidi(2016)}
F.~Morbidi, R.~Cano, and D.~Lara.
\newblock Minimum-energy path generation for a quadrotor uav.
\newblock In \emph{2016 IEEE International Conference on Robotics and
  Automation (ICRA)}, pages 1492--1498, 2016.

\bibitem[Morbidi et~al.(2018)Morbidi, Bicego, Ryll, and Franchi]{Morbidi(2018)}
F.~Morbidi, D.~Bicego, M.~Ryll, and A.~Franchi.
\newblock Energy-efficient trajectory generation for a hexarotor with dual-
  tilting propellers.
\newblock In \emph{2018 IEEE/RSJ International Conference on Intelligent Robots
  and Systems (IROS)}, pages 6226--6232, 2018.

\bibitem[Patle et~al.(2019)Patle, {Babu L}, Pandey, Parhi, and
  Jagadeesh]{Patle2019}
B.K. Patle, Ganesh {Babu L}, Anish Pandey, D.R.K. Parhi, and A.~Jagadeesh.
\newblock A review: On path planning strategies for navigation of mobile robot.
\newblock \emph{Defence Technology}, 15\penalty0 (4):\penalty0 582--606, 2019.

\bibitem[Popović et~al.(2024)Popović, Ott, Rückin, and
  Kochenderfer]{Popovic2024}
M.~Popović, J.~Ott, l~J. Rückin, and M.~J. Kochenderfer.
\newblock Learning-based methods for adaptive informative path planning.
\newblock \emph{Robotics and Autonomous Systems}, 179:\penalty0 104727, 2024.

\bibitem[Preparata and Shamos(1985)]{Preparata1985}
F.~P. Preparata and M.~I. Shamos.
\newblock \emph{Computational Geometry: An Introduction}.
\newblock Springer New York, 1985.

\bibitem[Prim(1957)]{Prim1957}
R.~Prim.
\newblock Shortest connection networks and some generalizations.
\newblock \emph{The Bell System Technical Journal}, 36\penalty0 (6):\penalty0
  1389--1401, 1957.

\bibitem[Quirynen et~al.(2024)Quirynen, Safaoui, and Di~Cairano]{Quirynen2024}
R.~Quirynen, S.~Safaoui, and S.~Di~Cairano.
\newblock Real-time mixed-integer quadratic programming for vehicle
  decision-making and motion planning.
\newblock \emph{IEEE Transactions on Control Systems Technology}, PP:\penalty0
  1, 01 2024.

\bibitem[Ryll et~al.(2015)Ryll, Bülthoff, and Giordano]{Ryll(2015)}
M.~Ryll, H.~Bülthoff, and P.~Giordano.
\newblock A novel overactuated quadrotor unmanned aerial vehicle: Modeling,
  control, and experimental validation.
\newblock \emph{IEEE Transactions on Control Systems Technology}, 23\penalty0
  (2):\penalty0 540--556, 2015.

\bibitem[Sampedro et~al.(2019)Sampedro, Rodriguez-Ramos, Bavle, Carrio, de~la
  Puente, and Campoy]{Sampedro(2019)}
C.~Sampedro, A.~Rodriguez-Ramos, H.~Bavle, A.~Carrio, P.~de~la Puente, and
  P.~Campoy.
\newblock A fully-autonomous aerial robot for search and rescue applications in
  indoor environments using learning-based techniques.
\newblock \emph{Journal of Intelligent \& Robotic Systems}, 95:\penalty0
  601--627, 2019.

\bibitem[Schouwenaars(2006)]{schouwenaars2006safe}
T.~Schouwenaars.
\newblock \emph{Safe Trajectory Planning of Autonomous Vehicles}.
\newblock PhD thesis, Massachusetts Institute of Technology, 2006.

\bibitem[Steich et~al.(2016)Steich, Kamel, Beardsley, Obrist, Siegwart, and
  Lachat]{Steich(2016)}
K.~Steich, M.~Kamel, P.~Beardsley, M.~Obrist, R.~Siegwart, and T.~Lachat.
\newblock Tree cavity inspection using aerial robots.
\newblock In \emph{2016 IEEE/RSJ International Conference on Intelligent Robots
  and Systems (IROS)}, pages 4856--4862, 2016.

\bibitem[Strobel(2016)]{Strobel(2016)}
A.~Strobel.
\newblock \emph{Verteilte nichtlineare modellpr{\"{a}}diktive Regelung von
  unbemannten Luftfahrzeug-Schw{\"{a}}rmen}.
\newblock PhD thesis, Technische Universit{\"a}t Darmstadt, 2016.

\bibitem[Tagliabue et~al.(2019)Tagliabue, Wu, and Mueller]{Tagliabue(2019)}
A.~Tagliabue, X.~Wu, and M.~Mueller.
\newblock Model-free online motion adaptation for optimal range and endurance
  of multicopters.
\newblock In \emph{2019 International Conference on Robotics and Automation
  (ICRA)}, pages 5650--5656, 2019.

\bibitem[{The MathWorks Inc.}(2023)]{Mathworks(2023)}
{The MathWorks Inc.}
\newblock {MATLAB version: R2023a}, 2023.
\newblock URL \url{https://www.mathworks.com}.

\bibitem[Wei and Shi(2022)]{Wei2022}
H.~Wei and Y.~Shi.
\newblock Mpc-based motion planning and control enables smarter and safer
  autonomous marine vehicles: Perspectives and a tutorial survey.
\newblock \emph{IEEE/CAA Journal of Automatica Sinica}, 09 2022.

\bibitem[Xiong et~al.(2019)Xiong, Hu, and Diao]{Xiong(2019)}
H.~Xiong, J.~Hu, and X.~Diao.
\newblock Optimize energy efficiency of quadrotors via arm rotation.
\newblock \emph{Journal of Dynamic Systems, Measurement, and Control},
  141\penalty0 (9):\penalty0 091002, 2019.

\bibitem[Yacef et~al.(2017)Yacef, Rizoug, Degaa, Bouhali, and
  Hamerlain]{Yacef(2017)}
F.~Yacef, N.~Rizoug, L.~Degaa, O.~Bouhali, and M.~Hamerlain.
\newblock Trajectory optimisation for a quadrotor helicopter considering energy
  consumption.
\newblock In \emph{2017 4th International Conference on Control, Decision and
  Information Technologies (CoDIT)}, pages 1030--1035, 2017.

\bibitem[Yacef et~al.(2020)Yacef, Rizoug, Degaa, and Hamerlain]{Yacef(2020)}
F.~Yacef, N.~Rizoug, L.~Degaa, and M.~Hamerlain.
\newblock Energy-efficiency path planning for quadrotor uav under wind
  conditions.
\newblock In \emph{2020 7th International Conference on Control, Decision and
  Information Technologies (CoDIT)}, volume~1, pages 1133--1138, 2020.

\end{thebibliography}
